\documentclass[amssymb,amsmath,prd,twocolumn,superscriptaddress,nofootinbib]{revtex4-1}

\usepackage{graphicx}
\usepackage{bm}
\usepackage{amssymb,amsmath}
\usepackage{mathrsfs}
\usepackage{latexsym}
\usepackage{color}
\usepackage[normalem]{ulem}
\usepackage{dcolumn}
\usepackage[colorlinks=true,citecolor=blue,urlcolor=blue]{hyperref}
\usepackage[usenames,dvipsnames]{xcolor}
\usepackage{xspace}
\usepackage{graphicx}
\usepackage{multirow}
\usepackage{mathtools}
\usepackage{diagbox}

\allowdisplaybreaks

\DeclareMathOperator{\tr}{tr}
\newcommand{\R}{\mathbb{R}}

\usepackage[english]{babel}
\usepackage{xspace}
\usepackage[normalem]{ulem} 
\usepackage{color,soul}

\newcommand{\new}[1]{{ #1}}

\newcommand{\quant}[1]{\left(#1\right)}

\newcommand{\deriv}[2]{\frac{d#1}{d#2}}

\begin{document}
\begin{abstract}

We reformulate the stellar structure equations in the language of dynamical systems and show that the maximum mass of stellar sequences arises from the existence of a fixed point in the relativistic regime. In an appropriate representation of the Tolman–Oppenheimer–Volkoff equations, this fixed point becomes manifest and is directly associated with the turnover of the mass–radius curve. The existence of a fixed point implies an effective reduction in dimensionality near the onset of instability, which provides a simple explanation for several equation-of-state–insensitive relations and predicts new ones. In the weakly relativistic limit, we identify a complementary universal structure shared by stellar sequences at their maximum mass, which we term the ``compressible limit,'' and derive distinct universal relations governing the maximum mass in the Newtonian and post-Newtonian regimes. Combining these theoretical results with current astrophysical constraints, we show that the J0740+6620 pulsar is unlikely to lie near the Tolman–Oppenheimer–Volkoff maximum mass unless the equation of state exhibits a strong first-order phase transition at densities just above its central density.

\end{abstract}

\title{Why Stellar Sequences Turn Over: \\
Fixed Points, Instability, and Equation-of-State Universality}

\author{Isaac Legred}
\email{ilegred2@illinois.edu}
\affiliation{Illinois Center for Advanced Studies of the Universe, Department of Physics, University of Illinois at Urbana-Champaign, \\Urbana, Illinois, 61801, USA}

\author{Nicol\'as Yunes}
\affiliation{Illinois Center for Advanced Studies of the Universe, Department of Physics, University of Illinois at Urbana-Champaign, \\Urbana, Illinois, 61801, USA}

\date{\today}
\maketitle
\section{Introduction}
\label{sec:Intro}

Understanding the structure of compact stars has long been intertwined with fundamental physics. At its core lies the problem of hydrostatic equilibrium: pressure gradients must balance gravity (or, in general relativity, pressure gradients must balance the curvature of spacetime, which is sourced by energy density and pressure), in order to sustain a static configuration. This balance determines the self-consistent internal structure of a non-rotating, spherically symmetric, self-gravitating fluid. Because the equilibrium equations couple pressure and energy density directly, their solutions are governed by the relation between them, i.e.~the equation of state (EoS). 

As one considers stars with increasingly large central densities, matter in their stellar cores is driven into increasingly extreme regimes, where the microphysics becomes uncertain and often strongly nonlinear. At sufficiently high compactness, relativistic effects become indispensable, introducing additional couplings between pressure, energy density, and geometry. The stellar structure problem thus becomes a \textit{nonlinear system} in which gravity and microphysics interact in a tightly constrained but highly nontrivial way. Understanding how global properties emerge from this interplay (and why they sometimes display unexpected ``quasi-universal'' or ``EoS-insensitive'' behavior) requires looking beyond individual solutions and toward the structure of the equations themselves.

The most extreme compact stars realized in nature are very likely neutron stars. Yet the composition and EoS of matter in their cores remain uncertain, and therefore so do their \textit{precise} macroscopic properties. For the purposes of this work, we adopt a broad definition: by ``neutron star'' we shall mean any stable, self-gravitating configuration of baryonic matter that solves the non-vacuum Einstein equations at densities exceeding those found in maximum-mass white dwarfs, i.e.~central energy densities $e_c \gtrsim 10^{10} \; {\rm{g}}\, {\rm{cm}}^{-3} \sim 5 \times 10^{-3} \; {\rm{MeV}} \, {\rm{fm}}^{-3}$. In this sense, we do not distinguish between ``conventional'' neutron stars, hybrid stars, quark stars, or strange stars unless explicitly noted. What unifies these objects is not their microscopic composition, but the structure of the equilibrium equations they obey. As we will show, viewing those equations through the lens of \textit{dynamical systems theory}~\cite{Nilsson:2000zf, Heinzle:2003ud} reveals organizing principles that transcend the details of the EoS\ and provide a unified explanation for the emergence and breakdown of quasi-universal behavior.

Stars in hydrostatic equilibrium in general relativity are governed by the Tolman–Oppenheimer–Volkoff (TOV) equations~\cite{Tolman:1939jz,Oppenheimer:1939ne}. In the limit of small compactness, $G M/(R c^2) \ll 1$ , and low pressure, $p \ll e$, these equations reduce to the familiar Newtonian equations of hydrostatic equilibrium (see, e.g.,~\cite{Kippenhahn:2012qhp}). The standard formulation of the TOV system, however, is numerically inconvenient. In this approach, one integrates the TOV equations outward in areal radius to an \textit{a priori} unknown location, where the pressure (one of the dynamical variables) vanishes and the stellar surface is reached. This introduces a free-boundary problem that is both conceptually and numerically awkward.

Lindblom reformulated the TOV equations in terms of the logarithm of an enthalpy-like variable, which serves as a natural monotonic parameter labeling the stellar interior~\cite{Lindblom:1998dp}. In this formulation the surface corresponds to a known value of the integration variable, and the central value uniquely labels the stellar model, eliminating the free-boundary difficulty. More generally, any formulation of the TOV equations that is based on a thermodynamic variable whose surface value is known in advance would suffice. The log-enthalpy variable, however, possesses additional structural advantages: it renders the equations manifestly regular at the center, provides a natural ordering of configurations by central density, and, as we will show, it reveals an underlying dynamical-systems structure that is largely hidden in the traditional radial formulation.

\begin{figure}
    \centering
    \includegraphics[width=.49\textwidth]{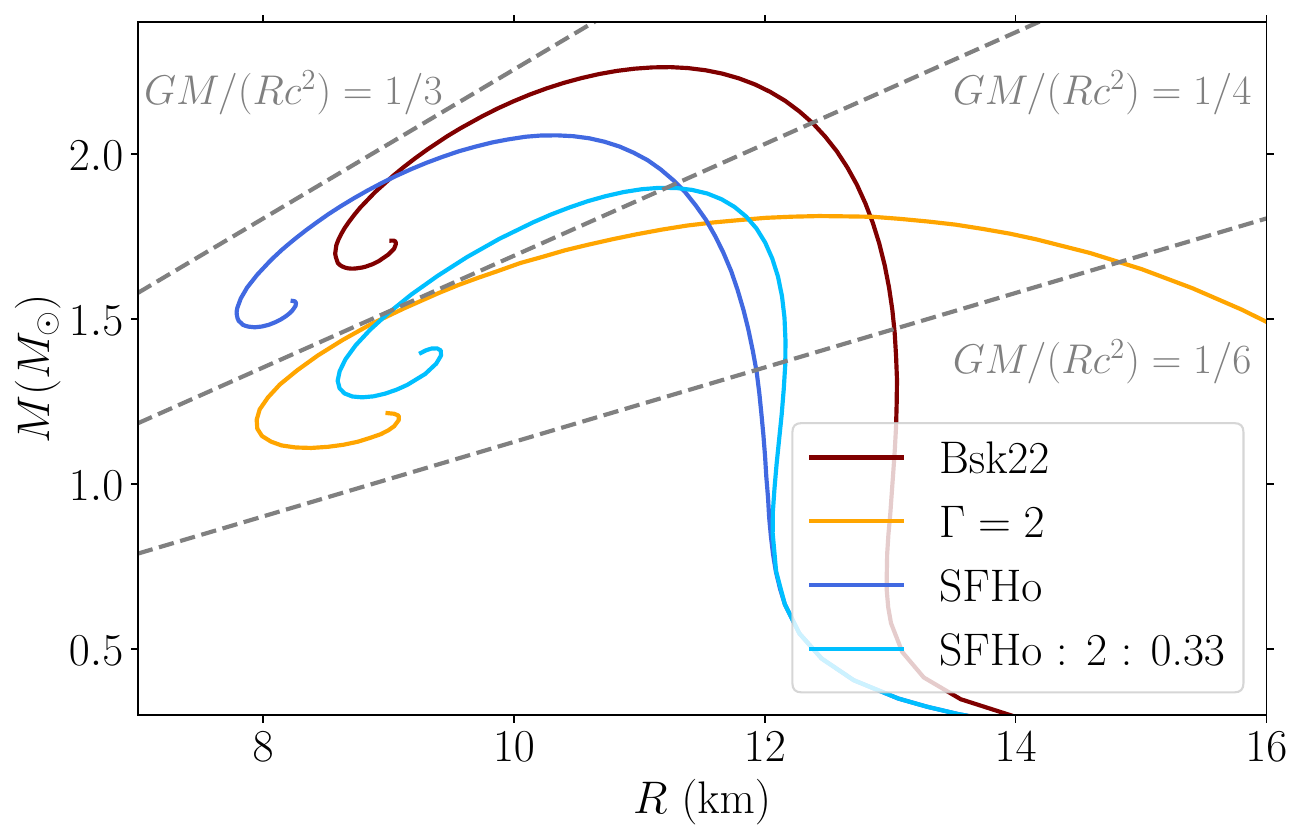}
    \caption{Spiraling behavior of the $M$--$R$ curve. Every EoS at extremely high densities has similar behavior, where it approaches a particular fixed point in $M$--$R$ space. In this paper, we discuss how this picture arises from the treatment of the stellar structure equations as a dynamical system, and how such a treatment explains certain quasi-universality. The SFHo~\cite{Steiner:2012rk} and BSk22~\cite{Pearson:2018tkr} EoSs are given a constant speed of sound $c_s^2=1$ above the densities where they become acausal. $\text{SFHo}:2:0.33$ transitions from SFHo to a $c_s^2 = .33$ constant speed of sound EoS at twice nuclear saturation density.  Gray dashed lines mark configurations of constant compactness. 
    Among other pathologies, each EoS displays regions of decreasing compactness with respect to increasing central density.
    }
    \label{fig:M-R fixed points}
\end{figure}

Some hints of this structure have been with us for decades, encoded in qualitative features of TOV solutions that were initially viewed as surprising. For example, it was noticed early on that in the ultra-relativistic, high-density regime, when the curvature scale set by the central energy density is much shorter than the stellar size, $Ge_c R^2/c^4 \gg 1$, the solutions exhibit an approximately periodic or ``spiraling'' behavior as one moves along sequences of increasing central density. Figure~\ref{fig:M-R fixed points} shows a few examples of this spiraling behavior for different EoSs. This feature was understood relatively early in the theory of neutron stars, even before the observational discovery of astrophysical neutron stars (see, e.g.,~\cite{HTWW:1965} and references therein), but it is less appreciated today. In part, this is because these spiraling solutions are violently unstable and are therefore often dismissed as astrophysically irrelevant. Nevertheless, their existence raises a more basic question: what parameters actually control the high-density behavior of the TOV equations?

Another clue, also recognized early on, is that this near-periodicity is closely tied to the appearance of a maximum mass. This connection became clear through the now well-known correspondence between turning points of the $M$--$R$ curve and changes in the number of unstable radial oscillation modes. That correspondence naturally leads to a further question. Is the underlying mechanism purely relativistic, or does an analogous organizing structure already appear in the Newtonian equations of hydrostatic equilibrium? It is tempting to attribute the existence of a maximum mass entirely to general relativity, but this is not correct. A termination of stable stellar sequences, and thus an effective maximum mass, can arise even within a Newtonian treatment once the equation of state softens sufficiently. Any satisfactory explanation of the high-density behavior of compact stars must therefore distinguish between instability driven by genuinely relativistic effects and instability that occurs while a Newtonian description is still adequate.

In this paper, we reformulate the stellar structure problem in the language of dynamical-systems theory, with the goal of isolating the pieces of the structure equations that control the approach to instability and the emergence of quasi-universal/EoS-insensitive behavior. The basic idea is to stop thinking of a stellar model as a single numerical solution. Instead, we should view an entire sequence of stars as a \textit{family of trajectories} in a low-dimensional phase space. In this picture, EoS dependence enters as a controlled ``driving'' of the flow; quasi-universal relations, then, arise when the dynamics becomes effectively insensitive to details of that driving over the interior part of the star that dominates the global observables.

We begin by working in the enthalpy-based formulation of the relativistic TOV equations and rewriting the system as a two-dimensional, non-autonomous flow in a compact set of variables. In that form, a distinguished structure becomes obvious. There is a \textit{fixed point} of the frozen system in the $(\tilde e,v)$ plane, where $\tilde e \propto r^2 e$ and $v = Gm(r)/(rc^2)$ are dimensionless versions of the energy density and the enclosed mass. The location of the fixed point is set by a small set of thermodynamic combinations, essentially the ratio of pressure to energy density, $w=p/e$, and the speed of sound squared, $c_s^2=dp/de$, evaluated in the high-density regime. The fact that a fixed-point appears and can be used to interpret the stellar structure  was first introduced by Ref.~\cite{Heinzle:2003ud}, but only under the hypothesis of a particular functional form of the EoS.\footnote{We also make a different choice for the independent variable of the dynamical system ($\ln h$, instead of a function of $r$ in~\cite{Heinzle:2003ud}), and, whereas ~\cite{Heinzle:2003ud} builds a 3-variable system of autonomous equations, we, in contrast, analyze a non-autonomous system of 2 variables.  We discuss our system setup in Sec.~\ref{sec:relativistic-mmax} and differences with previous works in Sec.~\ref{sec:discussion}.} 

We show that high-mass relativistic stars \emph{generically} evolve near this fixed point over the region where the bulk of the mass is accumulated. This produces the familiar spiral and turnover behavior of the $M$--$R$ curve and provides a dynamical-systems explanation for why the termination of relativistic stellar sequences is remarkably insensitive to EoS details, once the EoS is forced to be physical. From this viewpoint, the maximum mass is not an accident of numerics, or a coincidence of the form of the EoS. It is controlled by the geometry of the flow near the fixed point and by a few EoS-controlled scales that determine where the star exits the high-density, near-fixed-point regime. This leads to simple scaling estimates and quasi-universal relations for $M_{\max}$ in terms of high-density stiffness, and it clarifies when and why those relations should fail. 

We then ask whether the same organizing structure is present in the Newtonian equations, appropriate for describing weakly-gravitating stars. We show that periodic or spiraling behavior can indeed appear in the Newtonian system, in agreement with results of~\cite{Heinzle:2002sk}\footnote{The recasting of the Newtonian stellar structure equations into the form a dynamical system is well-established, usually related to the study of structural \emph{homology}. See, for example,~\cite{StellarStructuresPrialnik}.}, but only under special conditions. In practice, it requires tuning the EoS so that it approaches an approximately constant sound speed in the relevant density range, which is not generic in Newtonian physics. This immediately explains why the ``spiral mechanism'' that organizes relativistic termination almost never controls the end of Newtonian stellar sequences (even though fundamentally it is the same mechanism). Nonetheless, Newtonian sequences still terminate in a quasi-universal way, but for a different reason. Near the onset of instability, we find that stellar configurations develop a common internal structure that is largely independent of microphysical details. We identify this as a useful limiting case in stellar structure, dual to the incompressible limit, and we refer to it as the ``compressible limit.'' In that limit, we derive a separate quasi-universal relation for the maximum mass in terms of central quantities, and we extend the construction to the post-Newtonian regime. The outcome is a clear distinction between two mechanisms for termination, one controlled by the relativistic fixed-point geometry and one controlled by near-critical structure in the weak-field regime, together with an explanation for why both mechanisms can give similar numerical predictions in the density range relevant for astrophysical neutron stars.

These results are new and important for two reasons. First, they turn a collection of empirical observations about quasi-universality into a statement about the structure of the differential equations themselves, with explicit control parameters. Second, they provide a clean diagnostic for strong phase transitions. In the Newtonian and post-Newtonian regimes, a sufficiently strong softening can terminate sequences abruptly and drive apparent violations of the compressible-limit relation. In the highly relativistic regime, a strong transition can shift the effective fixed point and substantially alter the termination of the spiral. We illustrate the utility of this viewpoint by combining our analysis with current astrophysical constraints. In particular, we argue that J0740+6620 is unlikely to be very near the TOV maximum mass unless there is a strong first-order phase transition at densities just above its core density. Because the same high-density stiffness that controls the maximum mass also controls the tidal response of high-mass stars, the dynamical-systems picture provides a useful bridge between mass constraints and gravitational-wave-based bounds on dense-matter stiffness.

The rest of the paper presents the details behind the summary above. In Sec.~\ref{sec:ds_primer}, we briefly review the dynamical systems concepts we will use and fix notation. In Sec.~\ref{sec:relativistic-mmax}, we rewrite the enthalpy formulation of the TOV equations as a two-dimensional flow, identify the fixed point, and show how it organizes the spiral and turnover of the $M$--$R$ curve. In Sec.~\ref{sec:newtonian-falling-sound-speeds}, we turn to the Newtonian and post-Newtonian limits, explain why the spiral mechanism is non-generic, and develop the compressible limit and its associated quasi-universal relation. We extend the compressible-limit analysis to the post-Newtonian regime, where it can be understood along with neutron star observations, and compare it to the highly relativistic fixed-point mechanism. We conclude in Sec.~\ref{sec:discussion} with implications for phase transitions and with an application to current constraints, and we collect technical material and supporting analyses in the Appendices.

\section{Dynamical systems primer}\label{sec:ds_primer}

The main technical goal of this paper is to re-express the stellar structure problem in a way that makes its qualitative behavior (as one varies the central enthalpy) as transparent as possible. The appropriate language for that is dynamical-systems theory. In this language, the TOV equations define a flow on a low-dimensional ``phase space,'' once one chooses convenient state variables and an evolution parameter. Because most readers may be somewhat unfamiliar with dynamical systems theory and some mathematical theorems we will make use of, we provide here a brief review of the concepts that we will repeatedly invoke in later sections. Those readers that are already experts in dynamical systems theory may wish to skip to the next section. 

Throughout this primer, we will use the generic notation
\begin{equation}
\frac{d\mathbf{x}}{d\tau}=\mathbf{F}(\mathbf{x},\tau)\,,
\label{eq:ds_generic}
\end{equation}
where $\mathbf{x}(\tau)\in\mathbb{R}^n$ is a vector of state variables and $\tau$ is an evolution parameter. In the next sections, we will take $\tau=\ln h$ and $\mathbf{x}=(\tilde e,v)$, where $h$ is the enthalpy per unit rest mass, $\tilde e\equiv 4\pi r^2 Ge/c^4$ is a dimensionless energy-density variable, and $v\equiv Gm(r)/(rc^2)$ is the compactness of the enclosed mass $m(r)$ at radius $r$. We will reserve $M:=m(R)$ for the total gravitational mass, and $R$ for the areal radius of the star.

\subsection{Phase space, flows, and what ``autonomous'' really means}\label{sec:ds_phase_space}

If $\mathbf{F}$ in Eq.~\eqref{eq:ds_generic} has no explicit dependence on $\tau$, 
\begin{equation}
\frac{d\mathbf{x}}{d\tau}=\mathbf{F}(\mathbf{x})\,,
\label{eq:autonomous}
\end{equation}
we say the system is \emph{autonomous}. Geometrically, $\mathbf{F}$ is a vector field on the state space and solutions are its integral curves.
One can think of Eq.~\eqref{eq:autonomous} as defining a family of maps (a ``flow'') $\phi_{\Delta\tau}$ that takes an initial condition
$\mathbf{x}(\tau_0)$ to its evolved value $\mathbf{x}(\tau_0+\Delta\tau)$.

If $\mathbf{F}$ depends explicitly on $\tau$, the system is \emph{non-autonomous}. In our application, this happens because the microphysics
enters through functions that depend on the EoS, such as
\begin{equation}
w(\tau)\equiv \frac{p}{e}\,,\qquad c_s^2(\tau)\equiv \frac{dp}{de}\,,
\label{eq:w_cs2_def}
\end{equation}
which vary with $\tau=\ln h$ as the density changes within the star. Non-autonomous systems still define unique trajectories for given initial data, but many of the simplest geometric statements (e.g.\ fixed
points that are time-independent) must be carefully.
A decent approximation that we will repeatedly employ is the \emph{frozen-time} (or \emph{quasi-autonomous}) picture: for a given $\tau$, we temporarily treat the parameters $w(\tau),c_s^2(\tau)$ as constants, analyze the resulting autonomous system, and then
track how that analysis evolves as $\tau$ changes. We will see that this is a good approximation near the fix point of our problem. 

\subsection{Fixed points, linearization, and the trace--determinant plane}\label{sec:ds_fixed_points}

For the autonomous system of Eq.~\eqref{eq:autonomous}, a \emph{fixed point} (or equilibrium point) $\mathbf{x}_*$ satisfies
\begin{equation}
\mathbf{F}(\mathbf{x}_*)=\mathbf{0}\,.
\label{eq:fixed_point_def}
\end{equation}
Near $\mathbf{x}=\mathbf{x}_*$, we use perturbation theory and write $\mathbf{x}=\mathbf{x}_*+\boldsymbol{\delta x}$ to obtain the linearized system
\begin{equation}
\frac{d\,\boldsymbol{\delta x}}{d\tau}
=
\mathbf{J}_0\,\boldsymbol{\delta x}
+\mathcal{O}(|\boldsymbol{\delta x}|^2)\,,
\qquad
(\mathbf{J}_0)_{ij}:=\left.\frac{\partial F_i}{\partial x_j}\right|_{\mathbf{x}=\mathbf{x}_*}\,.
\label{eq:linearization}
\end{equation}
Thus, the local behavior is governed by the eigenvalues of the Jacobian $\mathbf{J}_0$, a square matrix of size $n \times n$ for a state vector $\mathbf{x}$ of length $n$.

In two dimensions, as will be applicable to us in the next sections, this all becomes especially concrete.
Let $\lambda_\pm$ be the eigenvalues of the $2\times 2$ Jacobian, which can be written in terms of the latter's trace and determinant via 
\begin{equation}
\lambda_\pm
=
\frac{\mathrm{tr}\,\mathbf{J}_0}{2}
\pm
\sqrt{\left(\frac{\mathrm{tr}\,\mathbf{J}_0}{2}\right)^2-\det \mathbf{J}_0}\,.
\label{eq:eigs_tr_det}
\end{equation}
The qualitative behavior of the solution near the fixed point is then classified by the signs of $(\mathrm{tr}\,\mathbf{J}_0,\det \mathbf{J}_0)$.
In particular, if $\det \mathbf{J}_0>0$ and $\left(\mathrm{tr}\,\mathbf{J}_0\right)^2<4\det \mathbf{J}_0$, then
\begin{equation}
\lambda_\pm=\sigma\pm i\omega\,,\qquad
\sigma=\frac{\mathrm{tr}\,\mathbf{J}_0}{2}\,,\qquad
\omega=\sqrt{\det \mathbf{J}_0-\frac{(\mathrm{tr}\,\mathbf{J}_0)^2}{4}}\,,
\label{eq:sigma_omega_def}
\end{equation}
\new{with $\sigma, \omega \in \R$} and $\mathbf{x}_*$ is a \emph{focus}: trajectories spiral inwards if $\sigma<0$ and spiral outwards if $\sigma>0$. This focus structure, transplanted into the TOV variables, will be the geometric engine behind the turnover of the $M$--$R$ curve and the appearance of spirals at high central density.

A caution that matters specifically for our application is the \emph{direction of integration}.We integrate the stellar structure equations from the center ($h=h_c$) to the surface ($h\to 1$), so $\tau=\ln h$ decreases as we move outward (and becomes 0 at the surface). Equivalently, if one defines an outward evolution parameter $s:=\ln h_c-\ln h$, so that $d/ds=-d/d\tau$, a fixed point that is \emph{unstable} in forward $\tau$ (e.g.\ $\sigma>0$) can act as an \emph{attractor} when one evolves outward in the star (forward $s$). We will use this ``backwards-in-$\ln h$'' viewpoint repeatedly in the next sections, and we will try to be as explicit as possible about it whenever the stability language could be confusing.

\subsection{A harmonic-oscillator-level example}\label{sec:ds_ho_example}

The simplest example we can think of that captures almost everything we need is (unsurprisingly) the damped harmonic oscillator:
\begin{equation}
\ddot q + 2\gamma \dot q + \Omega^2 q = 0\,,
\label{eq:damped_ho}
\end{equation}
with damping rate $\gamma$ and natural frequency $\Omega$.
Let us introduce the first-order variables $x_1=q$ and $x_2=\dot q$, so that
\begin{align}
\frac{d}{dt} \mathbf{x} &= \mathbf{A} \mathbf{x}\,,
\nonumber \\
\frac{d}{dt}\begin{pmatrix}x_1\\ x_2\end{pmatrix}
&=
\begin{pmatrix}
0 & 1\\
-\Omega^2 & -2\gamma
\end{pmatrix}
\begin{pmatrix}x_1\\ x_2\end{pmatrix}\,.
\label{eq:ho_first_order}
\end{align}
The origin $\mathbf{x}_* = (x_{1,*},x_{2,*})=(0,0)$ is a fixed point because $\mathbf{A} \mathbf{x}$ obviously vanishes there. The eigenvalues of $\mathbf{A}$ are
\begin{equation}
\lambda_\pm=-\gamma\pm \sqrt{\gamma^2-\Omega^2}\,.
\label{eq:ho_eigs}
\end{equation}
If $\gamma<\Omega$, then $\lambda_\pm=-\gamma\pm i\sqrt{\Omega^2-\gamma^2}$ and the origin is a \emph{stable focus}: phase-space trajectories spiral into the origin with an envelope $|x|\propto e^{-\gamma t}$ while rotating with angular speed $\omega_d=\sqrt{\Omega^2-\gamma^2}$.

Two features of this example can be directly imported to our stellar problem of the next sections. First, spiraling is controlled by the pair $(\sigma,\omega)$. The real part $\sigma$ sets the decay/growth per unit ``time,'' while the imaginary part $\omega$ sets the phase advance. Second, a discrete ``return map'' compresses the long-time behavior into one number. The amplitude at successive peaks satisfies $A_{n+1}=A_n e^{-\gamma T}$, where $T\simeq 2\pi/\omega_d$ is the oscillation period. Thus, the spiral can be characterized by a multiplier $\mu:=e^{-\gamma T}$ without solving the ODE in closed form. In our stellar application, an analogous multiplier will quantify how quickly trajectories approach the quasi-fixed point as one integrates from
the core to the surface. Finally, note the role of time reversal. If one evolves Eq.~\eqref{eq:ho_first_order} backward in $t$, the stable focus becomes an unstable focus: the spiral runs outward rather than inward. 

\subsection{Non-autonomous systems and frozen-time equilibria}\label{sec:ds_nonautonomous}

Consider now the non-autonomous system of Eq.~\eqref{eq:ds_generic}. A time-dependent equilibrium in the strict sense is generally not available because $\mathbf{F}(\mathbf{x},\tau)$ changes with $\tau$. Nevertheless, for each fixed value of $\tau=\tau_{\rm fr}$ one may define the \emph{frozen} fixed point $\mathbf{x}_*(\tau_{\rm fr})$ by
\begin{equation}
\mathbf{F}(\mathbf{x}_*(\tau_{\rm fr}),\tau_{\rm fr})=\mathbf{0}\,,
\label{eq:frozen_fp}
\end{equation}
and study the corresponding Jacobian $\mathbf{J}_0(\tau_{\rm fr}):=\partial_{\mathbf{x}}\mathbf{F}|_{\mathbf{x}=\mathbf{x}_*(\tau_{\rm fr})}$. In the limit that the parameters of the system drift slowly or adiabatically with $\tau$, solutions can ``track'' the moving equilibrium and inherit the same local spiral structure as the frozen system. This treatment is analogous to that of osculating orbits in celestial dynamics, or osculating geodesics in extreme mass-ratio modeling (see e.g.~\cite{PoissonWill2014}).  

A convenient way to see this is to define $\mathbf{y}:=\mathbf{x}-\mathbf{x}_*(\tau)$ and expand in small $|\mathbf{y}|$ to obtain
\begin{equation}
\frac{d\mathbf{y}}{d\tau}
=
\mathbf{J}_0(\tau)\,\mathbf{y}
-\frac{d\mathbf{x}_*}{d\tau}
+\mathcal{O}(|\mathbf{y}|^2)\,.
\label{eq:tracking_equation}
\end{equation}
The additional term $-d\mathbf{x}_*/d\tau$ is a forcing that measures how quickly the state solution drifts from the equilibrium point. If the drift is slow compared to the local linear rates set by the eigenvalues of $\mathbf{J}_0(\tau)$, then $\mathbf{y}$ remains small and the trajectory remains close to the frozen equilibrium.

\section{The maximum mass of relativistic stars: a dynamical systems approach}
\label{sec:relativistic-mmax}

The maximum mass of a stellar configuration can provide a global diagnostic of high-density microphysics. In this section, we recast the TOV system as a two-dimensional dynamical system in $(\tilde{e},v)$ driven by the enthalpy $\ln{h}$. This makes a quasi-fixed-point structure explicit and gives a simple explanation for the generic spiral/turnover behavior of the $M$--$R$ curve, as well as a compact scaling estimate for the maximum mass $M_{\rm max}$.

\subsection{Reinterpreting the TOV equations}
\label{sec:lindblom-newtonian}

Let us start our analysis from the Lindblom formulation of the TOV equations~\cite{Lindblom:2010bb}, but in slightly different variables. Let us define
\begin{align}
\label{eq:v-def}
u \equiv r^2, \qquad v \equiv \frac{G m(r)}{r c^2}\,,
\end{align} 
where $r$ is the (areal) radial coordinates of the usual spherically-symmetric spacetime metric (Schwarzschild-like, OV coordinates~\cite{Oppenheimer:1939ne}) and $m(r)$ is the enclosed mass function. Let us further define the following thermodynamic variables 
\begin{equation}
\label{eq:thermo-vars}
\left.
\begin{aligned}
e &:= \text{energy density} \\
p &:= \text{pressure} \\
\rho &:= \text{rest-mass density}
\end{aligned}
\right\}
\;\Longrightarrow\;
\left\{
\begin{aligned}
h &:= \frac{e+p}{\rho c^{2}},\\
w &:= \frac{p}{e}.
\end{aligned}
\right.
\end{equation}
The last thermodynamic quantity, $h$, is colloquially called the ``enthalpy'' in general relativity, although, in reality, it is the relativistic specific enthalpy (i.e.~the enthalpy per baryon mass). 

The TOV equations in their standard form are\footnote{There is a third equation of structure that relates the rate of change of a metric variable to other thermodynamic variables. We will not make use of this equation here.}
\begin{align}
\label{eq:tov-standard-dimful}
\frac{dp}{dr} &= -\,\frac{G}{c^{2}}\,
\frac{(e+p)\left(m + 4\pi r^{3}p/c^{2}\right)}
     {r^{2}\left[1-2Gm/(rc^{2})\right]}\,,
\\
\label{eq:tov-standard-dimful2}
\frac{dm}{dr} &= 4\pi r^{2}\,\frac{e}{c^{2}} \,,
\end{align}
but in our variables, they become
\begin{align}
\label{eq:tov-lindblom-dimful}
\frac{du}{d\ln h} &=
\frac{-2u\,(1-2v)}{4\pi G \, u \,p/c^{4} + v},
\\
\frac{dv}{d\ln h} &=
-(1-2v)\,
\frac{4\pi G \,u\,e/c^{4} - v}{4\pi G \,u\,p/c^{4} + v}.
\end{align}
Let us further define the dimensionless thermodynamic variable 
\begin{align}
\label{eq:etilde-def}
\tilde e \equiv \frac{4\pi G}{c^4} r^2 \, e = \frac{4\pi G}{c^4} u \, e\,,    
\end{align}
so that the TOV equations now become 
\begin{align}
    \label{eq:tov-reformulated-etilde}
    \deriv{\tilde e}{\ln h} &= \tilde e \left[\frac{-2(1-2 v)}{w \tilde e + v} + \deriv{\ln e}{\ln h}\right]\\
    \label{eq:tov-reformulated-v}
    \deriv{v}{\ln h} &= \frac{-(1-2v)}{w \tilde e + v}\quant{\tilde e - v}\,,
\end{align}
where we have used the definition for $w$ in Eq.~\eqref{eq:thermo-vars}. The right-hand side of the TOV equations then depend on the EoS only through the  $w$ and the $d\ln e/d\ln h$ factors, the latter of which is controlled by $c_s^2$:
\begin{equation}
\label{eq:dlnedlnh}
    \deriv{\ln e}{\ln h} = \frac{1}{c_s^2} (1 + w), \qquad {\rm{with}} \qquad c_s^2 := \frac{dp}{de}\,.
\end{equation}
We consider only barotropic EoSs here, so $p$ will always be a function of $e$ only and not other thermodynamic variable.

Before solving these equations numerically, let us consider the structure of the TOV equations (and their solutions) near the center of the star. The dimensionless quantity $w=p/e$ is small in the low-density region (sub-saturation density) for any realistic EoS. We can intuitively understand this in the Newtonian limit, where $e \sim \rho c^2$ and $p \sim \rho v_F^2$, where $v_F$ is the typical Fermi particle velocity, so then $w \sim (v_F/c)^2 \ll 1$. The quantity $d{\ln e}{d \ln h}$ of Eq.~\eqref{eq:dlnedlnh}, however, is never small, because typically $c_s^2$ is not close to unity for realistic EoSs. 

Nonetheless, near the center of the star, we have that 
$(\tilde e_c,v_c)\to 0$ as $\ln h\to \ln h_c$.
Since $(d\ln e/d\ln h)=(1+w)/c_s^2$ remains finite for finite $c_s^2$, while
$2(1-2v)/(w\tilde e+v)$ diverges as $w\tilde e+v\to 0$, the latter term dominates
the near-center dynamics. Moreover, over this small neighborhood we may freeze the
EoS function $w(\ln h)$ to its central value,
$w(\ln h)=w_c+\mathcal{O}(\ln h_c-\ln h)$.
With these approximations, we then have
\begin{align}
\deriv{\tilde e}{\ln h} &\sim -2\,\frac{\tilde e}{w_c \tilde e + v}\,,
\\
\deriv{v}{\ln h} &\sim -\,\frac{\tilde e - v}{w_c \tilde e + v}\,,
\end{align}
near the center of the star.
Taking as an ansatz $\tilde{e}(\ln h) = \tilde{e}_1 (\ln h_c - \ln h)$ and $v(\ln h)=v_1(\ln h_c-\ln h)$, the above equations can be solved to obtain
\begin{align}
    \label{eq:bc-u}
    \tilde e(\ln h) & = \frac{6}{1+3w_c}(\ln h_c - \ln h) + {\cal{O}}(\ln h_c-\ln h)^2\\
    \label{eq:bc-v}
    v(\ln h) &= \frac{2}{1+3w_c}(\ln h_c - \ln h) + {\cal{O}}(\ln h_c-\ln h)^2\,,
\end{align} 
where note $v = \tilde e/3 + {\cal{O}}(\ln h_c-\ln h)^2$. In summary, the leading-order, near-center behavior is insensitive to the EoS; it depends only on the local value $w_c = p_c/e_c$ (and remains regular for finite $c_s^2$). 

\begin{figure*}[htb]
    \centering
    \includegraphics[width=0.49\textwidth]{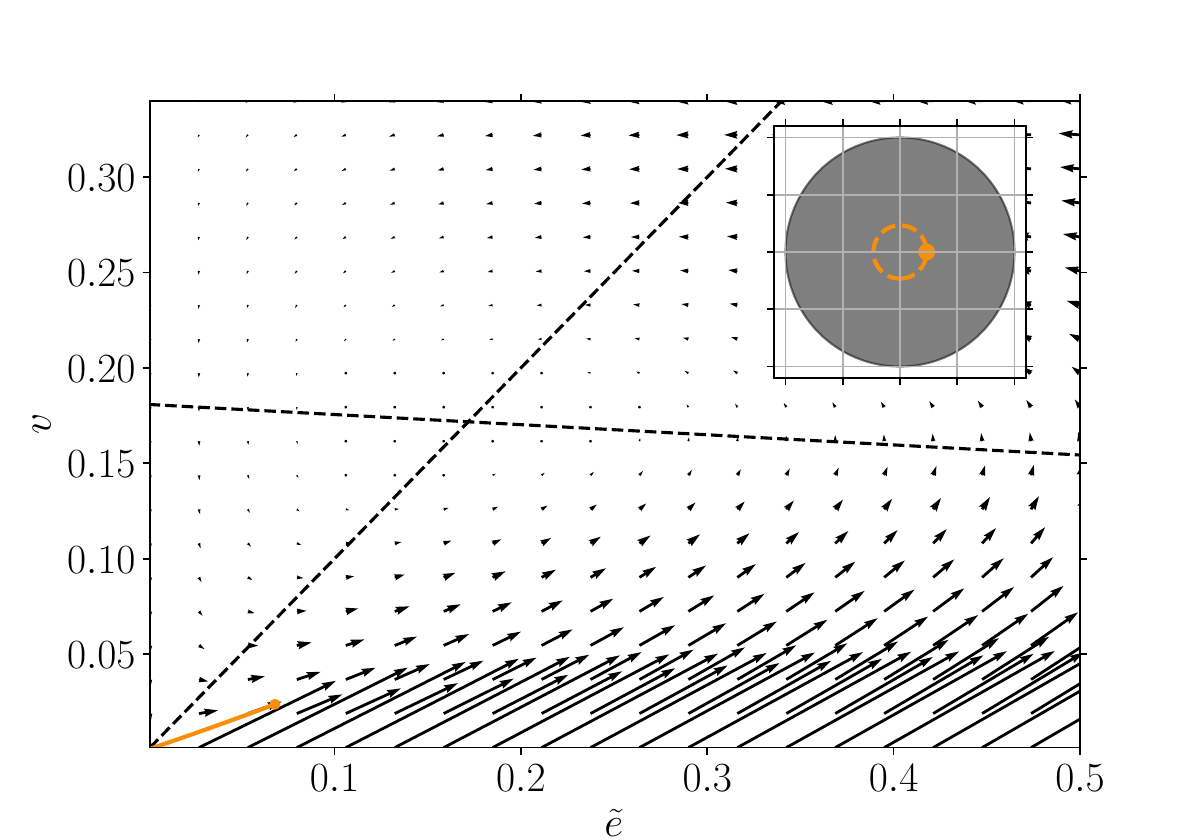}
    \includegraphics[width=0.49\textwidth]{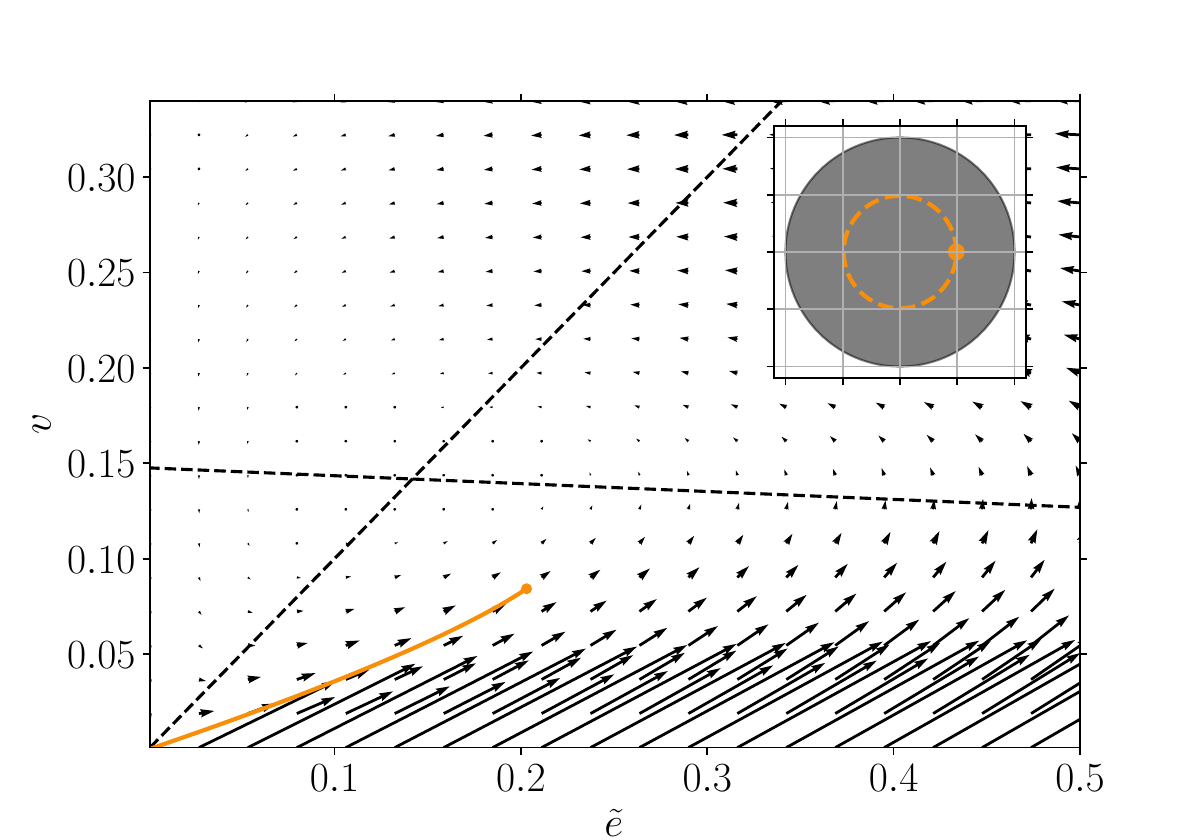} \\
    \includegraphics[width=0.49\textwidth]{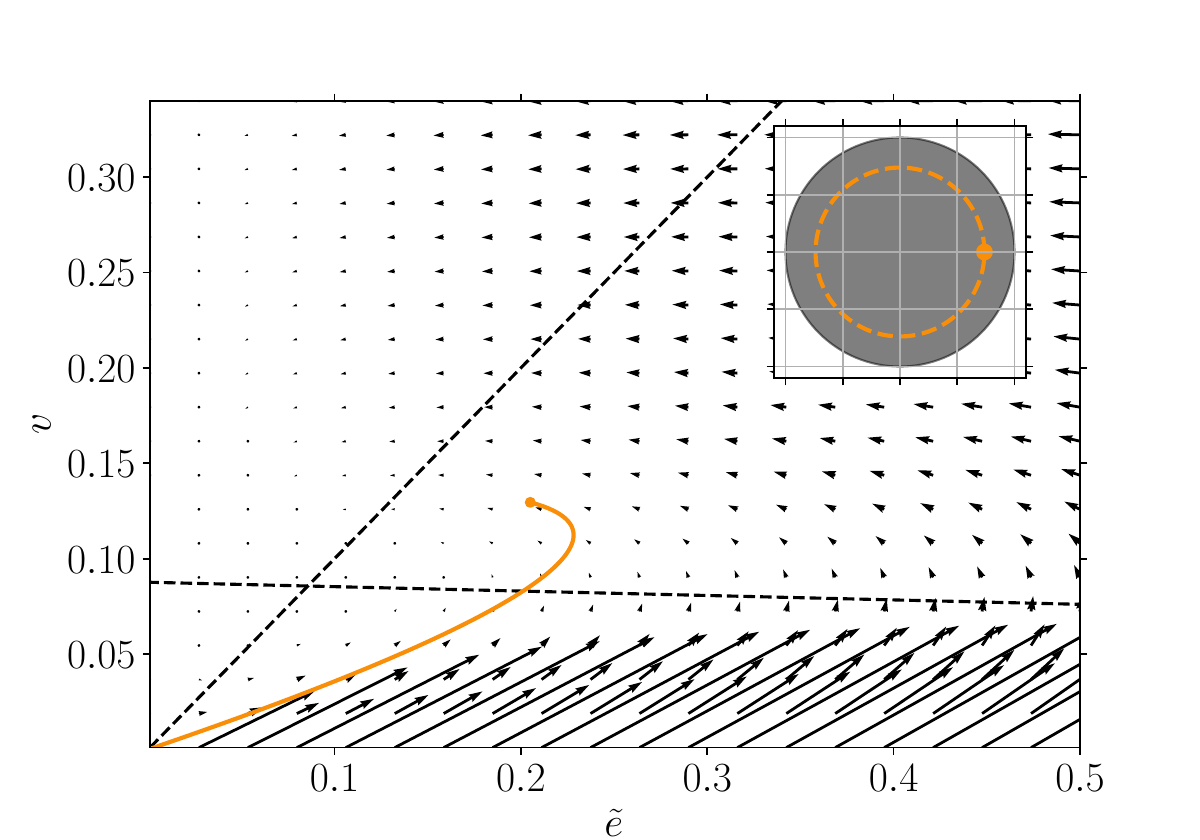}
    \includegraphics[width=0.49\textwidth]{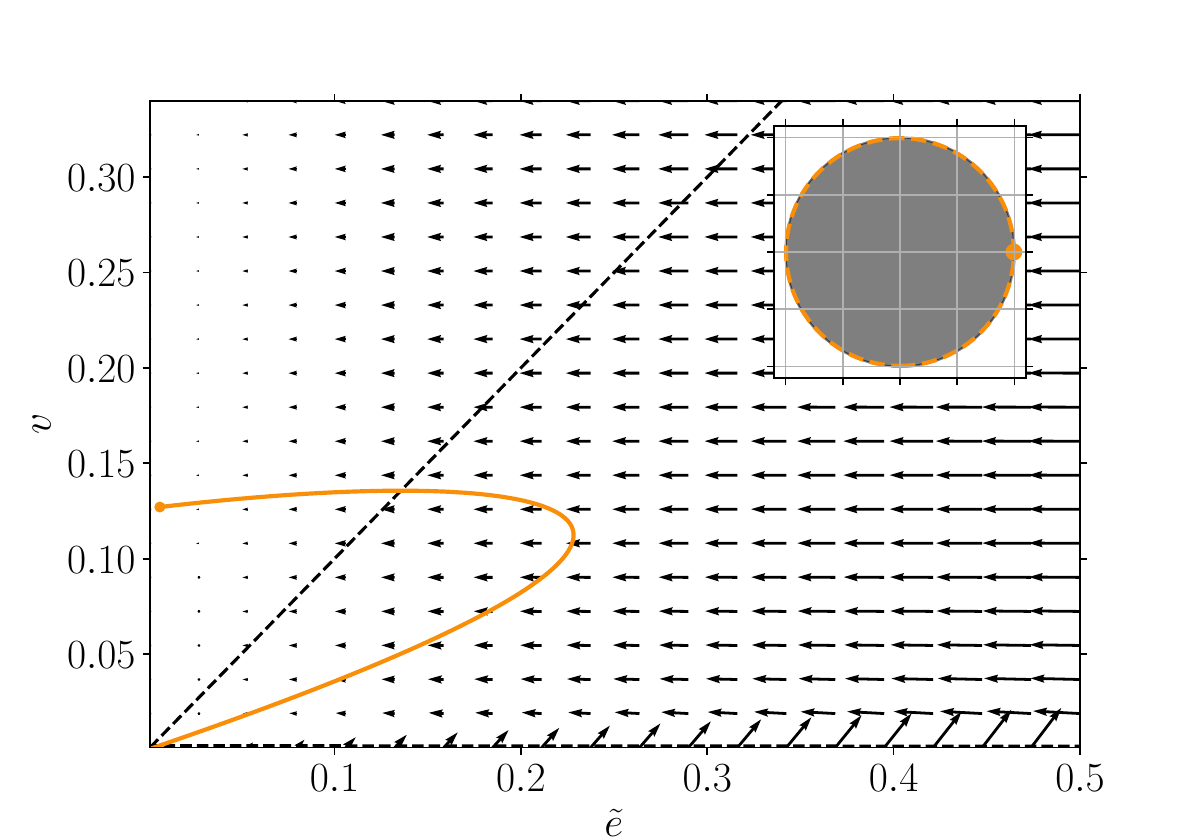}
    \caption{Partial solutions to the TOV equation (orange lines), integrated \new{from the central value, $\ln h_c$ until a particular} $\ln h$ or radius (orange circle), together with vector flows that represent the evolution of the TOV solution (i.e.~the right-hand side of Eqs.~\eqref{eq:tov-reformulated-etilde} and~\eqref{eq:tov-reformulated-v}).
    The inset on the top, right corner of each panel indicates the maximum radius to which each partial integration goes relative to the size of the star (shaded in gray). 
    Two dashed lines demarcate the regions where $\tilde e$ and $v$ change behavior (they change from increasing to decreasing or vice-versa).     
    \emph{Top, left:} when $\ln h$ is very close to $\ln h_c$, the functions $\tilde e(\ln h)$ and $v(\ln h)$ are nearly universal, and they are determined by the near-core expansion of Eqs.~\eqref{eq:bc-u} and~\eqref{eq:bc-v}. 
    \emph{Top, right:} the trajectory begins to subtly deflect from the near-core ($v = \tilde e/3$) expansion.   
    \emph{Bottom, left:} the solution is near the maximum value of $\tilde e=4\pi r^2 \, e$. This is the region of the star where most of the mass is accumulated.  The system is complicated, depending sensitively on $c_s^2$ and $w$, which determine the structure of the slope field when both $\tilde e$ and $v$ are not small compared to unity.   
    \emph{Bottom, right:} the solution is very near the surface of the star. The equation for $\tilde e$ is nearly homogeneous and independent of $v$.  The variable $v$ changes very little relative to $\tilde e$, since the right-hand side of the $\tilde e$ equation contains $1/c_s^2$, which is very large near the surface.
    }
    \label{fig:demo-near-core}
\end{figure*}

Let us now explore some solutions to the TOV equations expressed in these new variables. 

Figure~\ref{fig:demo-near-core} shows four solutions to Eqs.~\eqref{eq:tov-reformulated-etilde} and~\eqref{eq:tov-reformulated-v}, using a polytropic, $p(\rho) = K \rho^{\Gamma}$, $\Gamma=2$ EoS for illustrative purposes\footnote{Note that, in the TOV equations as formulated above, the polytropic constant $K$ is irrelevant for the variables $v$ and $\tilde e$, and it only enters as a ``post-processing step", when $r^2 = \tilde e/(4 \pi e(\rho))$ (see ~\cite{ShapiroAndTeukolsky, baumgarte_shapiro_2010} for related polytrope rescaling relations).}.  We display ``partial solutions" to the TOV equations in the sense that the integrations do not reach the stellar surface \textit{per se}. Since $\ln h$ and $r$ are in one-to-one correspondence inside the star, we can think of the different panels as snapshots of the TOV integration that have reached a certain radius or a certain $\ln h$.  

This figure also shows a vector flow, defined by the right-hand sides of Eqs.~\eqref{eq:tov-reformulated-etilde} and~\eqref{eq:tov-reformulated-v}. This flow indicates the direction in which the solution evolves as the density or enthalpy decrease. 

Let us now emphasize a couple of key points.  First, as seen in Fig~\ref{fig:demo-near-core}, up to a certain value of $\ln h$ or $r$, the solutions $\tilde e(\ln h)$ and $v(\ln h)$ are determined almost entirely by the near-core behavior of Eqs.~\eqref{eq:bc-u} and~\eqref{eq:bc-v}.  

Second, looking at the bottom row of Fig.~\ref{fig:demo-near-core}, very near the surface of the star, the equations are again ``universal" in some sense, as the homogeneity in $\tilde e$ and the rapid growth of $1/c_s^2$ in the right-hand side of the equation for $\tilde e$ mean the system generically approaches the surface $\tilde e=0 $ in qualitatively the same way.  On the other hand, in between the near core and the surface behavior, there is a turnover in $\tilde e$ depending on the relative sizes of $v, w, \tilde e,$ and  $c_s^2$.  This region is evidently where most of the complexity of the solution comes from. In the next section, we will tackle this complexity from a dynamical-systems approach.

\subsection{A dynamical-systems approach to TOV equations and their fixed point}
\label{sec:fixed-point}

The discussion of Sec.~\ref{sec:lindblom-newtonian} already hints at the key simplification: once written in terms of $(\tilde e,v)$ and parametrized by $\ln h$, the stellar structure problem becomes a two-dimensional flow whose explicit EoS dependence enters through only a small set of thermodynamic combinations. In this subsection, we make that statement precise, identify the relevant frozen-time fixed point, and show how its local geometry controls the high-density behavior of relativistic stellar sequences.

Let us begin by showing explicitly how the TOV equations can be cast as a non-autonomous flow problem. Let us define the state vector and time variable as
\begin{equation}
\mathbf{x}(\tau)=\big(\tilde e(\tau),v(\tau)\big)\,,\qquad \tau=\ln h\,.
\label{eq:state_vector_tov}
\end{equation}
The TOV equations then become Eq.~\eqref{eq:ds_generic}, or more precisely, 
\begin{equation}
\label{eq:TOV-as-autonomous}
\frac{d\mathbf{x}}{d\tau}=\mathbf{F}(\mathbf{x},\tau)
= 
\begin{pmatrix}
\tilde e\left[-\dfrac{2(1-2v)}{w\,\tilde e+v} +
\dfrac{1 + w}{c_s^2},
\right]\\[1.2ex]
-\dfrac{(1-2v)(\tilde e-v)}{w\,\tilde e+v}\,.
\end{pmatrix}
\end{equation}
For a barotropic EoS, the remaining logarithmic derivative is fixed by the sound speed, as shown in Eq.~\eqref{eq:dlnedlnh}.
The TOV equations recast in this way are therefore a driven (non-autonomous) two-dimensional system: the vector field is universal, and the EoS enters only through the ``time-dependent parameters'' $w(\ln h)$ and $c_s^2(\ln h)$. Such a non-autonomous flow is exactly the setting of Sec.~\ref{sec:ds_nonautonomous}. 

Whenever $w$ and $c_s^2$ drift slowly or adiabatically with $\ln h$, we say the system is \emph{approximately autonomous}, and a stability analysis yields approximate solutions. This is the case at large densities ($n_{B}\gg n_{\rm sat}$), in the inner core where one may find  asymptotically free quarks~\cite{Kurkela:2009gj, Alford:2013aca}. In these cases, it is useful to analyze the associated frozen system obtained by temporarily holding $w$ and $c_s^2$ fixed at some frozen $\tau_{\rm fr}$. The resulting system then admits a frozen equilibrium or fixed point $x_\star(\tau_{\rm fr})$, defined by
\begin{align}
\mathbf{F}(\mathbf{x}_\star,\tau_{\rm fr})=0\,.
\end{align}
Away from the unphysical surface $v=1/2$, corresponding to $r = 2 G m(r)/c^2$, the second component of Eq.~\eqref{eq:TOV-as-autonomous} implies immediately that the equilibrium lies on the diagonal,
\begin{equation}
\tilde e_\star=v_\star\,.
\label{eq:fixedpoint-diagonal}
\end{equation}
Substituting this into the first component of Eq.~\eqref{eq:TOV-as-autonomous}, and using Eq.~\eqref{eq:dlnedlnh}, yields the
second condition,
\begin{equation}
\frac{2(1-2v_\star)}{[1+w(\tau_{\rm fr})]\,v_\star}=\frac{1+w(\tau_{\rm fr})}{c_s^2(\tau_{\rm fr})}\,,
\end{equation}
which can be solved in closed form:
\begin{equation}
v_\star(\tau_{\rm fr})=\tilde e_\star(\tau_{\rm fr})
=\frac{2c_s^2(\tau_{\rm fr})}{4c_s^2(\tau_{\rm fr})+[1+w(\tau_{\rm fr})]^2}\,.
\label{eq:fixedpoint-location}
\end{equation}

Let us pause here to emphasize several points. First, the fixed point is ``relativistic'' in a very concrete sense: $v_\star$ and $\tilde e_\star$
become order unity (and thus, the enclosed compactness becomes of order unity) only if the EoS is sufficiently stiff, so that $w(\tau_{\rm fr})$ and $c_s^2(\tau_{\rm fr})$ are not everywhere
small. In particular, if $(w,c_s^2)\ll1$, then $v_\star\simeq 2c_s^2\ll1$, and the equilibrium sits
near the origin, where the flow is dominated by the universal near-core structure discussed in
Sec.~\ref{sec:lindblom-newtonian}.

Second, the location of the equilibrium is controlled by only a couple of thermodynamic numbers at the density scale being probed. For example, in the maximally stiff (but not particularly physical) limit $c_s^2\to1$ and $w\to0$, one finds $v_\star = e_\star \to 2/5$. Since $v$ at the surface is the stellar compactness $C=G M/(R c^2)$, this suggests an upper scale on the compactness of relativistic stars of less than approximately $2/5$, which is similar to the Tolman-Buchdahl bound~\cite{Tolman:1939jz, Buchdahl:1959zz}. This configuration is unphysical though; because $w$ and $c_s^2$ are related to each other, typically $w$ and $c_s^2$ are the same in order of magnitude. For more typical ultra-relativistic stiffness, where both $w$ and $c_s^2$ are order unity, one gets $v_\star\sim 0.2$--$0.3$; e.g.~$w=c_s^2=1$ gives $v_\star=\tilde e_\star=1/4$, which the limiting case of many relativistic hadronic EoS models~\cite{Steiner:2012rk, Pearson:2018tkr}. These are roughly the values encountered in the region of the star where $\tilde e$ is largest and where the bulk of the mass is accumulated (cf.~Fig.~2).

\subsection{Spiral evolution toward the fixed point}
\label{sec:jacobian-hessian}

The fixed point becomes dynamically relevant because, for the frozen system, it is generically a
focus, but for the dynamical systems, spiraling occurs around that focus. Linearizing about $\mathbf{x}_\star$ via $\mathbf{x}=\mathbf{x}_\star+ \mathbf{\delta x}$ as in Sec.~\ref{sec:ds_fixed_points}, we find
\begin{equation}
\frac{d}{d\ln h} \mathbf{\delta x} = \mathbf{J}_\star \, \mathbf{\delta x}+\mathcal{O}(|\mathbf{\delta x}|^2)\,,
\label{eq:linearized-flow}
\end{equation}
where the Jacobian matrix at equilibrium can be written purely in terms of $w$ and $c_s^2$:
\begin{equation}
\mathbf{J}_\star=
\begin{pmatrix}
\dfrac{w(\tau_{\rm fr})}{c_s^2(\tau_{\rm fr})} &
\dfrac{1 + w(\tau_{\rm fr}) + 4c_s^2(\tau_{\rm fr})}{c_s^2(\tau_{\rm fr}) [1+w(\tau_{\rm fr})]}\\[10pt]
-\dfrac{1+w(\tau_{\rm fr})}{2c_s^2(\tau_{\rm fr})} &
\dfrac{1+w(\tau_{\rm fr})}{2c_s^2(\tau_{\rm fr})}
\end{pmatrix}
\,.
\label{eq:jacobian-fixedpoint}
\end{equation}
The trace and determinant are
\begin{equation}
\tr \mathbf{J}_\star=\frac{3w(\tau_{\rm fr})+1}{2c_s^2(\tau_{\rm fr})}\,,\qquad
\det \mathbf{J}_\star=\frac{4c_s^2(\tau_{\rm fr})+[1+w(\tau_{\rm fr})]^2}{2c_s^4(\tau_{\rm fr})}\,,
\end{equation}
so $\det J_\star>0$ for any physical (stable) EoS with $c_s^2>0$. Moreover, for the range relevant to
compact-star cores ($w\lesssim1$ and $c_s^2\lesssim1$), the discriminant is negative and the
eigenvalues are a complex conjugate pair, $\lambda_\pm(\tau_{\rm fr})=\sigma(\tau_{\rm fr})\pm i\omega(\tau_{\rm fr})$, with 
\begin{align}
& \qquad \qquad \quad \sigma(\tau_{\rm fr})=\frac{3w(\tau_{\rm fr})+1}{4c_s^2(\tau_{\rm fr})}\,,
\nonumber
\\
\omega(\tau_{\rm fr})&=\frac{1}{4c_s^2(\tau_{\rm fr})}\sqrt{32c_s^2(\tau_{\rm fr})-w^2(\tau_{\rm fr})+10w(\tau_{\rm fr})+7}\,.
\label{eq:eigs-fixedpoint}
\end{align}
Thus, the frozen-time equilibrium is an \emph{unstable} focus when evolved forward in $\ln h$. 

But recall that, for stellar models, we integrate from the core to the surface, so toward \emph{decreasing} $\ln h$, as we had mentioned already in Sec.~\ref{sec:ds_fixed_points}. From this viewpoint, then the frozen-time equilibrium should really be a \emph{stable} focus when evolved from the core to the surface. Let us make this explicit by defining the outward evolution variable $s:=\ln h_c-\ln h$ (so that $d/ds=-d/d\ln h$). The linearized system then becomes $d/ds \mathbf{\delta x} = - \mathbf{J}_\star \, \mathbf{\delta x}$, and the real parts of the eigenvalues flip sign. In other words, the same focus that is unstable in forward $\ln h$ acts as an \emph{attractor} along the physical (outward) integration direction. As a concrete example, consider an EoS such that, at some frozen time $\tau_{\rm fr}$, we have $w(\tau_{\rm fr})=c_s^2(\tau_{\rm fr})=1$. Then, the Jacobian matrix is
\begin{equation}
\mathbf{J}_\star=
\begin{pmatrix}
1 & 3\\
-1 & 1
\end{pmatrix},
\qquad
\lambda_\pm=1\pm i\sqrt{3}\,,
\end{equation}
so trajectories spiral toward $\mathbf{x}_\star = (\tilde e_\star,v_\star)=(1/4,1/4)$, as one evolves outward in the
star.

We have thus far held $w$ and $c_s^2$ fixed at some frozen value $\tau_{\rm fr}$, but in true stellar problems, these quantities must drift with $\ln h$, and the equilibrium $x_\star(\ln h)$ must move accordingly. The relevant question is then whether solutions can \emph{track} this moving equilibrium, in the sense of Sec.~\ref{sec:ds_nonautonomous}, or whether the evolution away from the fixed point is fast enough to leave the frozen system in the dust. A quick diagnostic is provided by the drift rates of the thermodynamic parameters. For example,
\begin{equation}
\frac{dw}{d\ln h}=(1+w)\left(1-\frac{w}{c_s^2}\right),
\label{eq:dwdlnh}
\end{equation}
while the evolution of $c_s^2(\ln h)$ depends on higher derivatives of the EoS, i.e.
\begin{equation}
\frac{d}{d\ln h}\!\left(\frac{1}{c_s^2}\right)
=
-\frac{1}{c_s^4}\frac{d c_s^2}{d\ln h}.
\label{eq:cs2-deriv-clean}
\end{equation}
When $w$ and $c_s^2$ vary moderately over $\Delta\ln h\sim\mathcal{O}(1)$---as is typically the case once the EoS enters a stiff, relativistic regime---Eq.~\eqref{eq:dwdlnh} shows that $dw/d\ln h$ is at most $\mathcal{O}(1)$ (and can be smaller when $w\simeq c_s^2$). Thus, over intervals $\Delta\ln h\lesssim 1$, the parameters change moderately, and the vector field may be approximated locally by that of the frozen system. In this regime, the focus structure discussed above provides an effective geometric description of the dynamics: trajectories are drawn toward the instantaneous focus and execute a damped rotation about it, with local rates set by $\sigma$ and $\omega$ in Eq.~\eqref{eq:eigs-fixedpoint}. By contrast, if the EoS contains a strong first-order phase transition, $c_s^2$ can change on a much shorter $\ln h$ scale; in that case $x_\star(\ln h)$ shifts rapidly and the tracking picture can break down.

The situation is different in the Newtonian limit (as we will see in more detail in the next section), where $w\ll1$ and $c_s^2\ll1$.  In that case, the relative variation of the coefficients entering Eq.~\eqref{eq:TOV-as-autonomous} can become large even when $dw/d\ln h$ itself is not parametrically large, because the flow depends on ratios such as $w/c_s^2$ and on inverse powers of $c_s^2$. Consequently, as $\ln h_c\to 0$ the frozen-time approximation is generically poor: there is no extended range of $\ln h$ over which the system can be treated as approximately autonomous. This explains why the spiral mechanism is not generic in Newtonian stellar sequences.

There is, however, an important exception.  If the EoS is tuned so that $c_s^2$ is (approximately) constant in $\ln h$---for example, in the isothermal limit corresponding formally to a polytrope with $n\to\infty$---the system becomes effectively autonomous and the fixed-point structure regains predictive power even outside the relativistic regime. Thus, general relativity is not strictly required for a fixed point to exist, but special relativity, through the requirement of a finite and eventually non-negligible sound speed, ensures that realistic high-density equations of state inevitably enter a regime where the fixed-point picture becomes applicable.

The linearized system of Eq.~\eqref{eq:linearized-flow} explains the existence of a local focus and provides the characteristic spiral rates set by the eigenvalues $\lambda_\pm(\tau_{\rm fr})$ of $\mathbf{J}_\star$. Let us then end this discussion by estimating how large a neighborhood around $\mathbf{x}_\star$ is reasonably described by this linear picture. To do so, let us expand the forcing function $\mathbf{F}$ to second order in $\mathbf{\delta x}$ to find
\begin{equation}
\mathbf{F}(\mathbf{x}_\star+\delta\mathbf{x},\tau_{\rm fr})
=
\mathbf{J}_\star\,\delta\mathbf{x}
+
\frac12
\begin{pmatrix}
\delta\mathbf{x}^{\mathsf T}\mathbf{H}_{\tilde e}\,\delta\mathbf{x}\\[0.3ex]
\delta\mathbf{x}^{\mathsf T}\mathbf{H}_{v}\,\delta\mathbf{x}
\end{pmatrix}_{\!\mathbf{x}=\mathbf{x}_\star}
+\mathcal{O}(|\delta\mathbf{x}|^3)\,,
\label{eq:tov_taylor_about_fp}
\end{equation}
where $\mathbf{H}_{\tilde e}$ and $\mathbf{H}_{v}$ are the Hessians of the two components of the frozen vector field with respect to $\mathbf{x}=(\tilde e,v)$,
\begin{equation}
\label{eq:Hessian-def}
(\mathbf{H}_{\tilde e})_{ij}:=
\left.\frac{\partial^2 F_{\tilde e}}{\partial x_i\,\partial x_j}\right|_{\tau=\tau_{\rm fr}},
\qquad
(\mathbf{H}_{v})_{ij}:=
\left.\frac{\partial^2 F_{v}}{\partial x_i\,\partial x_j}\right|_{\tau=\tau_{\rm fr}}\,.
\end{equation}
Because the only nonlinear dependence on $(\tilde e,v)$ of $\mathbf{F}$ in Eq.~\eqref{eq:TOV-as-autonomous} enters through rational factors of the form $(w\tilde e+v)^{-1}$, the Hessians scale schematically as $(w\tilde e+v)^{-3}$. However, this nonlinearity also introduces factors of $\tilde e$ and $v$ in the numerators of the Hessians, so a na\"ive $(w\tilde e+v)^{-3}$ estimate overestimates their magnitude near the fixed point. 

Let us then make a more careful estimate. At $x_\star$, one has $w\tilde e_\star+v_\star=(1+w)v_\star$, and the relevant small parameter controlling truncation is
\begin{equation}
\frac{\|H\|\,|\delta x|^2}{\|J\|\,|\delta x|}
\sim \frac{\|H\|}{\|J\|}\,|\delta x|\,,
\end{equation}
rather than $\|H\|/\|J\|$ by itself. The Hessians of Eq.~\eqref{eq:Hessian-def} evaluate to 
\begin{align}
\mathbf{H}_{\tilde e}
&=
\frac{2}{(w\tilde e+v)^3}
\begin{pmatrix}
-2vw(2v-1)
&
4\tilde e v w-\tilde e w+v
\\[4pt]
4\tilde e v w-\tilde e w+v
&
-2\tilde e(2\tilde e w+1)
\end{pmatrix},
\label{eq:hessian_etilde}
\\
\mathbf{H}_{v}
&=
\frac{1+w}{(w\tilde e+v)^3}
\begin{pmatrix}
-2vw(2v-1)
&
4\tilde e v w-\tilde e w+v
\\[4pt]
4\tilde e v w-\tilde e w+v
&
-2\tilde e(2\tilde e w+1)
\end{pmatrix}.
\label{eq:hessian_v}
\end{align}
For representative relativistic values (e.g.\ $w=c_s^2=1$ so that $v_\star=\tilde e_\star=1/4$ and $w\tilde e_\star+v_\star=1/2$), the explicit Hessians evaluate to coefficients of $\mathcal{O}(10)$, while the Jacobian has coefficients of $\mathcal{O}(1)$. Thus, the linear focus picture is self-consistent for $|\mathbf{\delta x}|\lesssim 0.1$, i.e.~for the moderate neighborhoods around $x_\star$ actually explored by the solutions near the $\tilde e$ turnover (cf.~Fig.~\ref{fig:demo-near-core}). This analysis therefore justifies treating the local flow as a damped rotation about a (slowly moving) focus when trajectories remain in a moderate neighborhood of $\mathbf{x}_\star$.

The above estimate also clarifies two limits in which the focus picture is less useful. First, sufficiently close to the center, $w\tilde e+v\to 0$ and the rational structure of Eq.~\eqref{eq:TOV-as-autonomous} dominates, consistent with the separate universal near-core analysis of Sec.~\ref{sec:lindblom-newtonian}. Second, in the weak-field regime, where $c_s^2\ll 1$ (and, hence, $v_\star\simeq 2c_s^2\ll 1$), the equilibrium is pushed toward the origin and the coefficients controlling the nonlinear corrections grow. In that case, a frozen-time/near-focus description is generically not an efficient way to organize the dynamics unless the EoS is tuned so that $w(\ln h)$ and $c_s^2(\ln h)$ drift exceptionally slowly. We will return to this point in Sec.~\ref{sec:newtonian-falling-sound-speeds}, where we explain why the spiral mechanism is non-generic in Newtonian stars and identify a different universal structure near their termination.

The main takeaway here is that the complicated intermediate region between the ``universal core behavior'' and the ``universal surface behavior'' is organized by a simple geometric object: a (moving) focus in the $(\tilde e,v)$ plane whose location and local linear rates are controlled by $w$ and $c_s^2$ in the high-density regime. In Sec.~\ref{sec:implications}, we will exploit this fixed-point structure to obtain scaling estimates and quasi-universal relations for the maximum mass.

\subsection{Understanding the $M$--$R$ curve through dynamical systems}
\label{sec:dyn-sys}
 
We now have all of the ingredients we need to understand the whole picture.  At large $\ln h$ (near the center of the star), the slope field changes only slowly, and all solutions share the same universal near-core behavior. Different EoS families of trajectories largely collapse onto one another at a fixed $\ln h$. As the integration moves to lower densities, the vector field drifts because $w(\ln h)$ and $c_s^2(\ln h)$ drift, and trajectories peel away depending on how long they spend in the neighborhood of the focus and with what phase. Once $\tilde e$ begins to fall and the integration approaches the surface, $\tilde e$ evolves nearly homogeneously, while $v$ changes comparatively slowly, so the ``phase'' accumulated near the maximum of $\tilde e$ is carried out to the surface. Because the surface values determine the $M$--$R$ curve, this inherited phase structure is what ultimately produces the turnover and spiral behavior at high central enthalpy.

\begin{figure*}[htb]
     \centering
     \includegraphics[width=.495\textwidth]{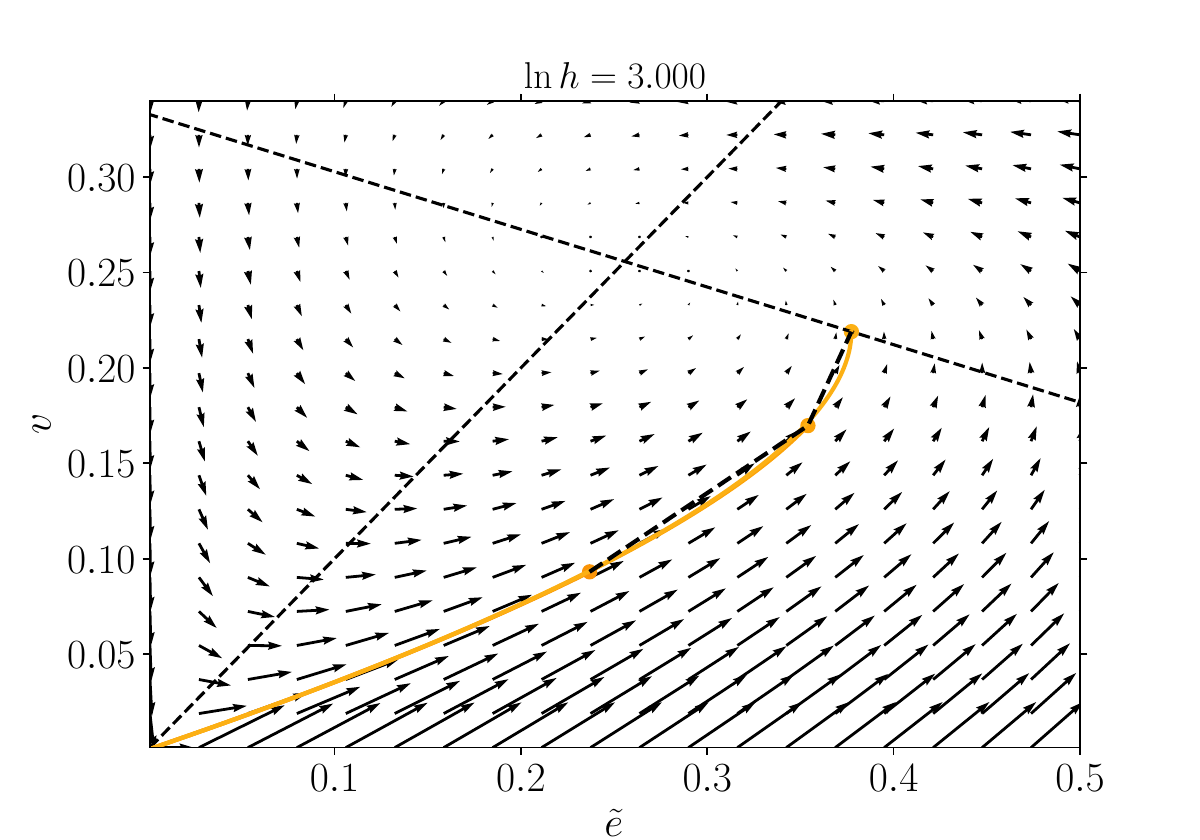}
     \includegraphics[width=.495\textwidth]{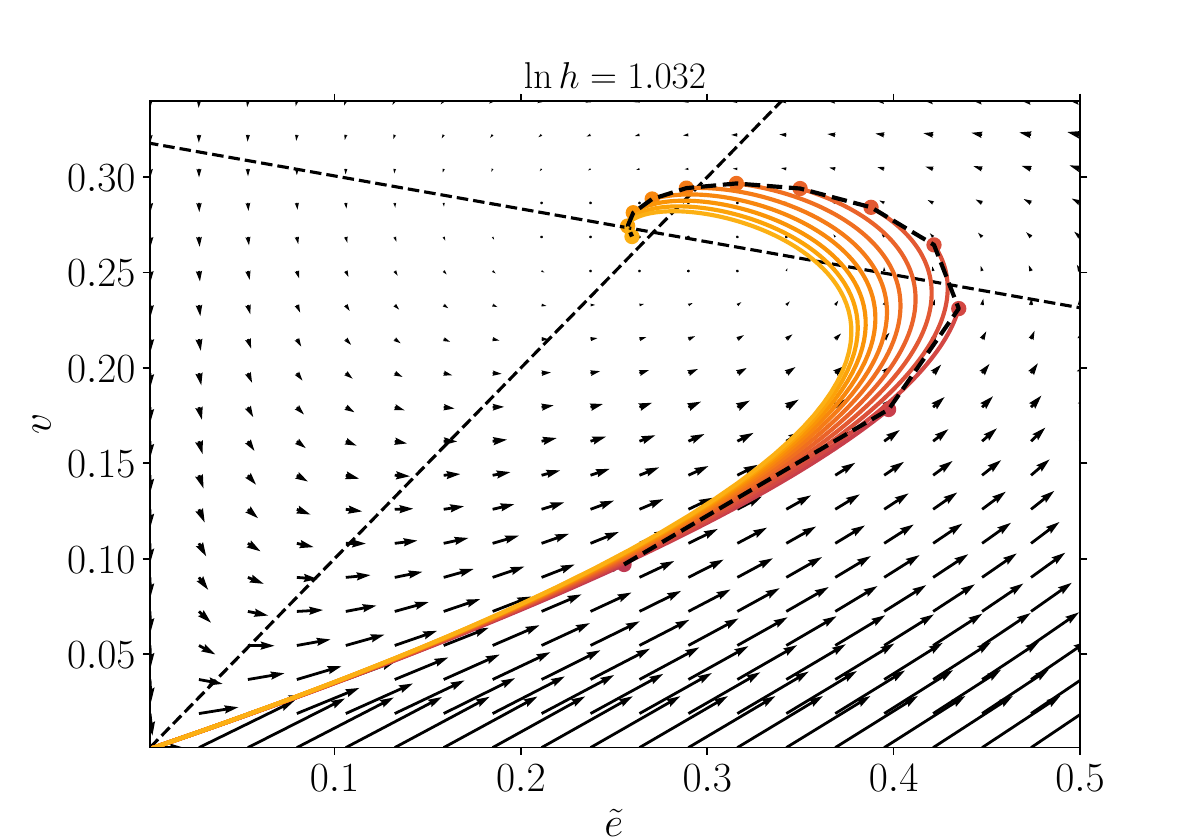}
    \includegraphics[width=.495\textwidth]{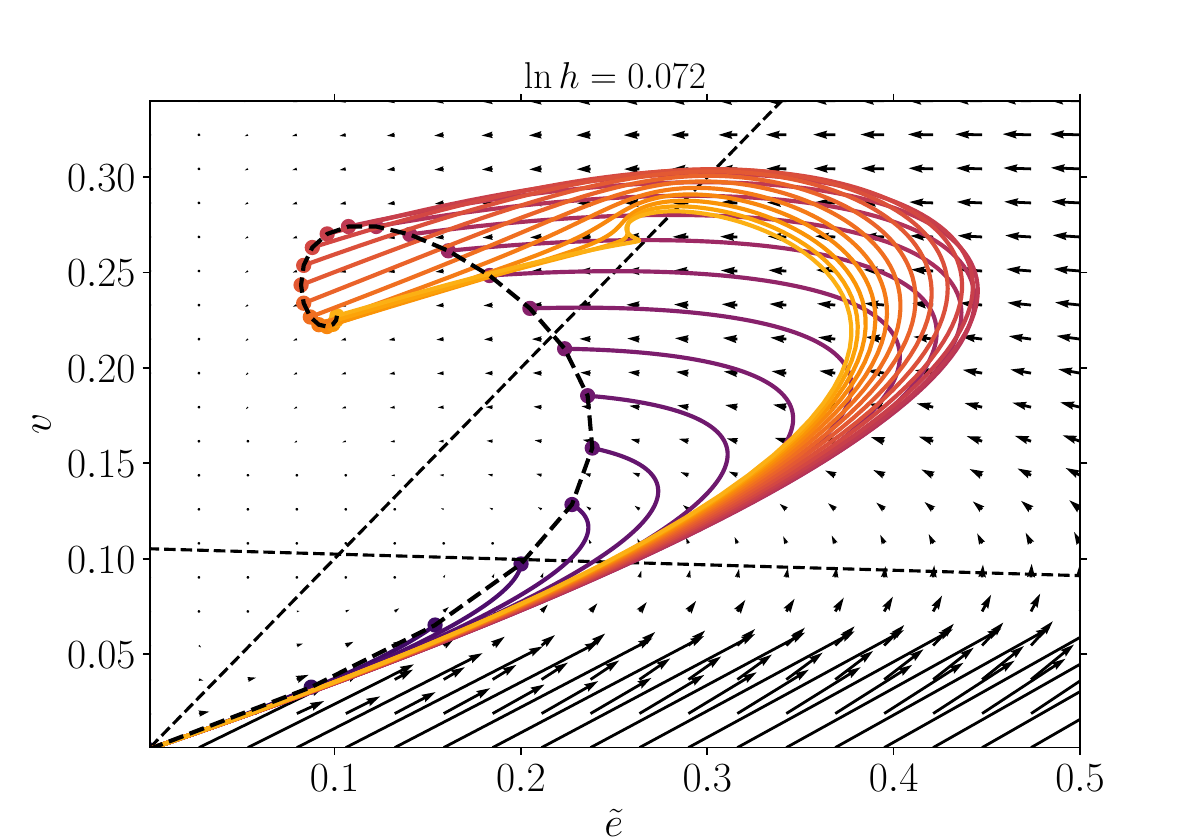}
     \includegraphics[width=.495\textwidth]{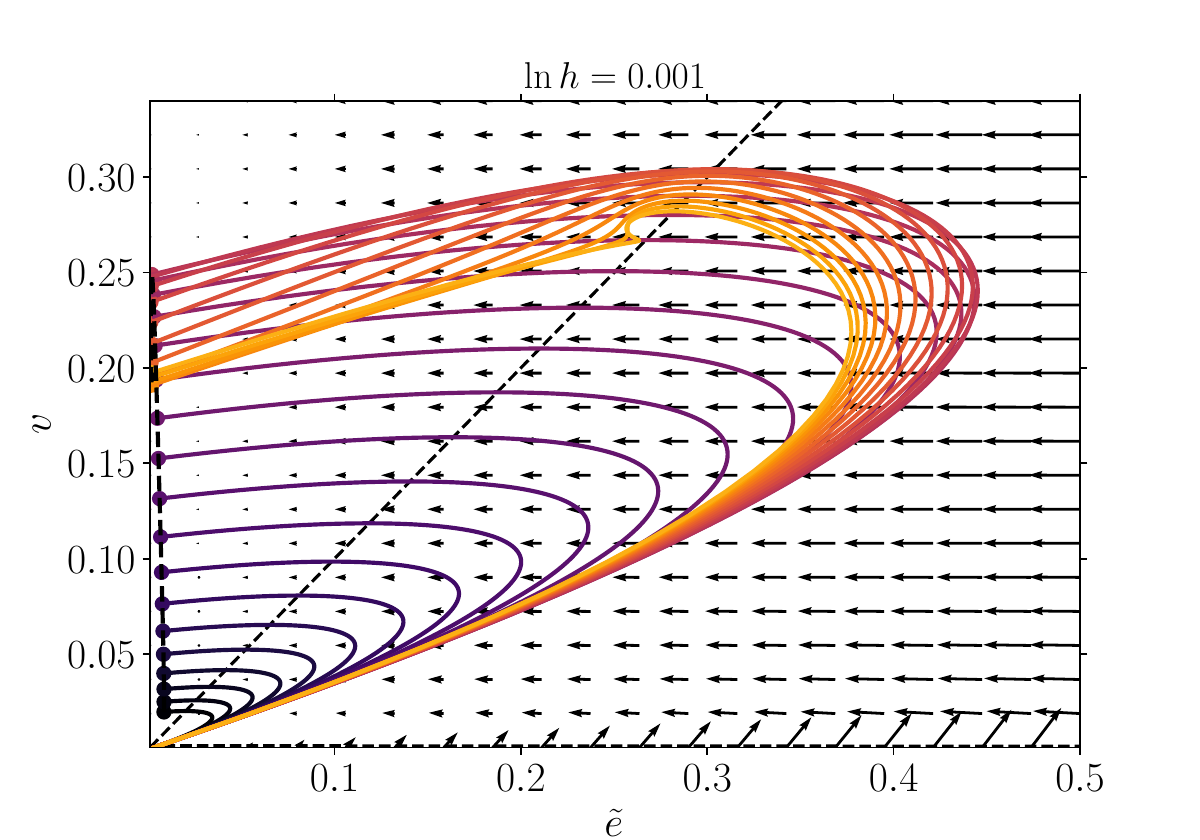}
     \caption{Partial solutions to the TOV equations and flow vectors (with a $\Gamma=2$ polytropic EoS for illustrative purposes). This figure is qualitatively similar to Fig.~\ref{fig:demo-near-core}), but carried out for many different central densities (different colored lines) and for snapshots that are integrated to different enthalpies (different final energy densities). Observe that the earliest snapshot (top, left panel) presents near-core universal behavior, with all solutions overlapping each other. As one integrates to lower energy densities (top, right panel), solutions with different central densities fan out, because the system stops being formally autonomous, although all trajectories are still attracted by the (slowly-drifting) fixed point. As one integrates further to even lower energy densities (bottom, left panel), the system becomes purely non-autonomous, and the spiral behavior becomes evident. When one integrates all the way to the surface (bottom, right panel), the spiral structure is maintained because the $\tilde e$ evolution is nearly homogeneous. The dashed curve is, to a good approximation, simply scaled horizontally.  \new{See Fig.~\ref{fig:final-state-rescaled} for a version rescaled by $e(\ln h)$.}     
     }    
     \label{fig:gamma-2-initial-system}
 \end{figure*}

We can visualize this story neatly in Fig.~\ref{fig:gamma-2-initial-system}. This figure shows the slope-field $\textbf {F}(\textbf{x}, \tau)$ of Eq.~\eqref{eq:TOV-as-autonomous} at several values of $\ln h$ (shown at the top of each panel). Overplotted are the snapshots of the trajectories of partial solutions to the TOV equations $\mathbf{x}(\tau')$ with $\tau'\in[\tau,\ln h_c]$, each with a different value of the central densities (represented with different colored lines).  The ``current state" of each solution is marked by a dot of the same color, which represents $\textbf{x}(\tau)$. We connect solutions with a black dashed line purely for visualization purposes.  Each partial solution reaches a different  ``phase" of the star, at a different fractional radius $r/R$, and energy density $e/e_c$, but each is locally governed by the same dynamical system. A star's central $\ln h$ value is equal to the value of $\ln h$ at which the solution appears at $\tilde e=v=0$, and thus, not all solutions are represented in all plots.

Now that we understand the basic features of this figure, let us dissect it panel by panel. The top, left panel shows the high enthalpy case ($\ln h = 3$), for which the current state has $c_s^2\approx1 \approx w$. All partial solutions have very high central density and present nearly universal behavior (i.e.~they all overlap each other in this panel). The system is dominated by the near-core, Taylor expansion of Eqs.~\eqref{eq:bc-u}~and~\eqref{eq:bc-v}.  Nonetheless, the solutions also show signs of initial-condition-independent autonomy, with each solution following a similar orbit to solutions at higher central density.  The system is not explicitly dependent on $\ln h$, a characteristic sign of this is that the black dashed line (interpolation of current state solutions) coincides with the history of each solution curve. 

The top, right panel presents a lower enthalpy case, where  the current state now has $c_s^2\approx 0.65$ and $w\approx 0.46$. This time we see explicit central density dependence of the solutions, as the different-colored lines cease overlapping with each other and fan out. The system is no longer completely autonomous, because $w$ and $c_s^2$ are not independent of $\ln h$. However, the fixed point is still an attractor and an effective description of the system. With sufficient evolution, the partial solutions \emph{re-converge} to the fixed point, despite having diverged from their shared initial trajectory.  
    
The bottom, left panel shows almost complete integrations, where at the current state $c_s^2\approx 0.1$ and $w \approx 0.05$. The system is strongly explicitly-dependent on $\ln h$, and is no longer nearly autonomous, except very close to the center of the star.  However, the degree to which autonomy is violated (i.e.~the degree to which solutions with different central densities diverge from each other) \new{creates nontrivial structure in the distribution of solutions, which will lead to nontrivial structure in the $M-R$ curve. If the system is (nearly) autonomous then $\tilde{e}(\ln h)$ and $\tilde v(\ln h)$ for solutions with central $\ln h$ are nearly coincident.  Consider solution 1 with $\ln h_{c,1}$ and 2 with $\ln h_{c, 2} = \ln h_{c, 1}+\delta \ln h$,
\begin{align}
    \tilde e_1(\ln h) = \tilde e_2(\ln h + \delta\ln h ) &\approx \tilde e_2(\ln h) + \tilde e'\delta \ln h\\
    v_1(\ln h) = v_2(\ln h+\delta\ln h)&\approx v_2(\ln h )+v'\delta \ln h,
\end{align}
so nearby solutions differ \emph{globally} only by terms that can be determined \emph{locally}, which limits complexity.  Conversely, non-autonomous evolution will lead to highly dissimilar behavior in $\tilde e-v$ space.  This makes this  region of parameter space the most challenging to analyze physically.}    
     
The bottom, right panel presents the full integrations, where the current state has  $c_s^2\approx 0 \approx w$.  The system is almost entirely homogeneous in $\tilde e$, and it is dominated by the attractor of the $\tilde e$-axis rather than the fixed point.\new{\footnote{This attractor only exists for the portion of the axis where the bracketed terms in Eq.\eqref{eq:tov-reformulated-etilde} sum to greater than zero, but when  $\ln h$ small this generically becomes the entire axis.} }  The overall pattern of the $\tilde e$--$v$ solution is ``frozen-in", making it possible to discuss the ``surface" properties of the star, such as the total mass and stellar radius, in terms of the solutions near the fixed point. \new{In particular, the spiral pattern in the $\tilde e$--$v$ plane for very high central density stars reflects the structure of the dynamical system in that region.  Such solutions spend most of their ``dynamical time" evolving in a nearly autonomous system, since we have (effectively) fixed the speed of sound to be constant at high enthalpies by using a polytrope.  The change to the regime where the sound speed scales like $\ln h$ is abrupt for these solutions, the structure in the $\tilde e$--$v$ plane is frozen in and the system undergoes nearly homogeneous evolution in $\tilde e$.  Therefore, $\tilde e /e\propto r^2$ remains nearly constant near the terminal value of $4\pi R^2$, and $v$ decreases by a small constant, but it is very near the terminal value $M/R$ for each solution. For the same configurations as bottom right panel of Fig.~\ref{fig:gamma-2-initial-system}, we plot the values of $\tilde e / e$ vs.~$v$ in Fig.~\ref{fig:final-state-rescaled}.  Qualitatively, a spiral in $\tilde e$--$v$ directly corresponds to a spiral in $\tilde e / e-v$ which is identical to a spiral in $M/R$--$R^2$, which then leads to a spiral in the $M$--$R$ plane. We display the $M/R$--$R^2$ and $M$--$R$ curves in Fig.~\ref{fig:spiral-gamma-2-macro}.  For the EoS we use here, $\Gamma=2$, the maximum mass point marks the onset where the fixed point becomes an effective description for the system.  For EoSs where the speed of sound rises more sharply just above saturation density, we expect the transition to the fixed-point description to become effective before the maximum mass is achieved.  

\new{
\begin{figure}
    \centering
    \includegraphics[width=0.99\linewidth]{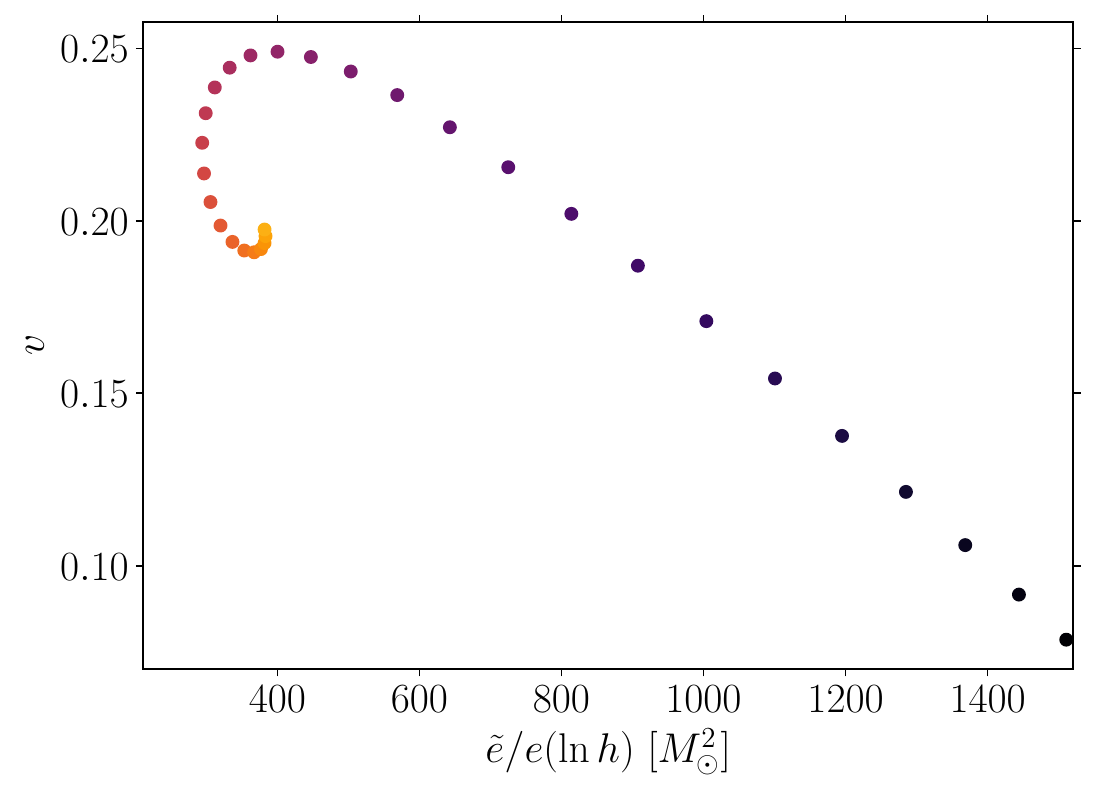}
    \caption{Same as the dots in the bottom right panel of Fig.~\ref{fig:gamma-2-initial-system}, but with the $x$--axis rescaled by $e(\ln h)$ for that panel. }
    \label{fig:final-state-rescaled}
\end{figure}
}

\begin{figure}
    \centering
        \includegraphics[width=0.99\linewidth]{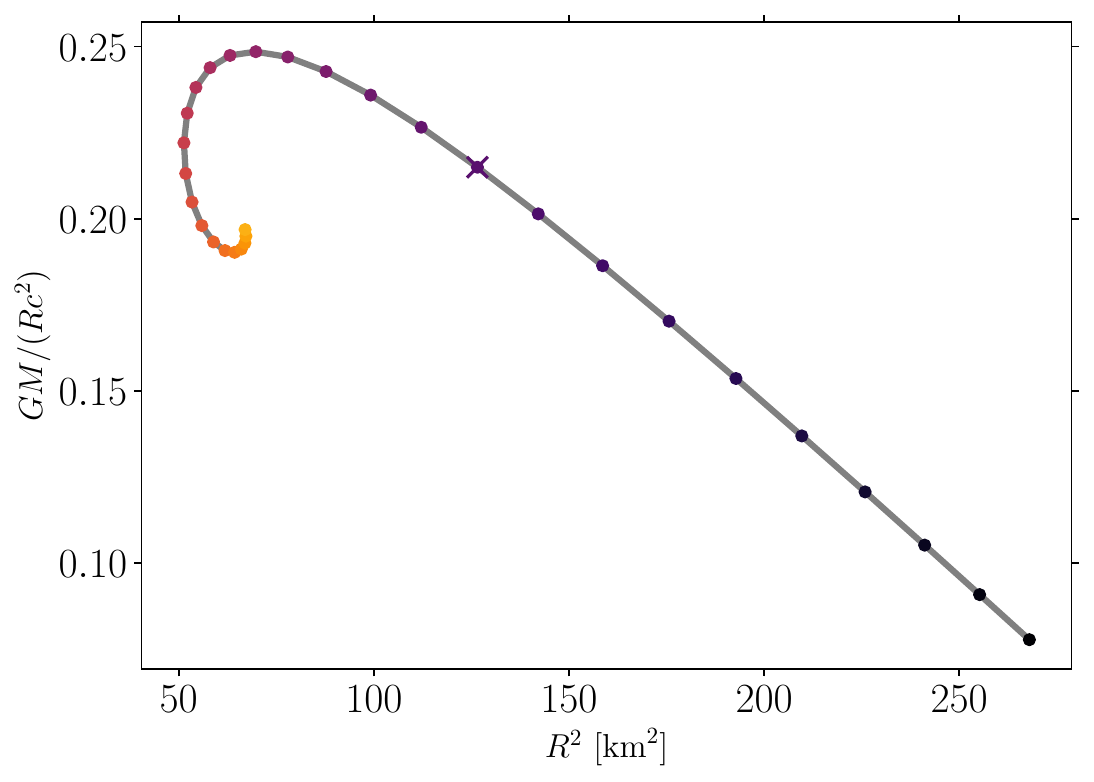}
    \includegraphics[width=0.99\linewidth]{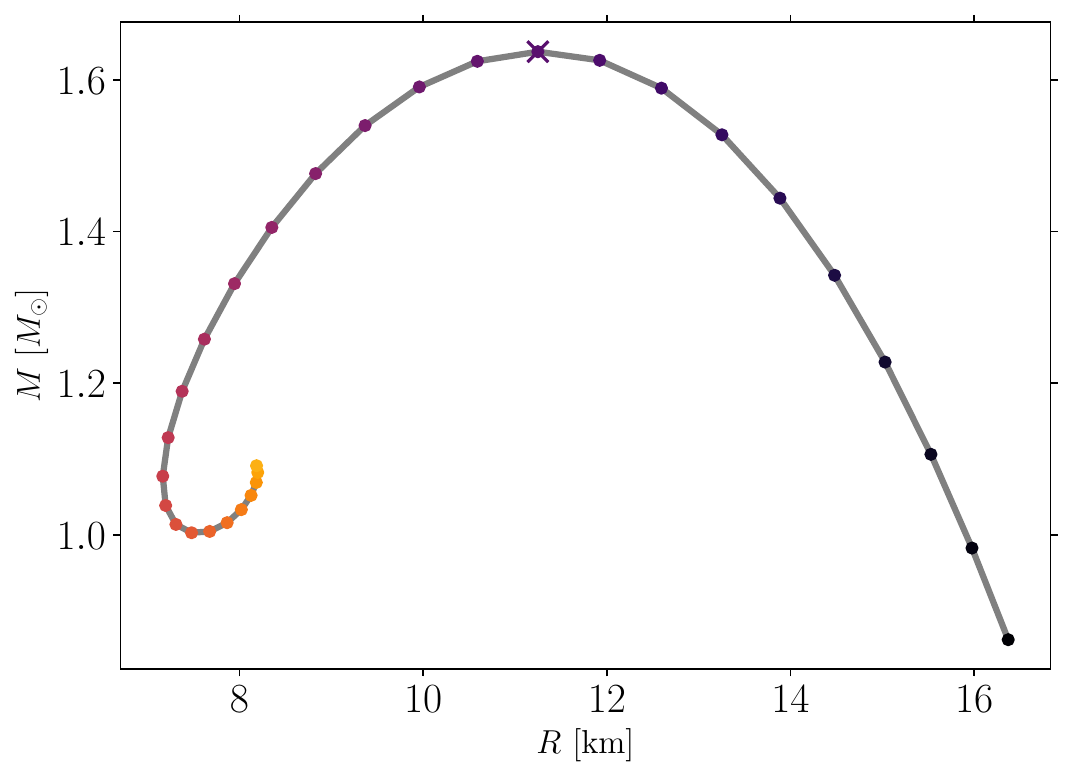}
    \caption{\emph{Top:} The $M/R$--$R^2$ curve for the TOV solutions from Fig.~\ref{fig:gamma-2-initial-system}.  \emph{Bottom:} The $M$--$R$ curve for the same solutions.  The maximum mass configuration is marked with a cross. }
    \label{fig:spiral-gamma-2-macro}
\end{figure}
}

\subsection{Implications of the Dynamical-Systems View on Astrophysical Observables}
\label{sec:implications}

Let us now connect the dynamical-systems picture of the previous subsections to two practical estimates: an approximate relation between high-density stiffness and the maximum mass of stars, and (more tentatively) an estimate of the radius at the maximum-mass point.  The logic is simple. For relativistic stellar sequences near their termination, \new{we find} the solution \new{typically} spends a substantial portion of its ``dynamical time'' in the neighborhood of the moving focus, identified in the previous subsections. The star’s macroscopic properties are then largely set by how and when the trajectory exits that near-fixed-point regime. 

Let us begin by connecting our state vector $\mathbf{x} = (\tilde e,v)$ to a more physical state vector $\mathbf{x}' = (m,r)$, which can be thought of as a coordinate transformation of phase space. Using the definitions of $(\tilde e,v)$ in Eqs.~\eqref{eq:v-def} and~\eqref{eq:etilde-def}, we immediately obtain
\begin{equation}
r^2 = \frac{c^4}{4\pi G}\,\frac{\tilde e}{e},
\qquad
m(r) = \frac{c^2}{G}\,v r
= \frac{c^4}{G}\,
v\,\sqrt{\frac{\tilde e}{4\pi G e}} .
\label{eq:m-from-v-etilde-units}
\end{equation}
This is then the coordinate transformation that will allow us to bridge the phase-space description to
the global quantities $(M,R)$.

Before building this bridge, let us first make a critical observation about the behavior of the dynamical system near the surface of the star. As discussed in Sec.~\ref{sec:dyn-sys}, once the star enters sufficiently low densities, the $\tilde e$ equation becomes nearly homogeneous. We can see this directly from  Eq.~\eqref{eq:tov-reformulated-etilde}, which, when $c_s^2$ becomes small and $v$ varies only slowly, reduces approximately to
\begin{equation}
\frac{d\tilde e}{d\ln h}
\simeq
\tilde e\,\frac{d\ln e}{d\ln h},
\qquad
\Rightarrow
\qquad
\tilde e \propto e .
\label{eq:etilde-propto-e-units}
\end{equation}
Hence, the ratio $\tilde e/e$ is approximately constant in the outer layers. By Eq.~\eqref{eq:m-from-v-etilde-units}, this implies that $r^2 \propto \tilde e/e$ changes only mildly as the surface is approached. The radial scale is effectively “frozen in’’ once the sound speed becomes small. Because very little mass resides in the outermost layers, the total mass is likewise largely determined at the density scale where the system exits the near-fixed-point regime.

\begin{figure}
    \centering
    \includegraphics[width=0.99\linewidth]{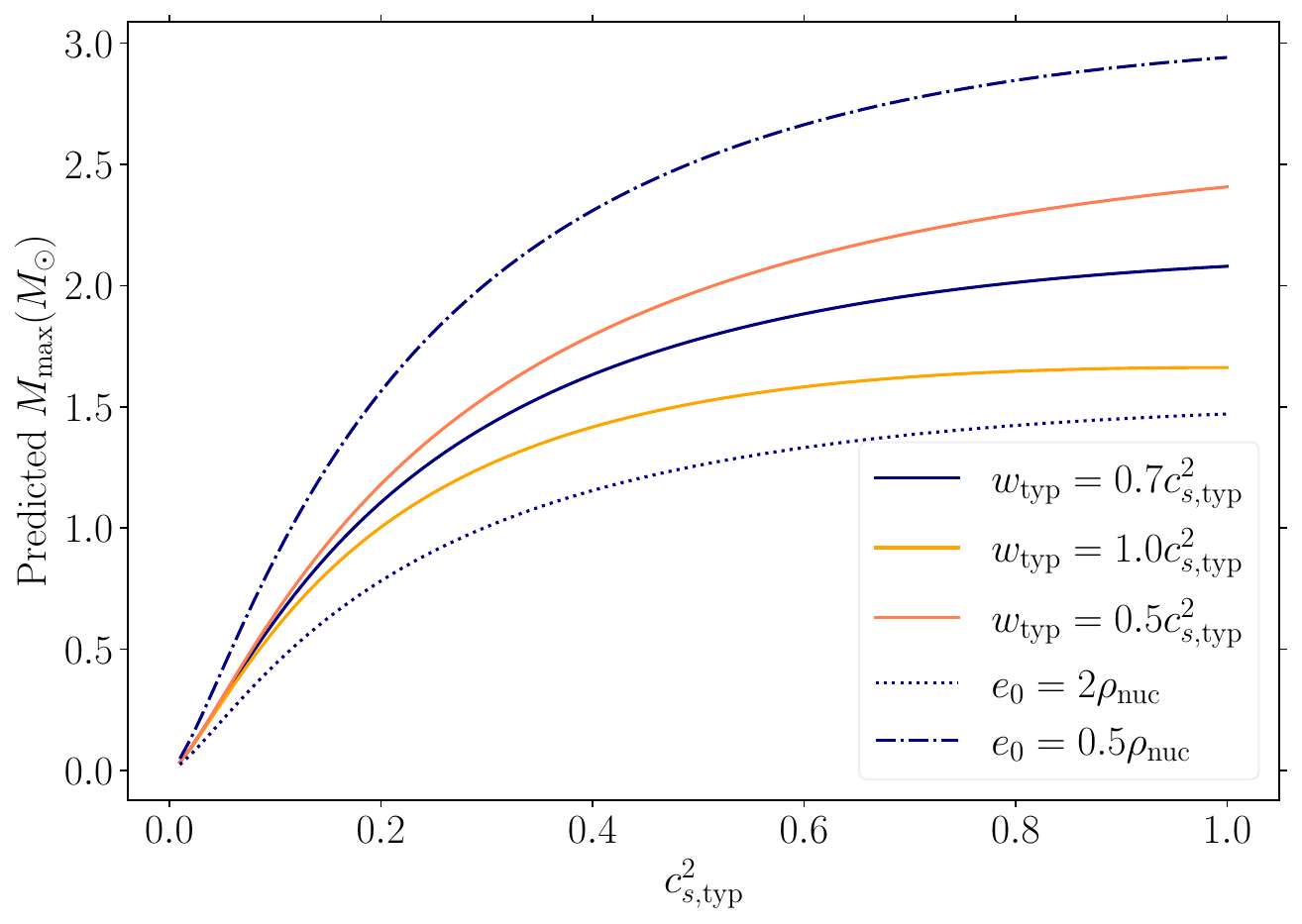}
    \caption{Parametric dependence of the fixed-point estimate for $M_{\max}$ [Eq.~\eqref{eq:Mmax-scaling-units}] as a function of $c_{s,\mathrm{typ}}^{\,2}$ for several representative choices of $w_{\mathrm{typ}}$ and $e_0$.  We consider ``neutron-star like" configurations with $e_0$ near $\rho_{\rm nuc}c^2$, the mass density of atomic nuclei ($\approx 2.8\times 10^{14}\,\rm{g}/{cm^3}$). Solid lines vary  $w_{\rm typ}$ while keeping $e_0=\rho_{\rm nuc} c^2$  fixed, while dash-dot and dotted lines vary $e_0$ while keeping $w_{\rm typ} = 0.7 c_{s, \rm typ}^2$ fixed. These results are qualitatively consistent with, \emph{e.g.}~\cite{Rhoades:1974fn, Kalogera:1996ci}.
    This plot is intended to illustrate the structure of the scaling, not to define a unique prescription for $(w_{\mathrm{typ}},e_0)$.
    %
    }
    \label{fig:Mmax_parametric}
\end{figure}

The density scale $e_0$ at which the system transitions out of the stiff, near-fixed-point regime into a much softer region is therefore important. Let us assume that the maximum-mass configuration exits this regime while still close to the diagonal $\tilde e \simeq v$.
At the fixed point of the frozen system, we showed that $\tilde e_\star = v_\star$, so evaluating Eq.~\eqref{eq:m-from-v-etilde-units} at $e=e_0$ gives
\begin{equation}
M_{\max}
\simeq
\frac{c^4}{G}
\frac{v_\star^{3/2}}{\sqrt{4\pi G e_0}}\,,
\label{eq:Mmax-leading-units}
\end{equation}
where we have used the fact that the enclosed mass does not increase much after the energy decreases drops below $e_0$. Using the fixed-point location of Eq.~\eqref{eq:fixedpoint-location}, we obtain the scaling relation
\begin{equation}
M_{\max}
\simeq
\frac{c^4}{G}
\left(
\frac{2 c_{s,\mathrm{typ}}^{\,2}}
{4 c_{s,\mathrm{typ}}^{\,2} + (1+w_{\mathrm{typ}})^2}
\right)^{3/2}
\frac{1}{\sqrt{4\pi G e_0}} .
\label{eq:Mmax-scaling-units}
\end{equation}
Here $c_{s,\mathrm{typ}}^{\,2}$ and $w_{\mathrm{typ}}$
represent typical stiffness parameters in the high-density
region where most of the mass accumulates. Figure~\ref{fig:Mmax_parametric} illustrates how this scaling varies with the representative stiffness parameters and with the choice of exit density $e_0$;
the dependence on $e_0$ simply reflects that it sets the curvature scale at which the trajectory leaves the near-fixed-point regime.

To make Eq.~\eqref{eq:Mmax-scaling-units} directly usable, one needs an operational definition of $c_{s,\mathrm{typ}}^{\,2}$ and $w_{\mathrm{typ}}$ in the high-density region, where $\tilde e$ is largest. A simple choice is to average $c_s^2(e)$ and $w(e)$ over the inner core of the maximum-mass configuration, for example over $e\in[e_{c,\max}/2,\,e_{c,\max}]$, or alternatively to evaluate them at the density where $\tilde e$ attains its maximum. In practice, these choices differ only at the $\mathcal{O}(10\%)$ level for the EoSs we consider. One may also adopt a more ``thermodynamic'' proxy motivated by quasi-universality arguments: recent work~\cite{Saes:2024xmv} shows that an appropriately-defined average sound speed in a star tracks the ratio $p_c/e_c$ (see Eq.~(1) in~\cite{Saes:2024xmv}), so for stiff cores one may take $c_{s,\mathrm{typ}}^{\,2}\sim \kappa\, w_c$ with $\kappa=\mathcal{O}(1)$. In this sense, $w_{\mathrm{typ}}$ and $c_{s,\mathrm{typ}}^{\,2}$ can be viewed as two closely related stiffness measures rather than independent inputs.

The leading-order estimate in Eq.~\eqref{eq:Mmax-scaling-units} assumes the trajectory sits exactly at the fixed point. In reality, the maximum mass solution executes roughly half a spiral before exiting the fixed-point regime. \new{This approximates the condition 
\begin{equation}
    \begin{pmatrix}
        \partial_{\tilde e}v\sqrt{\tilde e} &\partial _{v}v\sqrt{\tilde e}
    \end{pmatrix} \perp 
    \begin{pmatrix}
        \deriv{\tilde e}{\ln h} & \deriv{\tilde v}{\ln h}
    \end{pmatrix}
\end{equation}
exactly when the system leaves the fixed-point regime.  In the very coarse approximation $c_s^2\ll 1/v, 2 v\ll 1$, and $w\ll 1$, this renders the critical point at $\tilde e\approx v,$ which is one half-cycle away from the 
the initial point, $(0, 0)$}

In the frozen approximation, linearized evolution in the outward variable $s = \ln h_c - \ln h$ yields
\begin{equation}
|\mathbf{\delta x(s)}|\sim e^{-\sigma s}
\times \text{(rotation with frequency $\omega$)},
\end{equation}
with $\sigma$ and $\omega$ given in Eq.~\eqref{eq:eigs-fixedpoint}. After approximately half a turn, $\omega s \simeq \pi$, so the amplitude is suppressed by
\begin{equation}
\label{eq:amplitude-supression}
\mu = \exp\!\left(-\sigma \frac{\pi}{\omega}\right).
\end{equation}
For representative relativistic stiffness—where both 
$w_{\mathrm{typ}}$ and $c_{s,\mathrm{typ}}^{\,2}$ are of order unity, as is typical in the high-density cores of maximum-mass configurations—this yields $\mu \sim 0.1$–$0.2$, with only mild variation across realistic equations of state (the ratio $\sigma/\omega$ varies by less than a factor of two for $0.3 \lesssim (w,c_s^2) \lesssim 1$). Motivated by Fig.~\ref{fig:gamma-2-initial-system}, we parametrize the resulting overshoot by
\begin{equation}
\tilde e(\ln h_0) \simeq \alpha \tilde e_\star,
\qquad
v(\ln h_0) \simeq \alpha v_\star,
\qquad
\alpha \approx 1.2,
\end{equation}
\new{with $\ln h_0$ the value of $\ln h$ where $e_0 = e(\ln h)$, or equivalently the $\ln h$ value where the speed of sound drops below the chosen threshold}. 
\new{  We display this process schematically in Fig.~\ref{fig:schematic-stitching-to-fixed-point}.  In the top panel we show the frozen system, for $c_s^2=0.2$, $w=0.1$, and its solution in blue, and the solution to the  Jacobian approximation in red.  The linearized, Jacobian approximation initially diverges from the exact solution, but then eventually converges as both solutions approach the fixed point.  This convergence serves to control the error in the approximation.  In the bottom panel, we show how  $c_s^2\approx0$ solutions are attached.  The location where this stitching occurs, as well as $\tilde e$ and $v$ at this point,  determines the macroscopic properties at leading approximation (marked by a star and diamond for the exact and approximate solutions respectively).  The maximum mass level set, marked by dotted lines, agree relatively well event though the solutions are distinct.    
\begin{figure}
    \centering
    \includegraphics[width=0.49\textwidth]{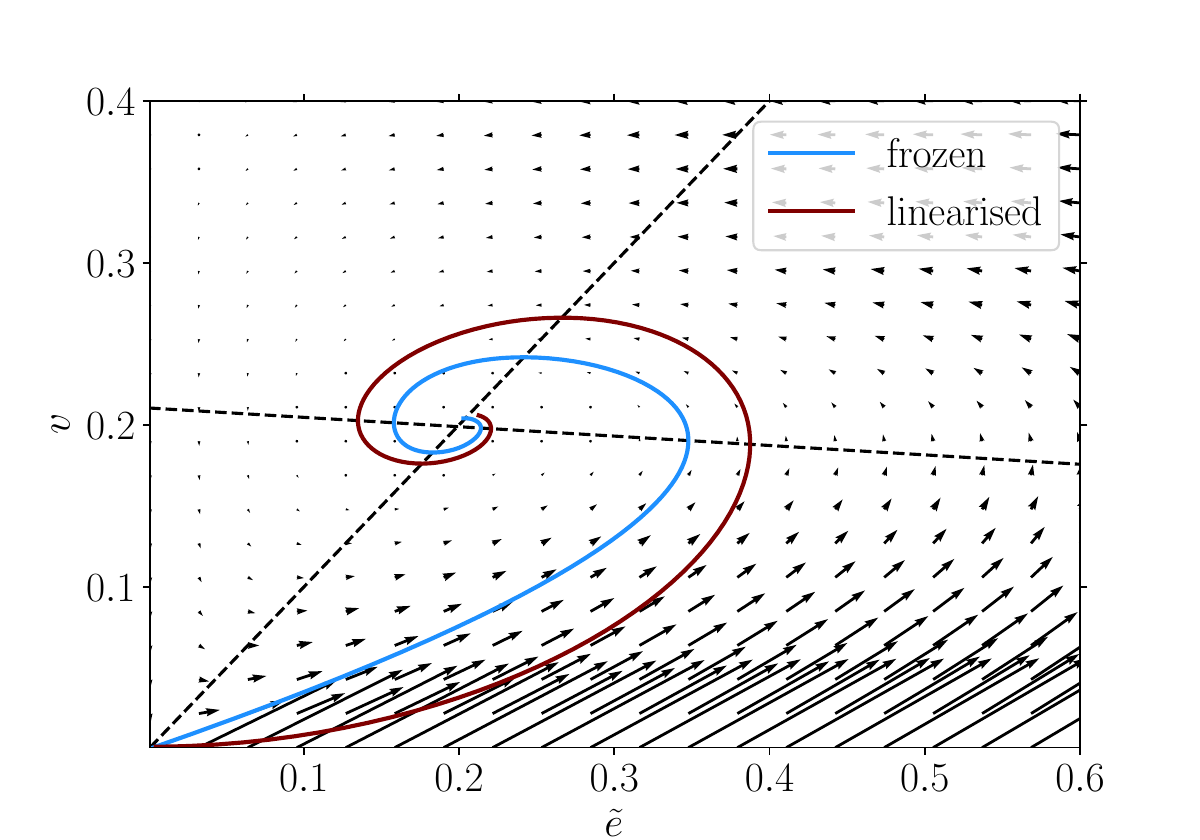}
    \includegraphics[width=.49\textwidth]{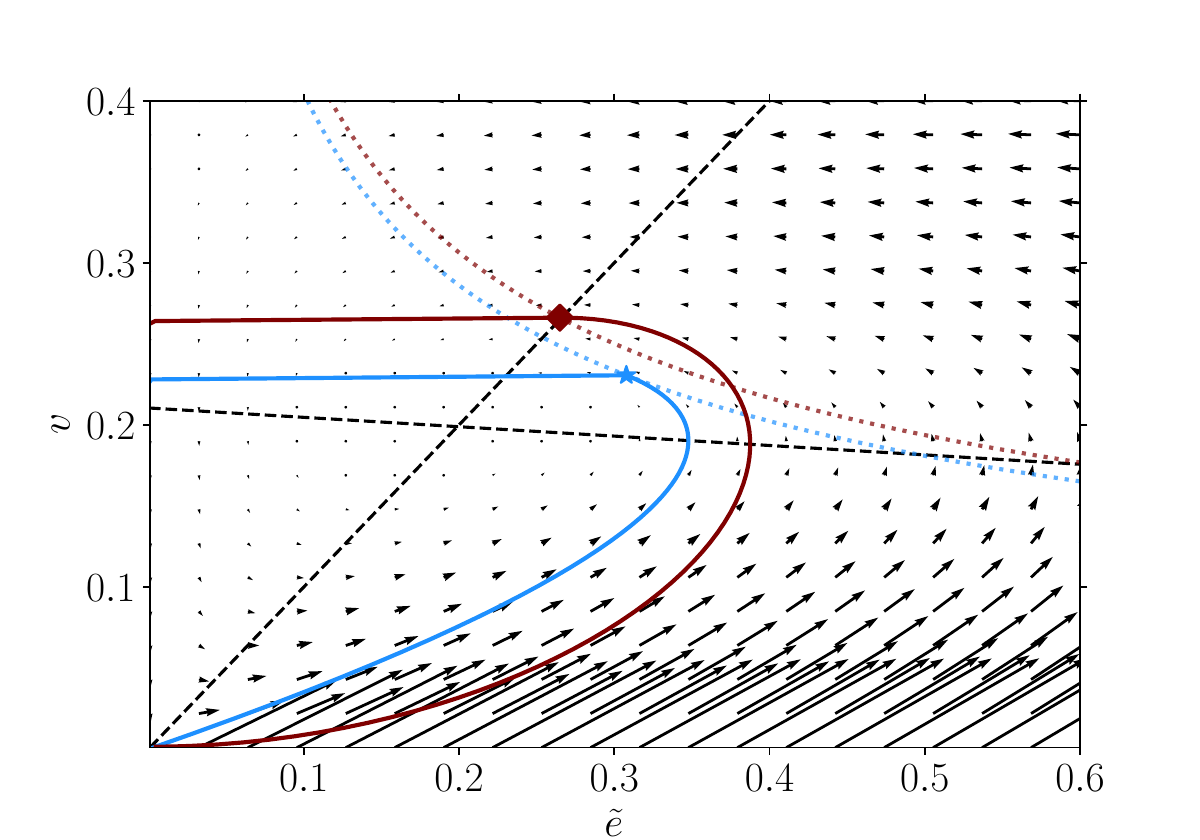}
    \caption{\new{Schematic diagram showing the approach to estimating the maximum mass for a system with constant sound speed.  \emph{Top}: The integrated solution under the ``frozen fixed point" approximation, and the same but approximating the system by the Jacobian.  \emph{Bottom:}  The integrated solution for a ``frozen fixed point"  transitioning at a given $\ln h$ to a $c_s^2=0$ solution (in blue, transition marked by a star).  The trajectory that gives the maximum mass is plotted.  In red, we plot the same but for the Jacobian approximation to the dynamical system, and transitioning to the $c_s^2=0$ solution at $\tilde e=v$, (\emph{e.g.} after a phase of $\pi$ from the initial condition at $(\tilde e, v)= (0,0)$. The level set of $\sqrt{\tilde e} v \propto M$ of the maximum mass is marked with a blue dotted line, and the corresponding estimate for the maximum mass that motivates Eq.~\ref{eq:mmax-ultrarel-fit} is given by the red dotted line.    } }
    \label{fig:schematic-stitching-to-fixed-point}
\end{figure}
}

With these we refine Eq.~\eqref{eq:Mmax-leading-units} to
\begin{equation}
M_{\max}
\simeq
\alpha^{3/2}
\frac{c^4}{G}
\frac{v_\star^{3/2}}{\sqrt{4\pi G e_0}} .
\label{eq:Mmax-final-units}
\end{equation}
The parameter $\alpha$ encodes a dynamical correction arising from the finite damping of the spiral, and should be viewed as \textit{distinct} from the $\mathcal{O}(10\%)$ variation associated with how $c_{s,\mathrm{typ}}^{\,2}$ and $w_{\mathrm{typ}}$ are operationally defined.  \new{Fundamentally, $\alpha\gtrsim1$ is a statement that the maximum mass configuration is associated with a trajectory that passes near to the fixed point, but exits the $c_s^2\approx\rm{const}$ regime while at a larger $v\sqrt{\tilde e}$ (in Fig.~\ref{fig:gamma-2-initial-system}, up and to the right). This is equivalent to the statement that the focus of the spiral  in the $M-R$ curve is consistently at a lower radius \emph{and, typically} lower compactness than the maximum mass solution (and, by definition, also has a lower mass, \emph{c.f.} Fig.~\ref{fig:M-R fixed points}).}  

Put another way, we have effectively estimated that the maximum-mass configuration has a trajectory $(\tilde{e}(\ln h), v(\ln h))$ that, by the time sound speed drops to a ``small" value, at $\ln h = \ln h_0$, has evolved to very nearly its maximum value of $v \sqrt{\tilde e}$, which is also the global maximum value of $v\sqrt{\tilde e}$.  This happens if there is sufficient time to undergo a ``half-cycle" of evolution, which requires $\Delta \ln h = \pi/\omega$, or \new{$\ln h_c-\Delta \ln h = \pi/\omega$}.  Therefore, we have implicitly identified the maximum central $\ln h_{c, \max} \approx \ln h_0 + \pi/\omega$. However, we also point out that the error in this approximation leads only to an error to the correction factor $\alpha$ in Eq.~\eqref{eq:Mmax-final-units}, which is why we do not seek to refine it further.

A similar estimate applies to the radius of the maximum mass stars. Using Eq.~\eqref{eq:m-from-v-etilde-units} and carrying out the same refinement through gives
\begin{equation}
R_{\max}
\simeq
\alpha^{1/2}
\sqrt{\frac{c^4}{4\pi G}\frac{v_\star}{e_0}}
\,
\left[1 - \dfrac{(1-2\alpha v_\star)}{\alpha v_\star}\ln h_0 \right]^{-1},
\label{eq:Rmax-units}
\end{equation}
where we kept the leading low-density correction only. 
Unlike the maximum mass, the radius depends explicitly on $\ln h_0$, and is therefore more sensitive to the details of the low-density EoS.  Despite this,  under the approximation to which we carry out the calculation, there are no additional ``free" parameters relative to the relation for $M_{\max}$.  Therefore, in practice, in the next sections, we will fit the $M_{\max}$ relation first, and then merely tolerate the comparatively large error in the radius relation that results from that choice of parameters.

In summary, the fixed-point geometry yields compact scaling relations
for $M_{\max}$ and $R_{\max}$ in terms of
(i) a high-density stiffness scale $(c_{s,\mathrm{typ}}^{\,2},w_{\mathrm{typ}})$
that determines the fixed point,
and (ii) an exit density $e_0$ that marks where the star leaves
the near-fixed-point regime.
The dimensionless correction factor $\alpha$ encodes
the finite spiral evolution before that exit.

\begin{figure*}
    \centering
    \includegraphics[width=0.495\linewidth]{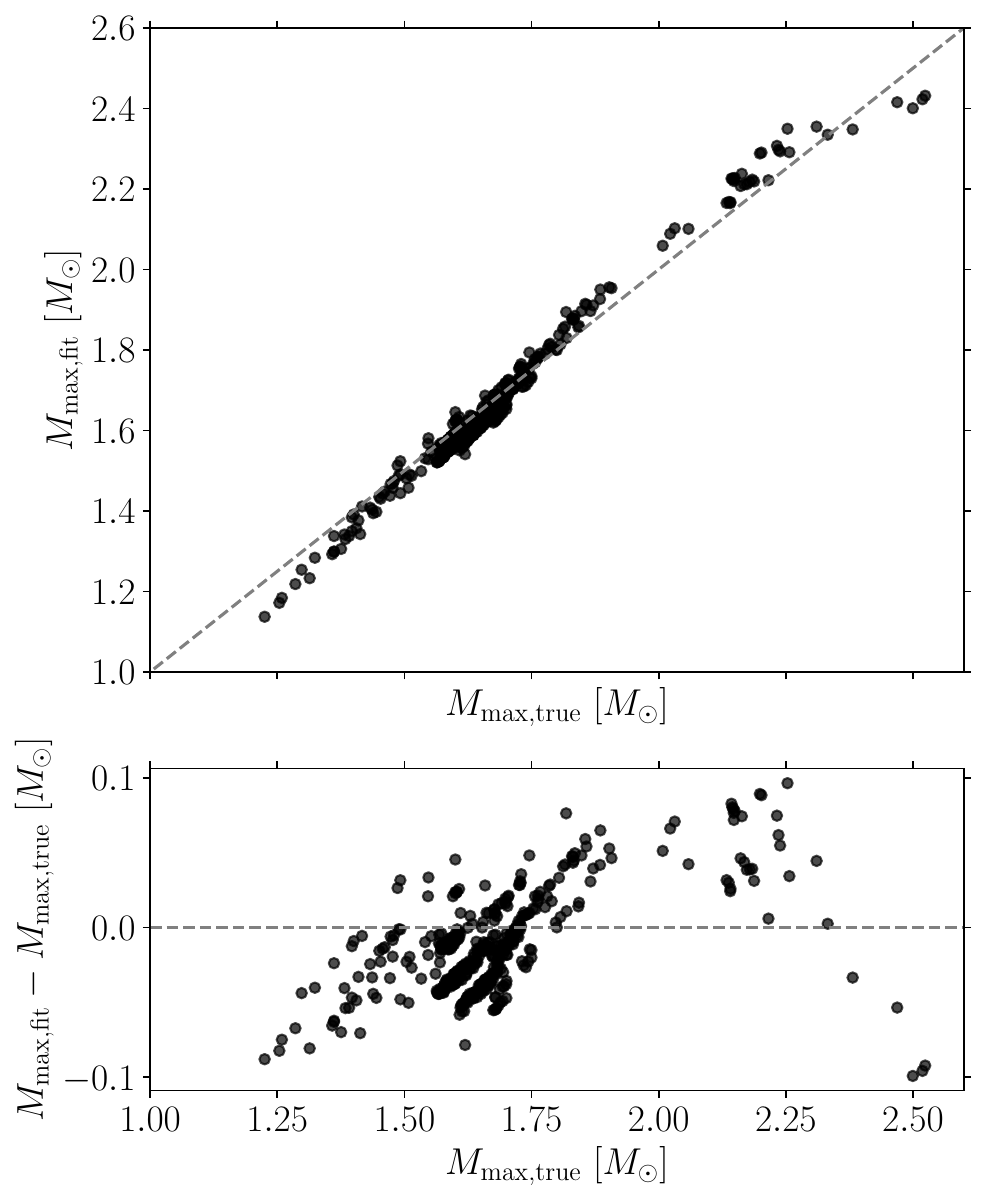}
    \includegraphics[width=0.495\linewidth]{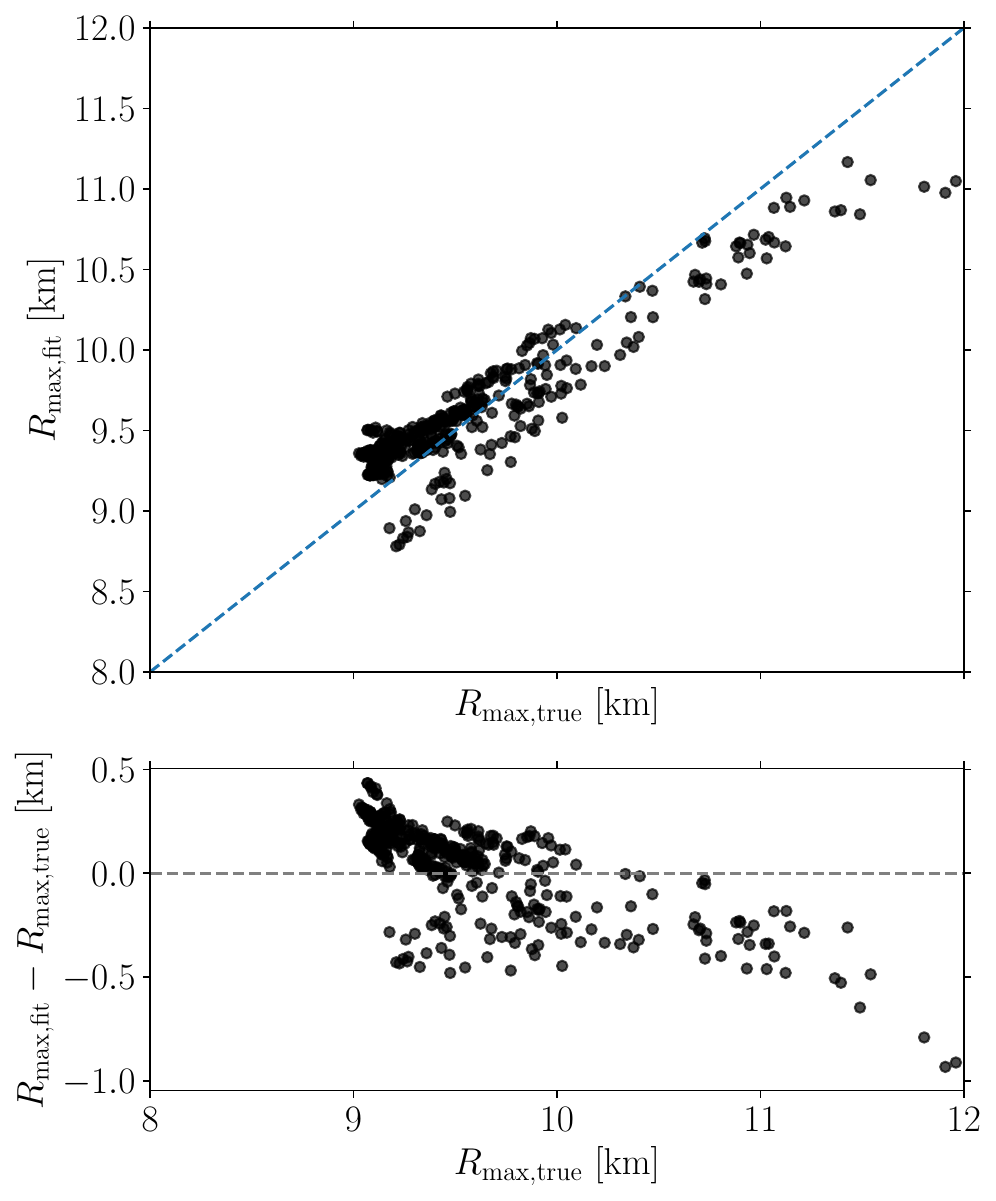}
    \caption{Comparison between the true maximum masses $M_{\max}$ (left panel) and the true radii of the maximum mass stars (right panel) of $\sim 500$ RMFT-informed Gaussian-process EoSs relative to the predictions of Eq.~\eqref{eq:Mmax-final-units} and~\eqref{eq:Rmax-units}, evaluated using a single operational prescription for $(c_{s,\mathrm{typ}}^{\,2},w_{\mathrm{typ}},e_0)$ (see text). The lower panels show fractional residuals. Observe the high accuracy of the predictions of the maximum mass, and the deterioration of this accuracy for the radii of those stars.
    }
    \label{fig:Mmax_and_Rmax_fit}
\end{figure*}

To assess the accuracy of the fixed-point estimate in a realistic setting, we can evaluate Eq.~\eqref{eq:Mmax-final-units} on an ensemble of candidate EoS using a \emph{single} operational prescription for the input quantities. Concretely, for each EoS we define the ``exit'' point $\ln h_0$ as the location where the sound speed falls below a small fixed threshold, $c_s^2(\ln h_0)=c_{s,0}^2$, and set $e_0 := e(\ln h_0)$. We then define $c_{s,\mathrm{typ}}^{\,2}$ and $w_{\mathrm{typ}}$ by averaging $c_s^2(e)$ and $w(e)$ over the high-density portion of the maximum-mass configuration, e.g.\ $e\in[e_{c,\max}/2,\,e_{c,\max}]$. With $\alpha$ fixed to the representative value inferred from the spiral geometry (cf.\ Fig.~\ref{fig:gamma-2-initial-system}), the resulting prediction recovers $M_{\max}$ at the $\sim 10\%$ level for the RMFT-informed Gaussian-process ensemble shown in the left panel of Fig.~\ref{fig:Mmax_and_Rmax_fit}. As anticipated, the corresponding radius estimate exhibits larger scatter, reflecting the enhanced sensitivity of $R_{\max}$ to low-density physics; see right panel of Fig.~\ref{fig:Mmax_and_Rmax_fit} and the discussion around Eq.~\eqref{eq:Rmax-units}.

\section{Instability in Newtonian Stars: The Compressible Limit}
\label{sec:newtonian-falling-sound-speeds}


In the previous section, we explained why, in relativity, stable stellar sequences generically terminate using a dynamical-systems approach: the $M$--$R$ curve is organized by the fixed-point structure of the dynamical system once the core becomes sufficiently relativistic.  In contrast, purely Newtonian sequences need not turn over at all.  When a Newtonian sequence \emph{does} terminate, the cause is different. 
If the EoS softens sufficiently at high density (or, equivalently, if the speed of sound 
does not increase quickly enough with density), equilibrium configurations become strongly centrally condensed, and, eventually, they violate the Newtonian (Chandrasekhar) stability criterion that the
appropriately-averaged effective 
polytropic 
exponent exceed
$4/3$~\cite{ShapiroAndTeukolsky}\footnote{Under certain assumptions this notion of an ``average" polytropic exponent can be made precise, see, for example, Ref.~\cite{ShapiroAndTeukolsky}, Eq.~(6.7.11).  For our purposes it suffices that the stellar structure of the configuration is sufficiently similar to a $4/3$ polytrope.}.  We refer to such a regime of highly centrally-concentrated configurations as the \emph{compressible limit}.  In this section, we recast this classic Newtonian instability in our enthalpy formulation, and use it to motivate a simple scaling relation between the termination point and the central pressure ratio $w_c\equiv p_c/e_c\simeq p_c/(c^2 \rho_c)$.

\subsection{The Newtonian equations}
\label{sec:newtonian-equations}

We continue to work with the Lindblom variables introduced in Sec.~\ref{sec:relativistic-mmax}, namely Eq.~\eqref{eq:v-def}, and we keep $\ln h$ as the independent variable.  The Newtonian limit corresponds to the weak-field and low-pressure regime, in which the enclosed compactness and the pressure-to-energy-density ratio satisfy $v\ll 1$ and $w\ll 1$.  In this regime, $4\pi G u\,p/c^4 \ll v$ and $1-2v\simeq 1$.  One may also use $e\simeq \rho c^2$ at leading order, but we will not do so systematically; the presentation is cleaner if we keep $e$ as an energy density and retain the same phase-space variables as in the relativistic discussion.

With this in mind, let us take the Newtonian limit of our structure equations. Expanding Eqs.~\eqref{eq:tov-standard-dimful}--\eqref{eq:tov-standard-dimful2} in $v\ll 1$ and $w\ll 1$ and keeping only the leading-order terms yields
\begin{align}
\label{eq:lindblom-newtonian-u}
\deriv{u}{\ln h} &= -\frac{2u}{v},\\
\label{eq:lindblom-newtonian-v}
\deriv{v}{\ln h} &= - \frac{4\pi G}{c^4}\,e\,\frac{u}{v} + 1.
\end{align}
In the second equation we have retained the leading self-gravity contribution from the energy density $e$ and dropped corrections that are higher order in $v$ and $w$.

For comparisons across EoS families, it is convenient to nondimensionalize lengths using the \emph{central} energy density $e_c$ of the configuration under consideration. Let us define
\begin{equation}
\bar u \equiv \frac{4\pi G e_c}{c^4}\,u,
\qquad
\bar e \equiv \frac{e}{e_c},
\end{equation}
so that $\bar u$ is dimensionless and $\bar e(\ln h_c)=1$ at the center.  Equations~\eqref{eq:lindblom-newtonian-u}--\eqref{eq:lindblom-newtonian-v} become
\begin{align}
\label{eq:lindblom-newtonian-reformulated-u}
\deriv{\bar u}{\ln h} &= -\frac{2\bar u}{v},\\
\label{eq:lindblom-newtonian-reformulated-v}
\deriv{v}{\ln h} &= - \bar e\,\frac{\bar u}{v} + 1.
\end{align}

Finally, the dimensionless combination that repeatedly appears in the Newtonian analysis is the mass scaled by $\sqrt{4\pi e_c}$.  With $G$ and $c$ explicit, the corresponding \emph{dimensionless} mass is
\begin{equation}
\label{eq:Mbar-def}
\bar M \equiv v(0)\sqrt{\bar u(0)}
      = \frac{G M}{c^4}\sqrt{4\pi G e_c},
\end{equation}
where $v(0)=GM/(Rc^2)$ and $u(0)=R^2$ are evaluated at the surface ($\ln h=0$).



 
\subsection{The compressible limit}
\label{sec:compressible-limit}
Given the form of Eqs.~\eqref{eq:lindblom-newtonian-reformulated-u} and~\eqref{eq:lindblom-newtonian-reformulated-v}, all of the impact of the EoS on the stellar structure can be encapsulated by a function $\bar e(\ln h)$, which can be thought of as the density profile of the star.  For a concrete example, note that for polytropes 
\begin{equation}
    \bar e(\ln h) = \quant{\frac{\ln h}{\ln h_c}}^n.   
\end{equation}
Since $\bar e(\ln h)$ is nondecreasing, and bounded between $0$ and $1$, there are two canonical limiting cases.  First, $\bar e(\ln h)=1$ everywhere, dropping to $0$ only at  $\ln h =0$.  This is the familiar case of a constant density star ($n=0$ polytrope), which can be thought of as the ``incompressible limit" of the equations.  However, there's an additional case which is ``dual'' to the previous one: $\bar e(\ln h)$ drops to zero immediately at $\ln h= \ln h_c$
and stays at zero all the way down to $\ln h = 0$.  This star is maximally centrally condensed, and can be associated with an $n=\infty$ polytrope, so here we will term this the ``compressible limit" of the equations.  This configuration is, of course, physically uninteresting because it is highly unstable.  However, we show below that by thinking of configurations \emph{near} this limiting case, the  structure of marginally stable stars can be better understood.

Newtonian configurations become radially unstable once the EoS is sufficiently soft that the star approaches the Chandrasekhar threshold, i.e.~an average polytropic exponent near $4/3$ \cite{ShapiroAndTeukolsky}.  
Restated in terms of $\bar e(\ln h)$, this implies that overall the density profile must behave like 
\begin{equation}
    \bar e(\ln h) \sim \quant{\frac{\ln h}{\ln h_c}}^3.    
\end{equation}
This is qualitatively similar to the compressible limit: most of the mass resides in a compact core, while an extended envelope carries little mass but a large fraction of the radius.  This separation of scales  motivates a simple core--envelope matching description that we will use to characterize the structure of marginally stable, Newtonian stars.

\subsubsection{Near-core expansion}
Let us use the outward evolution variable introduced in Sec.~\ref{sec:ds_primer}, $s := \ln h_c - \ln h$, so that $s=0$ at the center and increases monotonically toward the surface. Expanding the Newtonian structure equations [Eqs.~\eqref{eq:lindblom-newtonian-reformulated-u}-\eqref{eq:lindblom-newtonian-reformulated-v}] about $s=0$  leads to the near-core solution 
\begin{align}
\bar u(s) &= 6s + \frac{9}{5\,c^{2}_{s,c}}\,s^2 + {\mathcal{O}}(s^3), \label{eq:compressible-core-ubar}\\
v(s) &= 2s - \frac{3}{5\,c^{2}_{s,c}}\,s^2 + {\mathcal{O}}(s^3), \label{eq:compressible-core-v}
\end{align}
where $c^{2}_{s,c}:=c_s^2(\ln h_c)$ is the sound speed squared evaluated at the center of the star.

A convenient quantity to track is the ratio
\begin{equation}
\frac{v}{\bar u}
= \frac{m(r)c^2}{4\pi e_c\,r^3}
= \frac{1}{3}\,\frac{\langle e\rangle_r}{e_c},
\label{eq:compressible-v-over-ubar}
\end{equation}
i.e.\ $3v/\bar u$ is the enclosed mean-density ratio.
Using Eqs.~\eqref{eq:compressible-core-ubar}--\eqref{eq:compressible-core-v},
we find near the center
\begin{equation}
3\frac{v}{\bar u}
= 1 - \frac{3}{5\,c^{2}_{s,c}}\,s + {\mathcal{O}}(s^2).
\label{eq:mean-density-expansion}
\end{equation}
Meanwhile, in the Newtonian regime $w\ll 1$,
Eq.~\eqref{eq:dlnedlnh} implies $d\ln \bar e/d\ln h \simeq 1/c_s^2$,
so that
\begin{equation}
\bar e(s) = 1 - \frac{s}{c^{2}_{s,c}} + {\mathcal{O}}(s^2).
\label{eq:compressible-ebar-core}
\end{equation}
Comparing Eqs.~\eqref{eq:mean-density-expansion} and \eqref{eq:compressible-ebar-core},
we are naturally led to the \textit{resummed} closure
\begin{equation}
3\frac{v}{\bar u} \simeq \bar e^{3/5},
\label{eq:compressible-closure}
\end{equation}
which reproduces the first nontrivial term in the near-core expansion.

For intuition, one may also combine
Eqs.~\eqref{eq:lindblom-newtonian-reformulated-u}--\eqref{eq:lindblom-newtonian-reformulated-v}
to obtain an evolution equation for the enclosed mean-density ratio
\begin{equation}
\deriv{}{\,\ln h}\!\left(\frac{v}{\bar u}\right)
= \frac{1}{v}\left(3\frac{v}{\bar u}-\bar e\right).
\end{equation}
This equation shows that the difference $(3v/\bar u-\bar e)$ controls the drift of the mean-density ratio.
Near the center, this difference is positive at ${\cal{O}}(s)$, consistent with the fact that the enclosed mean density must lie above the local density for any decreasing profile.

The closure of Eq.~\eqref{eq:compressible-closure} cannot hold all the way to the surface. This is because $\bar e\to 0$ near the surface, while $3v/\bar u$ approaches the finite mean-density ratio $3M/(4\pi R^3 e_c/c^2)$.  A natural point to transition to an envelope description is where the ``self-gravity'' term in Eq.~\eqref{eq:lindblom-newtonian-reformulated-v} becomes ${\mathcal{O}}(1)$, i.e.~$\bar e\,\bar u/v \sim 1$.  Using Eq.~\eqref{eq:compressible-closure}, this occurs at a characteristic density
\begin{equation}
\bar e_s \simeq 3^{-5/2} \approx 6.4\times 10^{-2}.
\label{eq:compressible-es}
\end{equation}
In the compressible limit, this provides a convenient and nearly EoS-independent matching scale.

\subsubsection{Stitching to a crust solution}
In the low-density envelope, where $\bar e \ll v/\bar u$ (equivalently $\bar e\,\bar u/v \ll 1$), Eq.~\eqref{eq:lindblom-newtonian-reformulated-v} simplifies to
\begin{equation}
\deriv{v}{\ln h} \simeq 1,
\end{equation}
so integrating inward from the surface ($\ln h=0$) gives
\begin{equation}
v(\ln h) = v_0 + \ln h,
\label{eq:compressible-envelope-v}
\end{equation}
where $v_0 := v(0) = GM/(Rc^2)$.  Using Eq.~\eqref{eq:lindblom-newtonian-reformulated-u}, one obtains
\begin{equation}
\bar u(\ln h) = \bar u_0 \left(\frac{v_0}{v_0+\ln h}\right)^2,
\label{eq:compressible-envelope-ubar}
\end{equation}
with $\bar u_0 := \bar u(0)$.

In this regime,
\begin{equation}
\bar u\,v^2 = \bar u_0\,v_0^2 = \bar M^2,
\end{equation}
where $\bar M \equiv v(0)\sqrt{\bar u(0)}$ is the dimensionless mass defined in Eq.~\eqref{eq:Mbar-def}.  Thus, once the star enters the low-density envelope, $\bar M$ is effectively frozen in: the envelope carries little additional mass while contributing substantially to the radius.

\subsubsection{Controlled regimes and a polytropic benchmark}
\label{sec:polytropic-benchmark}

The stitching picture above becomes controlled in two limits.  In the incompressible limit, the density profile is nearly uniform and the core expansion remains accurate essentially to the surface.  In the compressible limit, the density drops rapidly enough that the envelope approximation becomes valid relatively deep inside the star (around $\bar e\sim \bar e_s$), giving an overlap region for matching. We want to extend the picture beyond the polytropic cases analyzed above in order to understand to what extent stars with arbitrary EoS can be analyzed in this picture.  To do this, we aim to approximate stars by the polytropic cases above, and control the error introduced by making this approximation by understanding neglected higher-order terms. 

\begin{figure}
    \centering
    \includegraphics[width=0.5\textwidth]{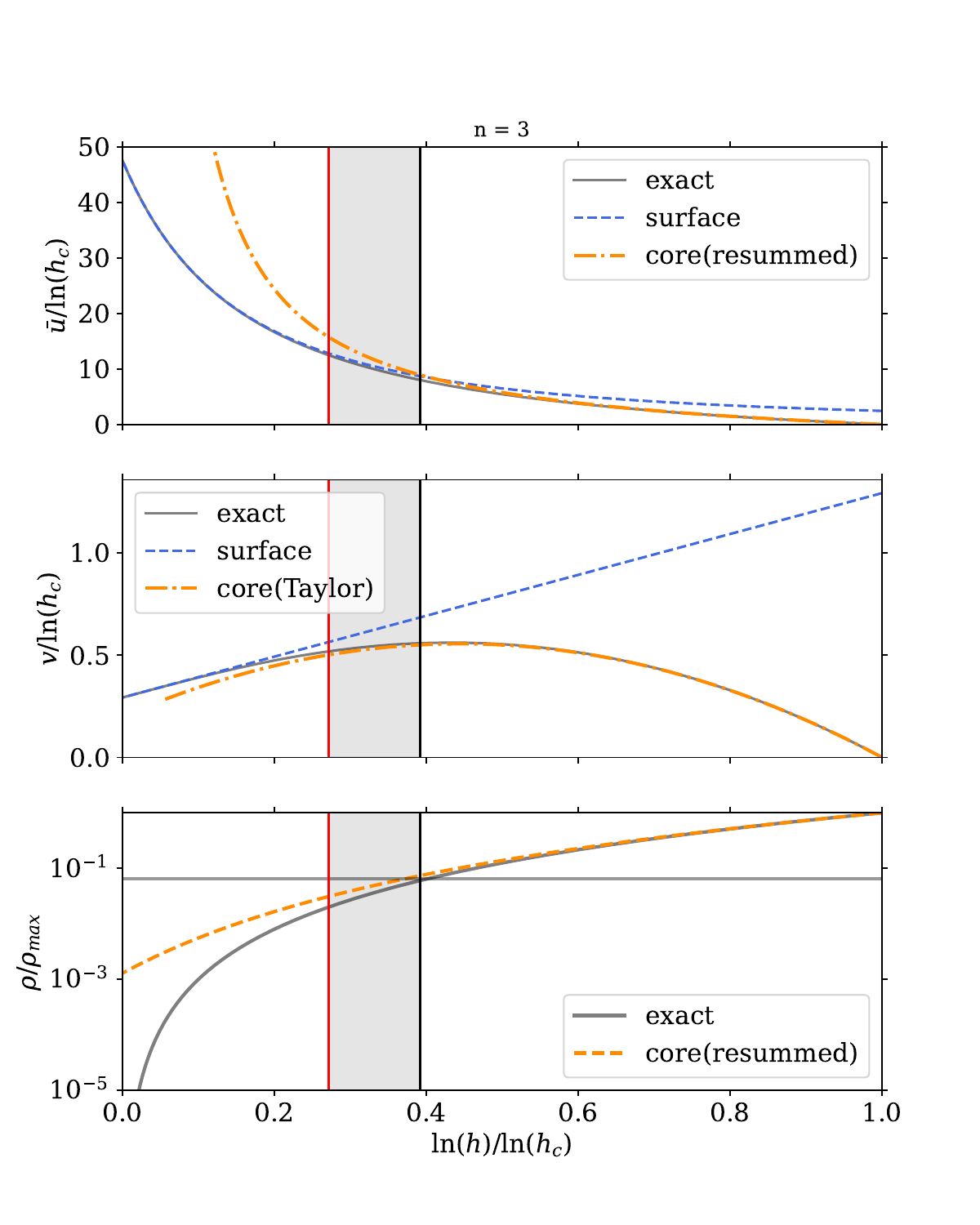}
    \caption{Illustration of the compressible-limit stitching strategy for a Newtonian $n=3$ polytrope. Top and middle panels compare the exact solution for $\bar u(s)$ and $v(s)$ to (i) the near-core expansion (both Taylor-expanded and resummed) and (ii) the low-density envelope approximation. The bottom panel shows the corresponding density profile (for a Newtonian polytrope, $\bar e$ and $\bar\rho$ coincide up to a factor of $c^2$).  \new{In a black vertical line, we mark where the density falls below the critical value we identified $\rho/\rho_{\max}=0.06$~[Eq.~\eqref{eq:compressible-es}],, where the surface solution may be expected to apply.  In red, we identify an estimate for where the density has likely become ``too small", $\rho/\rho_{\max}=.02$, so that the resummed expression for $v/\bar u$ is a poor approximation at lower densities.}  We expect matching to be optimal between these two lines, which we mark in shaded gray. This overlap region underlies the quasi-universal scaling of $\bar M_{\max}$ in Eq.~\eqref{eq:compressible-universal}.
    }
    \label{fig:newtonian-stitching}
\end{figure}

An obvious way forward is to identify a star by a polytrope with the same core sound speed, which leads us to define
\begin{equation}
n \equiv \frac{\ln h_c}{c^{2}_{s,c}},
\end{equation}
which coincides with the usual Lane--Emden index when the EoS is truly polytropic. 
For \textit{polytropes}, in terms of the normalized variable $\hat s := s/\ln h_c$, the near-core expansion of $v$ takes the form
\begin{equation}
v = \ln h_c\left[2\hat s - \frac{3n}{5}\hat s^2 + \frac{n(17n-50)}{175}\hat s^3 + {\mathcal{O}}(\hat s^4)\right].
\label{eq:compressible-v-polytrope}
\end{equation}
The cubic coefficient is proportional to $(17n-50)$ and is therefore anomalously small near $n\simeq 50/17\simeq 2.94$, i.e.~close to the $n\simeq 3$ regime associated with marginal Newtonian stability.  This helps explain why low-order core expansions, combined with the envelope solution, can provide a surprisingly accurate description near the terminal configuration.  We display the stitching, with the Taylor expansion in $v$ carried out to second order in $\hat s$, in Fig.~\ref{fig:newtonian-stitching} for an $n=3$ polytrope.  We contrast this with examples where the stitching does not produce an effective solution in Appendix~\ref{sec:when-expansion-doesnt-work}.


If the EoS is not polytropic, however, then higher-order terms may not follow the same pattern as they do for a polytropic EoS.  Generically, higher derivatives of the sound speed $dc_s^2/d\ln h$ enter the higher-order coefficients of the expansion.  These terms may be very large if the EoS has a quickly-varying sound speed. Even worse, if the sound speed drops sharply, but only over a small range of $\ln h$ (such as in a first-order phase transition), then the quadratic term in Eq.~\eqref{eq:compressible-v-polytrope} could become arbitrarily large in absolute value, while the physical structure of the star is essentially unaffected.   

Instead, we choose a different definition of $n$ to insert in Eq.~\eqref{eq:compressible-v-polytrope}, which serves to ``average'' over the entire star:
\begin{equation}
    n \equiv \frac{
    \ln h_c
    }{ w_c} - 1.
\end{equation}
This expression is also exact for polytropes, but because 
\begin{equation}
    w_c = \int_0^{\ln h_c} \bar e(\ln h) d\ln h,
\end{equation}
this definition incorporates information about the EoS over the entire stellar structure.  The cost to this assignment is that our approximation to $v$ is no longer formally correct even at second order in $\hat s$ if the EoS is not polytropic. Nonetheless, we find this scheme has certain desirable features. For example, under this approximation, it can easily be shown that the coefficients of higher-order terms are all bounded in magnitude by unity (see the Appendix~\ref{sec:approximation-justification}).   With this identification of $n$, we can approximate the stellar structure given only a single pair of numbers for each EoS, $(\ln h_c, w_c)$.  The ability to extract realistic stellar properties from this approximation, in certain cases, will lead to universal behavior.

\subsubsection{Scaling of the maximum mass}
\label{sec:scaling-of-mmax-newtonian}
In the compressible limit, the matching between the near-core solution and the envelope solution occurs at a density $\bar e_s$ that is approximately fixed across EoS families (Eq.~\eqref{eq:compressible-es}).  Since $\bar u_s \sim s_s$ and $v_s\sim s_s$ at the matching point, we obtain the scaling
\begin{equation}
\bar M \propto (\ln h_c)^{3/2}.
\end{equation}
Near the onset of Newtonian instability, one has $\ln h_c \simeq 4w_c$ (exactly so for an $n=3$ Newtonian polytrope). This reduces our two bits of data ($\ln h_c, w_c$) down to a single number, which we take to be $w_c$.  So, the terminal configuration satisfies
\begin{equation}
\bar M_{\max} \simeq \gamma\, w_c^{3/2},
\label{eq:compressible-universal}
\end{equation}
where $\gamma$ is an ${\mathcal{O}}(10)$ constant that depends only weakly on the detailed EoS within the compressible class (with variability of about 10\%, as we will show in the next section). As a point of reference, for the \emph{marginally stable} Newtonian polytrope ($\Gamma=4/3$, i.e.\ $n=3$),
the Lane--Emden solution gives $\gamma \simeq 16.15$
(with $\gamma=8[-\xi_1^2\theta'(\xi_1)]$ in standard notation). Realistic neutron-star EoSs are not polytropes and, in practice, relativistic corrections become important before a purely Newtonian $n\simeq 3$ description applies; we quantify those effects in Sec.~\ref{sec:post-newtonian-extension}.

\section{Post-Newtonian Extension of the compressible limit: Astrophysical Implications}
\label{sec:post-newtonian-extension}

We now discuss how the compressible-limit picture extends into the \emph{post-Newtonian} regime.  This gives a simple quasi-universal relation for neutron stars that connects the maximum mass to the central stiffness of the maximum-mass configuration.  We then use this relation to interpret astrophysical constraints, and we also clarify how the post-Newtonian regime connects to the highly relativistic fixed-point picture of Sec.~\ref{sec:relativistic-mmax}.

\subsection{Relativistic Corrections}
It is perhaps not too surprising that the compressible limit extends \textit{qualitatively} to relativistic stars, but the \textit{quantitative} treatment must change.  In the Newtonian compressible limit of the previous section, we found that the terminal configuration satisfies $\ln h_c\simeq 4w_c$, leading to the universal scaling $\bar M_{\max}\simeq \gamma w_c^{3/2}$ [see Eq.~\eqref{eq:compressible-universal}].  Since the leading relativistic corrections to the stellar-structure equations enter through the dimensionless combinations $(1+w)$, $(1+3w)$, and $(1-2v)$, one expects the first post-Newtonian deformation of this relation to be largely EoS-insensitive when expressed in terms of $w_c$.  In practice, these corrections steepen the enthalpy gradient and suppress the dimensionless size of the star at fixed $w_c$, so the terminal value of $\bar M$ should fall \textit{below} the Newtonian prediction.

More precisely, it is straightforward to estimate the leading post-Newtonian corrections to the quantities computed in Sec.~\ref{sec:newtonian-falling-sound-speeds} by starting from the exact reformulation of the TOV system, Eqs.~\eqref{eq:tov-reformulated-etilde} and~\eqref{eq:tov-reformulated-v}, which we slightly simplify below:
\begin{align}
    \label{eq:tov-reformulated-etilde-reproduced}
    \deriv{\tilde e}{\ln h} &= \tilde e \quant{\frac{-2(1-2 v)}{w \tilde e + v} +\frac{1+w}{c_s^2}};\\
    \label{eq:tov-reformulated-v-reproduced}
    \deriv{v}{\ln h} &= \frac{-(1-2v)}{w \tilde e + v}\quant{\tilde e - v}.
\end{align}
Per the boundary conditions, the near-center solution has the universal form
\begin{align}
    \tilde e &\sim \bar u \sim \frac{6}{1+3w_c}\,(\ln h_c -\ln h),\\
    v &\sim \frac{2}{1+3w_c}\,(\ln h_c - \ln h),\
    \label{eq:v-exp}
\end{align}
where we used $\tilde e=\bar u\,\bar e$ with $\bar e\simeq 1$ at leading order near the origin.
Rather than Taylor expanding in the small parameter $w_c$, we find it more robust to keep the factor $(1+3w_c)$ in the above equations unexpanded, i.e.\ to treat the dominant post-Newtonian correction as a \textit{homogeneous rescaling} of the Newtonian core solution.  Additional post-Newtonian effects---including the geometric factor $(1-2v)$---introduce a further ${\mathcal O}(1)$ suppression in the mildly relativistic regime; we bundle these into a single phenomenological coefficient and write, schematically,
\begin{align}
    \bar u_{\rm rel} &\approx \frac{\bar u_{\rm Newt}} {1+ \beta w_c},\\
    v_{\rm rel}&\approx \frac{v_{\rm Newt}}{1+\beta w_c}.
\end{align}
Here $\beta$ depends mildly on how one defines the ``representative'' post-Newtonian correction across the star, but it should be at least $\sim 3$ (since the core expansion alone already produces the factor $1/(1+3w_c)$).  The larger fitted values that we find below reflect the fact that all post-Newtonian corrections at leading order increase the magnitude of $d\ln h/dr$.

With this in hand, we can estimate the maximum mass appropriate for a given EoS when its maximum-mass configuration is only mildly relativistic, by deforming the Newtonian compressible-limit relation of Eq.~\eqref{eq:compressible-universal}.  Denoting by $w_{c,\max}\equiv p_{c,\max}/e_{c,\max}$ the central ratio evaluated at the maximum-mass star, we obtain
\begin{equation}
    \label{eq:newtonian-mmax-prescription}
     \bar M_{\max} \equiv \frac{G M_{\max}}{c^4}\sqrt{4\pi G e_{c,\max}}
     \simeq \gamma 
     \frac{(w_{c,\max})^{3/2}}{(1+\beta w_{c,\max})^{3/2}}.
\end{equation}
The functional dependence presented above is analytically motivated, but $\gamma$ and $\beta$ are phenomenological; below we determine them by an iterative fit to the Gaussian-process EoS sample, which yields $\gamma=19.8$ and $\beta = 9$. 

\begin{figure*}
    \centering
    \includegraphics[width=0.99\linewidth]{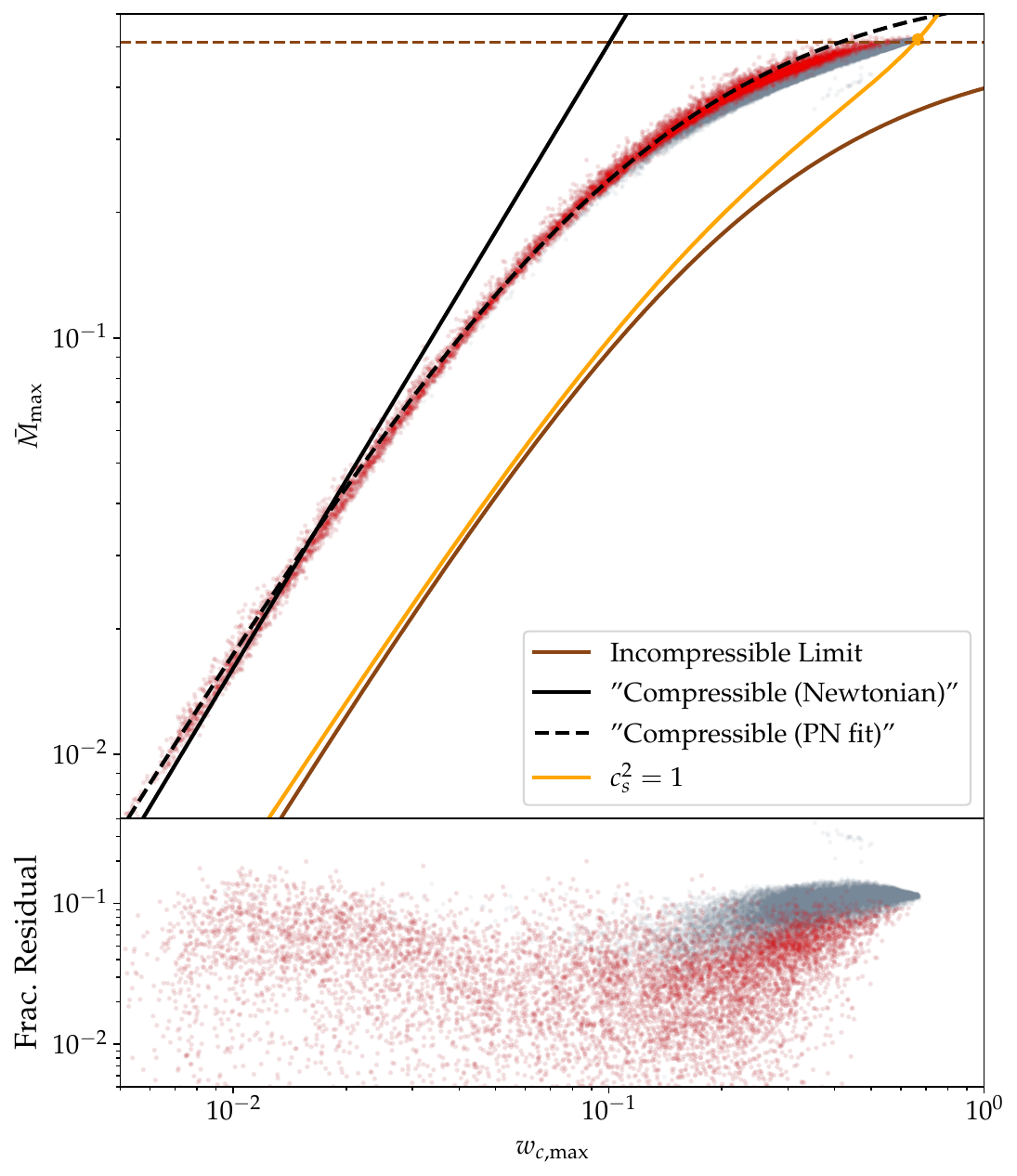}
    \caption{Dimensionless maximum mass $\bar M_{\max}$ versus central stiffness $w_{c,\max}=p_{c,\max}/e_{c,\max}$ of the maximum-mass configuration for $\gtrsim 20{,}000$ Gaussian-process EoSs~\cite{Legred:2021}.  EoSs whose maximum-mass configuration has a small (large) \emph{central} sound speed ($c_{s,c}^2 < 0.9\,w_{c,\max}$) are marked in red (blue). \new{Each dot represents the parameters of the maximum mass star for the given EoS as determined by solving for the entire TOV sequence with that EoS.} The black solid curve shows the Newtonian compressible-limit estimate [Eq.~\eqref{eq:compressible-universal}], while the black dashed curve shows the post-Newtonian relation [Eq.~\eqref{eq:newtonian-mmax-prescription}].  The brown curve is the exact incompressible sequence in general relativity (see App.~\ref{app:tolman-bound}); the horizontal brown dashed line marks the incompressible maximum.  The gold curve shows a representative maximally causal constant-sound-speed model with $c_s^2=1$.  Empirically, all stable configurations lie between the \new{brown and black solid lines}, motivating their use as ``incompressible'' and ``compressible'' limits. The bottom panel shows the fractional residual between the $\bar{M}$--$w_c$ relations of a given EoS and the post-Newtonian expression. 
    }
    \label{fig:wc-vs-mmax}
\end{figure*}

Let us now compare this analytical estimate to the maximum-mass configurations of a large array of EoS samples drawn from the model-agnostic Gaussian-process prior of
Refs.~\cite{Landry:2018prl, Essick:2019ldf, Legred:2021}\footnote{We exclude EoSs for which the maximum mass is attained over a range of central densities, since in that case it is not possible to unambiguously define either $w_{c,\max}=p_{c,\max}/e_{c,\max}$ or $\bar M_{\max} = M_{\max}\sqrt{4\pi e_{c,\max}}$ (in $G=c=1$ units).  See Sec.~\ref{sec:discussion} and App.~\ref{app:when-universal-relations-fail}.}. These EoSs are designed to probe a large array of possible behavior, including potential phase transitions~\cite{Essick:2023fso}, and they are causal and stable by construction.
We display the post-Newtonian fit (black dashed) together with the Newtonian estimate (black solid) in Fig.~\ref{fig:wc-vs-mmax}.  For each EoS, a single point marks the values of $(w_{c,\max},\,\bar M_{\max})$; these points are colored according to whether the \emph{central} sound speed is large compared to $w_{c,\max}$ ($c_{s,c}^2 > 0.9\, w_{c,\max}$, blue) or small ($c_{s,c}^2 < 0.9\,w_{c,\max}$, red) in the maximum-mass configuration.  We also display the full track $\bar M(w_c)$ for a representative maximally causal constant-sound-speed model with $c_s^2=1$ throughout most of the star (except for a small crust) in gold; for Newtonian stars this track lies close to the corresponding incompressible sequence (gray). The two main points of this figure are that (i) the $\bar M_{\max}$--$w_c$ relation appears EoS insensitive (to order 10\%), and (ii) the approximately universal $\bar M_{\max}$--$w_c$ follows the analytical functional form of Eq.~\eqref{eq:newtonian-mmax-prescription}. 

Figure~\ref{fig:wc-vs-mmax} also motivates our naming of the ``compressible limit.'' For all EoSs we studied, stable stars have $(w_c,\bar M)$ values that lie between the incompressible (gray) and compressible (black solid) envelopes.  While the scatter about the post-Newtonian fit is nontrivial (of order $10\%$), it is small compared to the separation between the compressible and incompressible limits (roughly an order of magnitude at fixed $p_c/e_c$).  The point is that stellar sequences explore a wide range of $\bar M$, but nonetheless terminate within a relatively narrow band.  This agreement may be enhanced by the fact that these EoSs share a common crust model, although this crust agreement only extends up to $\lesssim 0.1\,\rho_{\rm nuc}$.  The fact that the termination points are not identical across EoSs further indicates that the crust is not the primary driver of the observed quasi-universality.

Above a certain central stiffness, nearly incompressible stars are excluded by causality.  Eventually this constraint excludes even the \emph{compressible} envelope, and at larger $w_c$ no stable, spherically symmetric perfect-fluid stars exist.  We conjecture that this limiting value, $p_c/e_c \sim 0.7$ (cf.\ Fig.~\ref{fig:wc-vs-mmax}), represents the highest relative pressure achievable in any spherically symmetric, static perfect-fluid solution of general relativity, in tension with results of Ref.~\cite{Cai:2026rzp}.  Whether or not this bound is nearly achieved depends on the true high-density EoS: for any causal barotrope, one has $p/e < \max(c_s^2)$, so only sufficiently stiff EoSs can approach $p_c/e_c\sim 0.7$.  This is what one might expect if the sound speed in neutron-star cores approaches the causal limit $c_s^2\to 1$ at high density (though this possibility may be disfavored~\cite{Hippert:2024hum}).  

On the other hand, if the speed of sound asymptotically approaches the conformal value $c_s^2\to 1/3$, then there is a limiting mass but, in Newtonian theory, it is not achieved at any finite central density; relativistic and compositional effects eventually destabilize the star at a termination density that is sensitive to the detailed microphysics.  This is, at zeroth order, the case of Chandrasekhar-mass white dwarfs.  In this case, small relativistic and compositional effects slightly modify the mass but largely determine the radius (and thus the density) at which the white-dwarf sequence terminates.  Interestingly, for a $\Gamma=4/3$ polytrope, one has $\rho \propto (\ln h)^3$ and $p/\rho = (\ln h)/4$, so at \emph{any} termination density one finds $M\sqrt{\rho_c}\propto w_c^{3/2}$.  Therefore, regardless of the mechanism that sets the actual termination point (e.g.\ electron capture effects or relativistic corrections; see~\cite{ShapiroAndTeukolsky, Bethe:1979} for a review), the scaling in Eq.~\eqref{eq:compressible-universal} (or, equivalently, Eq.~\eqref{eq:newtonian-mmax-prescription} in the $w_c\ll 1$ limit) is satisfied for an appropriate choice of $\gamma$.  We find empirically for the Chandrasekhar EoS that this constant is $\gamma \simeq 15.8$, and we can also estimate it analytically:
\begin{equation}
    M_{\max}\sqrt{4\pi \rho_c} = \gamma\, w_c^{3/2}
    = \gamma \,(K\rho_c^{1/3})^{3/2},
\end{equation}
since $p/\rho = K\rho^{\Gamma-1}$ for a polytrope with $\Gamma=4/3$.  Thus,
\begin{equation}
    \gamma = \frac{M_{\max}\sqrt{4\pi}}{K^{3/2}}.
\end{equation}
Taking $M_{\max} \approx 1.4\,M_{\odot}$ and $K$ to be the value predicted for a relativistic, degenerate electron gas with $2$ baryons per electron ($K\simeq 0.46\,M_{\odot}^{-2/3}$) gives $\gamma \approx 15.9$, consistent with the numerical value above.  This is smaller than the $\gamma\simeq 19.8$ that fits the neutron-star EoS sample in Fig.~\ref{fig:wc-vs-mmax} by about $\sim 20\%$.  That mismatch is not surprising: Chandrasekhar white dwarfs sit in the extreme $w_c\ll 1$ regime and are described (to leading order) by an exactly marginal $\Gamma=4/3$ polytrope, whereas the neutron-star sample in Fig.~\ref{fig:wc-vs-mmax} spans a much broader range of microphysics and includes genuinely relativistic structure.

\subsection{Astrophysical Constraints}

\begin{figure*}
    \centering
    \includegraphics[width=0.98\textwidth]{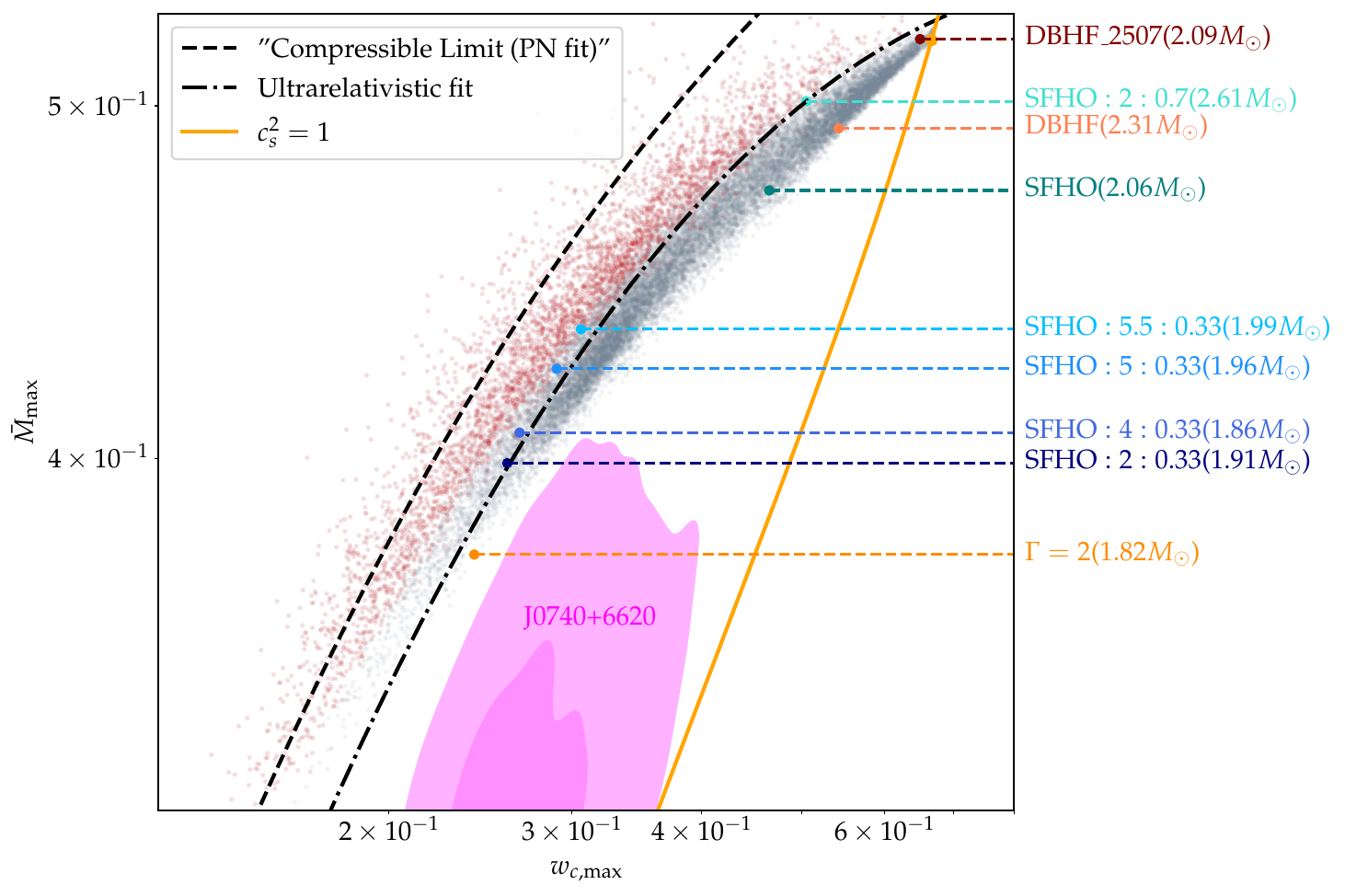}
    \caption{Same as Fig.~\ref{fig:wc-vs-mmax}, but restricted to EoSs consistent with current astrophysical constraints (red and gray dots).  The post-Newtonian compressible-limit fit [Eq.~\eqref{eq:newtonian-mmax-prescription}] is shown as a black dashed curve, while the dash-dot curve shows the ultra-relativistic comparison described in the text, and the causal limit EoS is shown in solid orange.  On the right, we display, for various EoS candidates, the maximum TOV mass $M_{\max}$ and the corresponding value of $\bar M_{\max}=GM_{\max}/c^4\sqrt{4\pi Ge_{c,\max}}$ (horizontal dashed line).  Each EoS is marked at its $(w_{c,\max},\bar M_{\max})$ location with a dot of the same color.  From top to bottom, the external EoSs shown are DBHF\_2507 (a maximally causal extension of DBHF)~\cite{Alford:2013aca, Chatziioannou:2019yko}, SFHo~\cite{Steiner:2012rk}, DBHF~\cite{vanDalen:2006pr}, and a $\Gamma=2$ polytrope with polytropic constant $K=100$ (see, e.g.,~\cite{ShapiroAndTeukolsky}).  For the remaining examples, constructions, such as $\text{SFHo}:2:0.33$, indicate an SFHo EoS transitioning to a constant-sound-speed extension with $c_s^2=0.33$ at $2\,\rho_{\rm nuc}$.  Several trends are evident: all else being equal, a larger high-density sound speed leads to larger terminal values of $p_c/e_c$, $M_{\max}$, and $\bar M_{\max}$ (cf.\ $\text{SFHo}:2:0.7$ versus $\text{SFHo}:2:0.33$).  However, a larger $\bar M_{\max}$ does not necessarily imply a larger $M_{\max}$ (cf.\ DBHF and DBHF\_2507), reflecting sensitivity of $M_{\max}$ to the lower-density EoS and to the density scale at which the core stiffens (see, e.g.,~\cite{Rhoades:1974fn, Kalogera:1996ci}).  By contrast, $\bar M_{\max}$ depends primarily on the high-density stiffness and is comparatively insensitive to the low-density envelope. Astrophysical constraints effectively impose a lower cut on the minimum value of $\bar{M}$.}
    \label{fig:wc-vs-mmax-posterior}
\end{figure*}

Using this framework, we now map astrophysical constraints into the $(w_{c,\max},\bar M_{\max})$ plane in Fig.~\ref{fig:wc-vs-mmax-posterior}.  The plot is qualitatively similar to Fig.~\ref{fig:wc-vs-mmax}, but instead of sampling EoSs directly from the prior, we sample EoSs from the posterior distribution of ~\cite{Legred:2021}, obtained by analyzing heavy pulsar radio data~\cite{Antoniadis:2013pzd, Cromartie:2019kug, Fonseca:2021wxt}, X-ray~\cite{Miller:2019cac, Riley:2019yda, Miller:2021qha, Riley:2021pdl}, and gravitational-wave~\cite{Abbott:2018exr, LIGOScientific:2020aai} data and using the same Gaussian process EoS prior~\cite{Landry:2018prl, Essick:2019ldf, Legred:2021} used in Fig.~\ref{fig:wc-vs-mmax}.   We display the value of $(w_c,\bar M_{\max})$ as a colored dot, with the same color scheme as Fig.~\ref{fig:wc-vs-mmax}; that is, EoSs with a ``small" central sound speed compared to $w_c$ are colored red, whereas those not meeting this criteria are colored gray. 

Together with these posterior samples, Fig.~\ref{fig:wc-vs-mmax-posterior} also plots some additional information. First, the figure includes a set of EoS candidates (to the right of the right axis) by displaying their maximum masses $M_{\max}$ (in parenthesis) and marking their locations in the $(w_{c,\max},\bar M_{\max})$ diagram with color-matched dots. Second, we overlay the same maximally causal ($c_s^2=1$) EoS from Fig.~\ref{fig:wc-vs-mmax} (gold), and the post-Newtonian compressible-limit fit of Eq.~\eqref{eq:newtonian-mmax-prescription} (black dashed). Third, we also overlay an ``ultra-relativistic'' comparison curve, obtained by rewriting the fixed-point estimate of Eq.~\eqref{eq:Mmax-final-units} as
\begin{equation}
    \label{eq:mmax-ultrarel-fit}
    \bar M_{\max} \approx \sqrt{\frac{e_{c,\max}} {e_0}}\,
    \alpha^{3/2}\quant{\frac{2c_s^2}{4 c_s^2 + (1+w)^2}}^{3/2},
\end{equation}
shown in black dash-dot.  Here, $(w,c_s^2)$ should be interpreted as representative stiff-core values (in the sense of Sec.~\ref{sec:relativistic-mmax}), and $e_0$ is the exit density from the near-fixed-point regime.  For the purposes of Fig.~\ref{fig:wc-vs-mmax-posterior}, we take the simple proxy $c_s^2 = w/0.7$, and we find that setting $({e_{c,\max}}/{e_0})\,\alpha^3 \simeq 13$ provides a reasonable fit.  Taking $\alpha=1.2$ as before then implies ${e_{c,\max}}/{e_0}\sim 7.5$.  This is broadly consistent with the Newtonian compressible-limit picture, in which the envelope becomes important at a density scale $e_s/e_c\simeq \bar e_s\sim 0.06$ [Eq.~\eqref{eq:compressible-es}], i.e.\ ${e_c}/{e_s}=\mathcal{O}(10)$; the somewhat smaller ratio inferred here is consistent with the idea that relativistic stars cannot sustain as extended an envelope at fixed central density.  

While the ultra-relativistic and post-Newtonian expressions share a similar leading dependence on stiffness, they differ in their coefficients because they ultimately arise from different limiting mechanisms (fixed-point exit versus Newtonian compressible instability), and because nonlinear relativistic effects are not parametrically small in the most compact configurations.  In practice, for the least relativistic EoS posterior samples (the red dots on the bottom left part Fig.~\ref{fig:wc-vs-mmax-posterior}) that remain viable, the post-Newtonian relation provides a good approximation; for the most relativistic posterior samples, the fixed-point scaling is more appropriate; and there remains an intermediate region where neither simple curve captures the full distribution particularly well.

Relative to Fig.~\ref{fig:wc-vs-mmax}, astrophysical constraints already shrink the allowed region of parameter space by orders of magnitude
This can be seen because the range of allowed values of $\bar M_{\max}$ and $w_{c, \max}$ are constrained to lie above $\sim 3\times 10^{-1}$ and $\sim 2\times 10^{-1}$ respectively, whereas in Fig.~\ref{fig:wc-vs-mmax} these quantities both extend down to $\sim 10^{-2}$.  Broadly, heavy pulsar mass measurements require a stiff EoS at high density (i.e.~at high $w_c$), while gravitational-wave observations and the comparatively well-constrained crust place upper bounds on the EoS stiffness at lower densities.  Together these imply that the dense-matter sound speed must rise rapidly between $\sim 2$--$4\,\rho_{\rm nuc}$, allowing a large $M_{\max}$ while still favoring comparatively small radii; in the $(w_{c,\max},\bar M_{\max})$ plane this imposes a nontrivial lower bound on $\bar M_{\max}$.  

Before ending with the discussion section, let us compare the posterior samples in the $(\bar M_{\max},w_{c, \max})$ plane to the marginalized posterior of $(\bar M,w_{c})$ for a given star, namely the J0740+6620 pulsar~\cite{Cromartie:2019kug, Fonseca:2021wxt,Miller:2021qha, Riley:2021pdl}. This marginalized posterior can be obtained from the same Bayesian hierarchical analysis of the data described above, but this time marginalizing over EoS instead of the masses and radii of the J0740+6620 pulsar, and then transforming to $(\bar M_{\max},w_{c, \max})$ variables\footnote{ 
The same X-ray and radio data which informs the posterior distribution on the properties of J0740+6620 is included in the posterior on the EoS, but each represents marginalization over different degrees of freedom.  To construct the EoS posterior, we marginalize over all single-observation parameters, see e.g.~Eqs.~(1), (5), (8), and (11) of ~\cite{Landry:2020vaw} (including properties, such as the mass, radius, and central energy density of J0740+6620).  To construct the estimate for the posterior on the properties of J0740+6620, we marginalize out the \emph{EoS} parameters,  see e.g.~Eq.~(9) of~\cite{Landry:2020vaw}.  In short, a density estimate is built using posterior samples on the mass and radius, which, by dividing out a stated prior, can be converted to a marginalized likelihood $\mathcal L(d|M,R)$. The radius is then required to be consistent with a sampled EoS to construct $\mathcal L(d| M, \varepsilon) = \mathcal L(d| M, R(M, \varepsilon))$.  This likelihood corresponds to the joint posterior on the mass, radius, and EoS, which is then marginalized over to get either the properties of the EoS, or the properties of the J0740+6620 pulsar. For an example of the difference between the parameters of J0740+6620 and the parameters of the maximum-mass star, see e.g.~Fig.~(4) of \cite{Legred:2021} and the surrounding discussion.}. This posterior is shown in Fig.~\ref{fig:wc-vs-mmax-posterior} with pink contours (roughly $50\%$ and $90\%$ credible regions). Note importantly that these pink contours are \textit{not} posteriors on the maximum values of $\bar{M}$ and $w_c$, but rather, posteriors on the $\bar{M}$ and $w_c$ of the J0740+6620 pulsar.


Comparing the pink posteriors to the posterior sample dots, we see clearly that the J0740+6620 pulsar does not have a $\bar{M}$ that is close to the maximum $\bar{M}$ of neutron stars. Moreover, the large inferred radius of J0740+6620~\cite{Miller:2021qha, Riley:2021pdl} \new{implies it does not have an exceptionally large mean density, which in practice limits its central density $e_c$.  This upper bounds the value of $\bar M$, which is proportional to $\sqrt{e_c}$, given the mass is well measured}.   Therefore, the pink posterior does not appear to be consistent with the posterior samples of $(w_{c,\max}, \bar M_{\max})$ for any EoS in our prior (\emph{i.e.} the displayed dots).  Attempting to quantify this is not straightforward, but a simple Bayesian estimate of the error in universal relation prediction for J0740+6620 gives $L^1(\bar M_{\rm true} -\bar M_{\rm pred})/\bar M_{\rm true} = 0.41^{+0.28}_{-0.22}$ for the relation Eq.~\eqref{eq:newtonian-mmax-prescription} and $L^1(\bar M_{\rm true} -\bar M_{\rm pred})/\bar M_{\rm true} = 0.31^{+0.26}_{-0.21}$ given the prediction of Eq.~\eqref{eq:mmax-ultrarel-fit}\footnote{This estimate only takes into account the uncertainty in the properties of J0740+6620 (which are substantial), not the uncertainty in the parameters of the universal relation, which are challenging to quantify, but are comparatively small.}. This is somewhat puzzling given that no higher-mass neutron stars have yet been definitively confirmed---one might have expected to see heavier systems if the observed population were far from the end of the TOV sequence---though an abrupt termination triggered by a rapid drop of the sound speed (e.g.\ a strong phase transition) could reconcile these facts. We discuss these possibilities further in the next section. 

\section{Discussion}
\label{sec:discussion}

In this paper, we have presented strategies for estimating the maximum mass of stellar sequences by recasting the equations of stellar structure in the language of dynamical systems.
We carried out this recasting in two controlled limits.
On the relativistic side, we identified a fixed point of the TOV flow that we associate with the universal emergence of a maximum mass in neutron-star sequences.
On the Newtonian side, we identified universal behavior in sequences that terminate because the effective stiffness decreases with density (or equivalently, because the relevant, averaged polytropic index falls through the instability threshold).
This Newtonian behavior appears to extend qualitatively into the relativistic regime, though extracting a clean quantitative mapping there is more challenging.
We leave that interpolation problem to future work.

In the ultrarelativistic regime, we recast the stellar structure equations in the enthalpy ($h$) formulation of~\cite{Lindblom:1998dp}, using
$\tilde e = 4\pi G r^2 e/c^4$ and $v = Gm/(c^2 r)$.
In these variables, the system exhibits an obvious fixed point. At sufficiently high densities, and under fairly general hypotheses, the flow spends enough ``time'' near this fixed point that it largely determines the mass and radius. When the sound speed is not varying too rapidly as a function of $\ln h$, we expect the mapping we derive to predict the maximum mass at the $\sim 10\%$ level. If the sound speed changes rapidly (e.g., by factors of several over a small range in $\ln h$), the prediction can degrade substantially. That failure is not necessarily a negative: it flags interesting microphysics, such as a sharp change in effective degrees of freedom.

The picture of the TOV equations as a dynamical system was also discussed in~\cite{Heinzle:2003ud} (and in~\cite{Heinzle:2002sk}, for the Newtonian limit).  Our approach is different from theirs primarily because of (i) our choice to use a microscopic variable ($\ln h$) as the independent variable, and (ii) our decision to keep the system 2-dimensional, rather than promoting a third variable to an evolution variable.  Ref.~\cite{Heinzle:2003ud} uses their approach to prove several interesting theorems relating the asymptotic behavior of the EoS to the global properties of the mass and radius.  We focus instead on astrophysical observables associated with  the dynamical system approach.  In our study, the termination condition for the integration is known ahead of time, because of the choice of independent variables.  We can therefore build Taylor expansions from both the core and the surface, which we can then stitch onto a dynamical systems picture in the intermediate regime.  

However, just because the dynamical systems picture exists for all central densities, it may not be a particularly useful description.  Seen from the point of view of our dynamical system approach, we can effectively quantify when this is the case by considering when the velocity of the fixed-point in phase space is small compared to the velocity of the solution.  Therefore, the non-autonomy of our dynamical system gives us useful information about what the correct modeling approach to the problem is. For generic NSs, we do not expect the dynamical systems picture to be particularly insightful because the orbit of the trajectory in $(\tilde{e},v)$ space is never truly well-described by the fixed-point.   Only for stars very near the maximum mass do dynamically stable configurations also show strong dimensionality reduction to just the location of the (2D) fixed point.

In the Newtonian regime, stellar sequences only terminate if an appropriate average polytropic index over the star drops below $4/3$~\cite{ShapiroAndTeukolsky}. We find that if the speed of sound falls ``slowly'' enough with increasing density, the termination mass is a universal (EoS-insensitive) function of the central pressure and density. However, if the speed of sound drops to (nearly) zero in the core, then stellar sequences may terminate abruptly (Appendix~\ref{app:when-universal-relations-fail}). In that case, the mass can be nearly constant over orders of magnitude in central density, so it is not straightforward to define the ``central density at maximum mass'' in a robust way. Nonetheless, the possibility that a strong first-order phase transition could lead to an apparent violation of the universal relation in Eq.~\eqref{eq:newtonian-mmax-prescription} is an interesting hypothesis, because it would amount to a clear signature of an extended region with extremely small $c_s^2$.

If the true EoS termination point is found to lie between the causal limit and the bulk of the posterior displayed in Fig.~\ref{fig:wc-vs-mmax-posterior}, it would point to the TOV sequence terminating ``early'' due to a strong first-order phase transition. In that sense, the question of whether the observed neutron-star population extends significantly above the $\sim 2\,M_\odot$ pulsars (e.g.\ J0740+6620) is key. Population studies can attempt to establish whether an apparent absence of heavier stars constrains the EoS~\cite{Alsing:2017bbc, FarrChatziioannou2020, Fan:2023spm, Biswas:2024hja}, though care must be taken to faithfully incorporate the large relative uncertainty in the inferred neutron-star population~\cite{Golomb:2024lds}.   It may be the case that $>2M_{\odot}$ pulsars are consistent with the dense-matter EoS, but are very rare due to limited possible astrophysical formation scenarios.   We will study this possibility more in future work.

Finally, it is useful to compare and contrast the two limits examined here. Interestingly, in both cases the maximum mass depends primarily on a characteristic sound speed in the core, schematically scaling like $\langle c_s^2\rangle^{3/2}$. In the Newtonian case $\langle c_s^2\rangle$ can be interpreted as a true average (see Refs.~\cite{Saes:2021fzr, Saes:2024xmv}), while in the ultrarelativistic case, it is better thought of as a ``typical'' value over the range in $\ln h$ where the flow is controlled by the fixed point. Both limits also share a common structural feature: the maximum-mass configuration is obtained by ``filling in'' between a near-core expansion and a near-surface expansion, and the act of targeting the maximum mass reduces the effective dimensionality of the problem.  Without the assumption that we are near the maximum mass (near the ultrarelativistic fixed point on the one hand, or near the instability threshold of the averaged polytropic index on the other), this reduction would not occur and the space of solutions would be much more complicated.

Heuristically, the success of the stitching strategy we use may be related to the curvature invariants inside the star.   The quadratic Riemann curvature invariant, sometimes called the Kretschmann scalar, is given by
\begin{align}
    K &\equiv R_{\alpha \beta \gamma \delta} R^{\alpha \beta \gamma \delta}  \nonumber \\ 
    &= C_{\alpha \beta \gamma \delta} C^{\alpha \beta \gamma \delta} + 128\pi^2 T_{\alpha \beta} T^{\alpha \beta} - \frac{64\pi^2T^2}{3}.
\end{align}
In a perfect fluid star this becomes 
\begin{multline}
    K = C_{\alpha \beta \gamma \delta} C^{\alpha \beta \gamma \delta} + 128\pi^2 (e^2  + 3p^2) - \frac{64\pi^2(e-3p)\new{^2}}{3} \\
    = C_{\alpha \beta \gamma \delta} C^{\alpha \beta \gamma \delta} + \quant{\frac{\tilde e}{3r^2}}^2 (60 + 72w + 108w^2),
\end{multline}
with 
\begin{align}
    C_{\alpha \beta \gamma \delta} C^{\alpha \beta \gamma \delta} &= \frac{1}{r^4}\quant{48 v^2 - 32 v \tilde e + \frac{16}{3}\tilde e^2}.
\end{align}
The Weyl curvature invariant $C_{\alpha\beta\gamma\delta}C^{\alpha\beta\gamma\delta}$ vanishes at the center of a regular star (which depends on the linear approximation, $\tilde e \approx 3v$, holding), while in the exterior (and thus arbitrarily near the surface) the Ricci curvature vanishes, implying $R_{\alpha\beta\gamma\delta}R^{\alpha\beta\gamma\delta}-C_{\alpha\beta\gamma\delta}C^{\alpha\beta\gamma\delta}=0$.  In the ``interesting region'' of the interior, neither quantity vanishes generically.  Furthermore, it is straightforward to see that the variables $(\tilde e, v)$ can be exchanged for the curvature invariants as dependent variables of the TOV integration.  From this point of view, the core solution is constructed by approximating $C_{\alpha\beta \gamma \delta}=0$, and the exterior solution is constructed by assuming $T_{\alpha\beta}=0$.  Again, when neither invariant is small, and the evolution of the system is governed by the fixed point, as long as the system has sufficient time to converge to the fixed point.  This convergence is characteristic of high density solutions, and therefore is applicable to stars near and above the central density of the maximum mass star.  Even though this discussion is framed in terms of curvature invariants, it is equally valid in the Newtonian limit as in the ultrarelativistic limit.  

There are also important differences between these two limits. In the ultrarelativistic case, the fixed point dominates the flow at high density, and the relativistic corrections enter in a way that is not simply perturbative. For example, the denominator of Eq.~\eqref{eq:fixedpoint-location} scales like $4c_s^2+(1+w)^2$, and since $w\sim \mathcal{O}(c_s^2)\sim \mathcal{O}(1)$ at relativistic densities, the dimensionless maximum mass $(GM/c^2)\sqrt{4\pi G e_c/c^4}$ 
is pushed to smaller values than would be expected from Newtonian intuition. Empirically, the ultrarelativistic relation between $(GM/c^2)\sqrt{4\pi G e_c/c^4}$
and $w_c=p_c/e_c$ is much shallower than in the Newtonian case (Fig.~\ref{fig:wc-vs-mmax-posterior}).
Relativity also introduces genuine upper limits on what can be stably attained for a given central density regardless of the EoS, as illustrated by the incompressible case (Fig.~\ref{fig:wc-vs-mmax}). At the same time, the incompressible limit is a poor proxy for realistic relativistic matter: special relativity enforces $c_s^2\le 1$ (in $c=1$ units), so the ``$c_s^2\to\infty$'' limit is not physically attainable. Unlike in Newtonian gravity, where it is meaningful to compare an ``incompressible'' and ``compressible'' limiting sequence, the more relevant relativistic limiting family is instead the maximally causal one, $c_s^2=1$.

Finally, the near-constancy of the fixed point in the ultrarelativistic case naturally leads to the familiar spiral structure, which is associated with progressive destabilization of higher radial modes~\cite{HTWW:1965, ShapiroAndTeukolsky}. An analogous feature exists in the Newtonian dynamical system, but it requires the sound speed to remain nearly constant over a parametrically large range of $\ln h$. In Newtonian gravity, this is fine-tuned~\footnote{In a star composed of an ideal gas, this would require a temperature profile which is increasing and then becomes effectively isothermal in the core.  In a solid it would require a careful engineering of the bulk modulus to eventually be proportional to the density.}, whereas in relativity the causal bound makes an order-of-magnitude plateau in $c_s^2$ comparatively natural. This is interesting on a theoretical level: turning over in the $M$--$R$ curve allows a sequence to attain static (though unstable) solutions of the Einstein equations with \emph{higher} total (gravitational plus internal) energy per particle than if the particles were dispersed to infinity~\cite{HTWW:1965}. The existence of analogous solutions around the ``Newtonian'' fixed point shows that such behavior is not exclusively tied to relativistic corrections, even if it is unlikely to arise for realistic Newtonian EoSs.

Nonetheless, despite the similarity between the static solutions, the \emph{dynamics} of highly unstable stars in Newtonian gravity and in general relativity are likely quite different. Critical gravitational collapse has been observed in relativistic evolutions at central densities above the TOV maximum-mass density for polytropes~\cite{Gundlach:2007gc, Radice:2010rw, Noble:2015anf}, though it is unclear how naturally such configurations are realized without external perturbations. It seems unlikely that Newtonian stars display truly analogous dynamics even with strong perturbations, though this has not been shown in a systematic way. Either way, understanding the nonlinear dynamics of configurations beyond first order in perturbations---especially beyond the first minimum in $M$ in a spiraling sequence---is warranted, and we leave that to future work.


\acknowledgements The authors would like to thank Nicholas Rui, Nicholas Corso, Yoonsoo Kim, Sarah Habib, Jayana Saes, Katerina Chatziioannou, Elias Most, and Saul Teukolsky for insightful conversations related to this work.  The authors acknowledge support from the Simons Foundation through Award No. 896696, the Simons Foundation International through Award No. SFI-MPS-BH-00012593-01, the NSF through Grants No. PHY-2207650 and PHY-25-12423, and NASA through Grant No. 80NSSC22K0806. This work was also supported by NSF within the framework of the MUSES collaboration, under grant number OAC-2103680.  The authors are grateful for computational resources provided by the LIGO Laboratory and supported by National Science Foundation Grants PHY-0757058 and PHY-0823459. This material is based upon work supported by NSF’s LIGO Laboratory which is a major facility fully funded by the National Science Foundation.  Equation of State samples to reproduce Figs.~\ref{fig:wc-vs-mmax} and \ref{fig:wc-vs-mmax-posterior} can be found at Ref.~\cite{legred_2022_6502467}.  EoS samples to reproduce Fig.~\ref{fig:Mmax_and_Rmax_fit} can be found at ~\cite{rmfteossamples}.  Additional EoS samples will be made available upon reasonable request.  Codes to reproduce certain expressions and figures are available at Ref.~\cite{plottingcode}.

\appendix

\appendix

\section{Review of Stability for Autonomous Systems}
\label{app:autonomous-stability}

Section~\ref{sec:ds_primer} already fixed notation and reviewed the basic local theory of fixed points.  Here we just collect the stability conventions that matter for our stellar-structure application, and record a convenient Lyapunov-function candidate for the frozen TOV fixed point.

Consider an autonomous system
\begin{equation}
    \frac{d\mathbf{x}}{d\tau} = \mathbf{F}(\mathbf{x}),
\end{equation}
with a fixed point $\mathbf{x}_*$ satisfying $\mathbf{F}(\mathbf{x}_*)=\mathbf{0}$.  Linearizing as in Sec.~\ref{sec:ds_fixed_points}, $\mathbf{x}=\mathbf{x}_*+\boldsymbol{\delta x}$, gives
\begin{equation}
    \frac{d\,\boldsymbol{\delta x}}{d\tau}
    =
    \mathbf{J}(\mathbf{x}_*)\,\boldsymbol{\delta x},
    \qquad
    \mathbf{J}(\mathbf{x}_*) \equiv \left.\partial_{\mathbf{x}}\mathbf{F}\right|_{\mathbf{x}_*}.
\end{equation}
In the usual forward-$\tau$ sense, $\mathbf{x}_*$ is (linearly) asymptotically stable if all eigenvalues of $\mathbf{J}(\mathbf{x}_*)$ have negative real parts.

For the TOV system written with $\tau=\ln h$, physical solutions are specified at $\ln h=\ln h_c$ and integrated outward to the surface at $\ln h=0$, i.e.\ toward \emph{decreasing} $\ln h$.  As emphasized in Sec.~\ref{sec:ds_fixed_points} and Sec.~\ref{sec:jacobian-hessian}, it is therefore cleaner to use the outward evolution variable
\begin{equation}
    s:=\ln h_c-\ln h,
    \qquad
    \frac{d}{ds}=-\frac{d}{d\ln h}.
\end{equation}
If the frozen system linearizes as $d\boldsymbol{\delta x}/d\ln h=\mathbf{J}\,\boldsymbol{\delta x}$, then along the physical direction one has
\begin{equation}
    \frac{d\,\boldsymbol{\delta x}}{ds} = -\mathbf{J}\,\boldsymbol{\delta x},
\end{equation}
so a focus that is unstable in forward $\ln h$ acts as an attractor along outward integration.

More global notions of stability are provided by Lyapunov functions.  A fixed point is (Lyapunov) stable if for any $\epsilon>0$ there exists a $\delta>0$ such that $|\mathbf{x}(0)-\mathbf{x}_*|<\delta$ implies $|\mathbf{x}(\tau)-\mathbf{x}_*|<\epsilon$ for all $\tau>0$.  A sufficient condition is the existence of a function $V(\mathbf{x})$ such that (i) $V(\mathbf{x}_*)=0$, (ii) $V(\mathbf{x})>0$ in a neighborhood of $\mathbf{x}_*$, and (iii) $\dot V\le 0$ in that neighborhood, where $\dot V\equiv (d\mathbf{x}/d\tau)\cdot \partial_{\mathbf{x}}V$.

For a linear system $d\boldsymbol{\delta x}/d\tau=\mathbf{A}\,\boldsymbol{\delta x}$, one may take
\begin{equation}
    V=\boldsymbol{\delta x}^{\,T}\mathbf{P}\,\boldsymbol{\delta x},
\end{equation}
with $\mathbf{P}$ a symmetric and positive-definite matrix that solves the (continuous-time) Lyapunov equation
\begin{equation}
    \mathbf{A}^T\mathbf{P}+\mathbf{P}\mathbf{A}=-\mathbf{Q},
\end{equation}
for any chosen positive-definite $\mathbf{Q}$ matrix.  In our application, it is natural to take $\mathbf{Q}=\mathbf{I}$, and to use $\mathbf{A}=-\mathbf{J}$ when we want stability in the physical direction $s$.  With that convention,
\begin{equation}
    \mathbf{J}^T\mathbf{P}+\mathbf{P}\mathbf{J}=\mathbf{I},
    \label{eq:lyap_equation_for_J}
\end{equation}
where $\mathbf{J}$ is the Jacobian of the original $\ln h$-flow.

Applying this to the frozen TOV fixed point $\mathbf{x}_\star$ of Sec.~\ref{sec:jacobian-hessian}, we take $\mathbf{J}=\mathbf{J}_\star$ from Eq.~\eqref{eq:jacobian-fixedpoint} (with $w$ and $c_s^2$ treated as frozen constants).  Solving Eq.~\eqref{eq:lyap_equation_for_J} yields the symmetric matrix
\begin{widetext}
\begin{equation}
    \mathbf{P} \;=\;
    \begin{pmatrix}
        \dfrac{2c_s^2\left(2c_s^2+(1+w)^2\right)}{(3w+1)\left(4c_s^2+(1+w)^2\right)} &
        \dfrac{c_s^2\left(w^2-1-4c_s^2\right)}{(3w+1)\left(4c_s^2+(1+w)^2\right)}
        \\[1.2em]
        \dfrac{c_s^2\left(w^2-1-4c_s^2\right)}{(3w+1)\left(4c_s^2+(1+w)^2\right)} &
        \dfrac{c_s^2\left(32c_s^4+4c_s^2 w^2+24c_s^2 w+20c_s^2+3w^4+8w^3+10w^2+8w+3\right)}
        {(1+w)^2(3w+1)\left(4c_s^2+(1+w)^2\right)}
    \end{pmatrix}.
\end{equation}
\end{widetext}
For example, when $c_s^2=w=1$ this reduces to
\begin{equation}
    \mathbf{P} \;=\;
    \begin{pmatrix}
        \frac{3}{8} & -\frac{1}{8}
        \\[1.1em]
        -\frac{1}{8} & \frac{7}{8}
    \end{pmatrix}.
\end{equation}

We stress that $V=\boldsymbol{\delta x}^{\,T}\mathbf{P}\,\boldsymbol{\delta x}$ is an \emph{exact} Lyapunov function for the linearized frozen flow about $\mathbf{x}_\star$ in the physical direction $s$, but for the full nonlinear and non-autonomous TOV system it is only a natural \emph{candidate}.  Numerically, we find it decreases throughout a sizable neighborhood of the fixed point for the cases we examined, but it is not globally decreasing over the full phase space (it can fail near the boundaries of the physically relevant domain, e.g.\ $\tilde e\to 0$ or $v\to 1/2$).  In any case, a truly global Lyapunov analysis is not the most useful framework here: (i) the global behavior is dominated by non-autonomous drift of $w(\ln h)$ and $c_s^2(\ln h)$, and (ii) physical stellar solutions are launched from a constrained family fixed by regularity at the center, not from arbitrary initial data.

\section{Constant speed of sound stars do not have zero-density surfaces}
\label{sec:css-no-surface}

The conjecture that constant-speed-of-sound cores generically yield solutions that approach the fixed-point spiral is closely related to a more concrete statement: no star described by a genuinely constant-$c_s^2$ EoS has a \emph{finite-radius} surface with $e=0$.   This statement is effectively proven as Theorem 5.1 of ~\cite{Heinzle:2003ud}.  We now relate this statement to our formulation of the dynamical system.  

To see the logic, it is helpful to distinguish two common ``constant-speed-of-sound core'' cases. If the EoS is of the self-bound form $p=c_{s,0}^2\,(e-e_0)$ with $c_{s,0}^2 = {\rm{const.}}$ and $e_0>0$, then $p=0$ occurs at $e=e_0\neq 0$. In other words, the star has a finite-radius surface but a nonzero surface density; one can interpret this as a constant-$c_s^2$ core stitched to an effectively zero-$c_s^2$ exterior (vacuum) at $e=e_0$.

If instead one insists on $p=c_{s,0}^2 e$ all the way down to zero density, then $w\equiv p/e=c_{s,0}^2$ everywhere and thermodynamics gives
\begin{equation}
    \frac{d\ln e}{d\ln h} = \frac{1+w}{c_{s,0}^2},
\end{equation}
which is bounded and positive. Integrating implies $e\to 0$ only if $\ln h\to -\infty$ (equivalently $h\to 0$), so one cannot place the stellar surface at $\ln h=0$ in this case.

In the dynamical-systems picture, if solutions launched from the regular center lie in the basin of attraction of the frozen fixed point, then evolving to $\ln h\to -\infty$ drives the trajectory arbitrarily close to that fixed point.  In particular, $\tilde e(\ln h)\to \tilde e_\star \neq 0$ as $\ln h\to -\infty$.  Since $e=\tilde e/[(4\pi G/c^4)r^2]$, this implies $e\propto r^{-2}$ asymptotically and therefore $e$ can only reach zero as $r\to\infty$.  In that sense, a strictly linear constant-speed-of-sound core EoS has no finite-radius zero-density surface.

\begin{figure}
    \centering
    \includegraphics[width=0.99\linewidth]{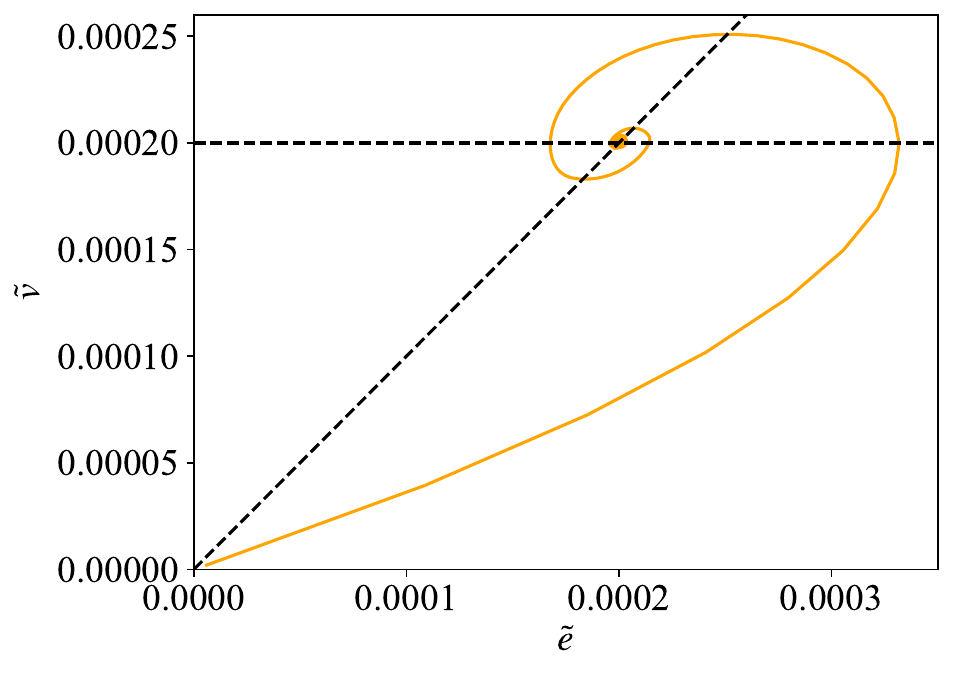}
    \caption{Solution to the Newtonian stellar structure equations for a star with constant speed of sound $c_s^2=10^{-4}$ (c.f. Fig. 9 of~\cite{Heinzle:2002sk}).
    The system clearly approaches the fixed point; once the trajectory is sufficiently close, its subsequent evolution is well approximated by the linearized flow about that fixed point (Sec.~\ref{sec:ds_fixed_points}).}
    \label{fig:newtonian-fixed-point}
\end{figure}

We conjecture that the fixed-point attractor picture indeed holds for the relativistic constant-speed-of-sound core system with physical (regular-center) initial conditions, although proving it rigorously would require a genuinely global Lyapunov function or a comparable argument. In the Newtonian case ($c_s^2\ll 1$), the corresponding constant-speed-of-sound core system enjoys a simple scaling symmetry: uniform rescalings of $(\tilde e,v,c_s^2)$ map solutions to solutions.
Consequently, it is enough to verify convergence to the fixed point for one fiducial value of $c_s^2$.
We have done this numerically for $c_s^2=10^{-4}$, and the solution is shown in Fig.~\ref{fig:newtonian-fixed-point}. Though this is not a formal proof, we have no doubt that all constant-$c_s^2$ solutions of the Newtonian stellar structure equations approach the fixed point as $\ln h\to -\infty$. We would consider it remarkable if a counterexample were found in the relativistic case.

\section{The intermediate regime between the incompressible and compressible limits}
\label{sec:when-expansion-doesnt-work}

When the average polytropic index inside a Newtonian star is not close to either $0$ (the incompressible limit) or $3$ (the highly compressible limit), the core--envelope matching strategy discussed around Fig.~\ref{fig:newtonian-stitching} does not generically work. The core-based expansion typically develops significant higher-order contributions before the crust-based (envelope) approximation becomes accurate.

The basic reason stitching worked particularly well in the $n=3$ case is that the third-order term in the near-core Taylor expansion of $v$ is anomalously small when $n\simeq 3$. This is not true generically: for intermediate $n$, the third-order term becomes important at relatively small $\ln h/\ln h_c$, and that is what prevents stitching ``too late.''

\begin{figure*}
    \centering
    \includegraphics[width=0.49\linewidth]{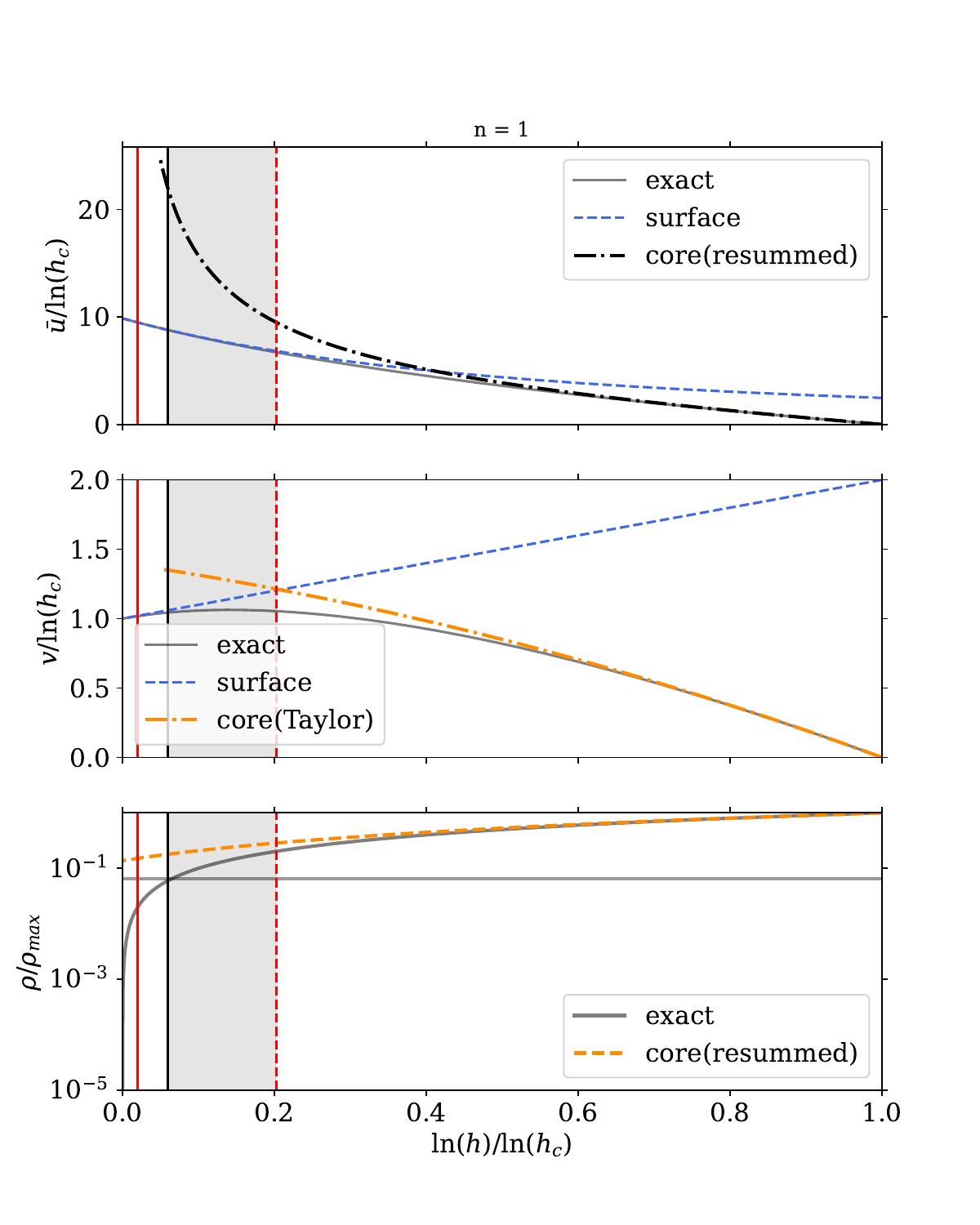}
    \includegraphics[width=0.49\linewidth]{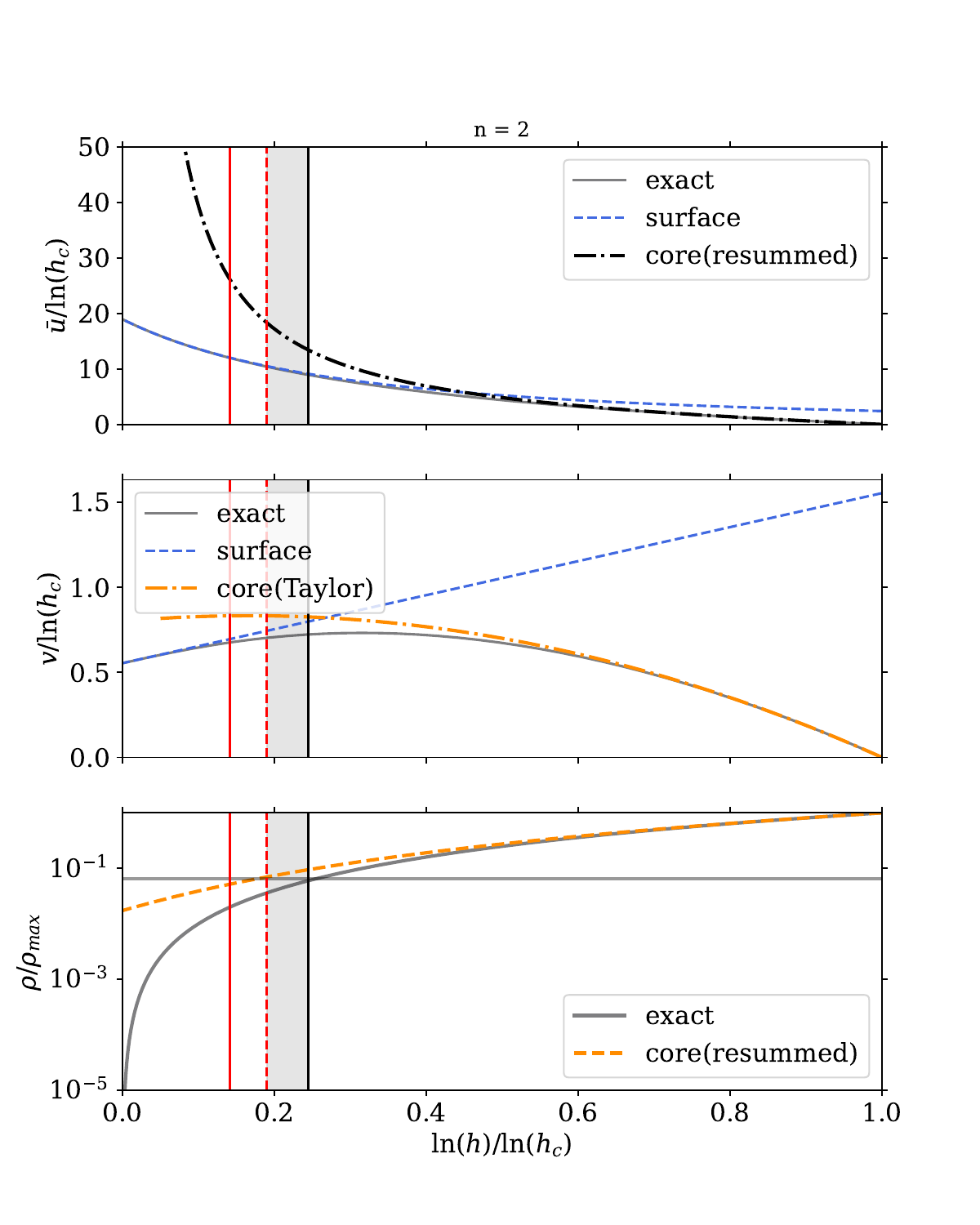}
    \caption{Similar to Fig.~\ref{fig:newtonian-stitching}, but for $n=1$ and $n=2$ polytropes.
    The vertical dashed red line indicates where the third-order term in the Taylor expansion of $v$ exceeds $6\%$ of the leading term, marking the point where the core expansion begins to break down.}
    \label{fig:stitching-not-working}
\end{figure*}

We show the analogs of Fig.~\ref{fig:newtonian-stitching} for different polytropic indices in Fig.~\ref{fig:stitching-not-working}, and indicate in each panel the point where the absolute value of the third-order term in the $v$ expansion exceeds $6\%$ of the leading term. In both examples, the crust expansion is only accurate at very small $\ln h/\ln h_c$, so identifying a clean stitching point is challenging. Physically, this indicates that stars which are neither near the incompressible nor compressible limits are more sensitive to intermediate-density physics in the EoS.

\section{Justification for the choice of approximation}
\label{sec:approximation-justification}

In Sec.~\ref{sec:newtonian-falling-sound-speeds} we argued that a na\"ive Taylor expansion can lead to poorly controlled higher-order terms. Here we briefly justify the expansion strategy used there: we want an approximation that is insensitive to rapid but structurally-insignificant variations in $c_s^2$.

Returning to Eqs.~\eqref{eq:lindblom-newtonian-reformulated-u} and \eqref{eq:lindblom-newtonian-reformulated-v}, we see that at the level of approximation of Eq.~\eqref{eq:compressible-closure} we can write schematically
\begin{equation}
    \frac{v(\hat s)}{\ln h_c}
    \;=\;
    \int_0^{\hat s}\!\!\left(3\,\bar e(\hat s')^{2/5}-1\right)\,d\hat s',
\end{equation}
where recall we have previously defined $\hat{s} := s/\ln{h_c}$. Taylor expanding the integrand around $\bar e=1$ by writing $\bar e = 1-\delta\bar e$ gives
\begin{align}
    3\,\bar e^{2/5}-1
    &= 3\,(1-\delta\bar e)^{2/5}-1
    \nonumber\\
    &= 2-\frac{6}{5}\,\delta\bar e + \mathcal{O}\!\left((\delta\bar e)^2\right),
\end{align}
so that
\begin{equation}
    \frac{v(\hat s)}{\ln h_c}
    \;=\;
    2\hat s-\frac{6}{5}\int_0^{\hat s}\delta\bar e(\hat s')\,d\hat s'
    +\mathcal{O}\!\left((\delta\bar e)^2\right),
    \qquad
    \delta\bar e \equiv 1-\bar e.
\end{equation}
This representation makes clear why the expansion used in Sec.~\ref{sec:newtonian-falling-sound-speeds} is better behaved than a purely local Taylor series: at a fixed order it depends only on \emph{integrals} of the density profile (e.g.\ $\int \delta\bar e\,d\hat s$), and is therefore less sensitive to rapid pointwise oscillations in $c_s^2$ that have little integrated effect.

It also motivates using the integrated quantity $w_c$ introduced in Eq.~(77) (Sec.~\ref{sec:polytropic-benchmark}),
\begin{equation}
    w_c = \int_0^{\ln h_c}\bar e(\ln h)\,d\ln h
\end{equation}
or equivalently
\begin{equation}
     \frac{w_c}{\ln h_c} = \int_0^1 {\bar{e}}( {\hat{s}}) \,d\hat s 
\end{equation}
as a natural ``averaged'' expansion parameter.  In the Newtonian regime, this coincides with the usual central stiffness measure $p_c/e_c$ up to post-Newtonian corrections, while remaining far less sensitive to sharply localized structure in the EoS.

\section{The Tolman--Buchdahl Bound}
\label{app:tolman-bound}

In the incompressible limit (formally $c_s^2 \to \infty$),
the TOV equation admits an analytic solution that yields
the Buchdahl compactness bound. Incompressibility means $e = \text{const}$. Recalling our definitions of $v$ and $\tilde{e}$ in Eqs.~\eqref{eq:etilde-def} and~\eqref{eq:v-def}, the TOV system (Eqs.~\eqref{eq:tov-reformulated-etilde} and~\eqref{eq:tov-reformulated-v}) becomes
\begin{align}
    \frac{d\tilde e}{d\ln h}
    &=
    -\tilde e\,
    \frac{2(1-2v)}{w \tilde e + v},
    \label{eq:buchdahl-de-Gc}
    \\
    \frac{dv}{d\ln h}
    &=
    -(\tilde e - v)\,
    \frac{(1-2v)}{w \tilde e + v}.
    \label{eq:buchdahl-dv-Gc}
\end{align}
Dividing Eq.~\eqref{eq:buchdahl-dv-Gc} by
Eq.~\eqref{eq:buchdahl-de-Gc} gives
\begin{equation}
    \frac{dv}{d\tilde e}
    =
    \frac{1}{2}
    -
    \frac{v}{2\tilde e},
\end{equation}
whose regular-center solution is
\begin{equation}
    v = \frac{\tilde e}{3}.
\end{equation}

We now trade $\ln h$ for $w \equiv p/e$.
Since $e$ is constant in the incompressible case, $dp = e\, dw$. Thermodynamics then gives
\begin{equation}
    d\ln h
    =
    \frac{dp}{e+p}
    =
    \frac{dw}{1+w},
\end{equation}
and using $v=\tilde e/3$ in
Eq.~\eqref{eq:buchdahl-de-Gc} yields
\begin{equation}
    \frac{d\tilde e}{dw}
    =
    -\frac{2\left(1 - \frac{2}{3}\tilde e\right)}
    {(w+1/3)(w+1)}.
\end{equation}
We can solve this equation to find
\begin{equation}
    \tilde e(w)
    =
    \frac{2w}{(w+1)^2}
    +
    \frac{4}{3(w+1)^2}
    +
    c_1 \frac{(3w+1)^2}{(w+1)^2},
\end{equation}
and at the surface $w=0$, we have
\begin{equation}
    \tilde e(0) = \frac{4}{3} + c_1.
\end{equation}
The integration constant $c_1$ is fixed by the
regular-center condition, $\tilde e(w_0)=0$,
where $w_0$ is the central value, which forces $c_1<0$.
The largest attainable surface value occurs in
the limit $w_0 \to \infty$, for which $c_1 \to 0$.
Therefore,
\begin{equation}
    \tilde e(0) = \frac{4}{3},
    \qquad
    v(0) = \frac{\tilde e(0)}{3} = \frac{4}{9},
\end{equation}
and using the definition of $v$,
\begin{equation}
    v(0)
    =
    \frac{G M}{R c^2}
    =
    \frac{4}{9}.
\end{equation}
We then arrive at the expected Buchdahl bound:
\begin{equation}
    \frac{2GM}{R c^2}
    =
    \frac{8}{9}.
\end{equation}

Finally, consider the dimensionless
combination appearing throughout this paper: $M \sqrt{4\pi e}$. Using $\tilde e = \frac{4\pi G}{c^4} R^2 e$ at the surface and $v = GM/(Rc^2)$,
we obtain
\begin{equation}
    M \sqrt{\frac{4\pi G}{c^4} e}
    =
    \sqrt{\tilde e(0)}\, v(0)
    =
    \sqrt{\frac{4}{3}} \times \frac{4}{9}
    =
    \frac{8}{9\sqrt{3}}.
\end{equation}
This is not a strict upper bound on
$M \sqrt{4\pi G e_c}/c^2$ for astrophysical
sequences (cf.\ Fig.~\ref{fig:wc-vs-mmax}), but it represents
the incompressible relativistic limit.
\new{
\section{Details of the Post-Newtonian Expansion}
\label{app:post-newtonian-stability-critieria}
In Sec.~\ref{sec:scaling-of-mmax-newtonian} we identified a natural approximate criteria for instability $w_c= \ln h_c / 4$ in Newtonian stars by examining polytropic solutions.  In post-Newtonian stars this criteria is modified.  This fact was first pointed out by Chandrasekhar, who showed that in the post-Newtonian regime, instability (or marginal stability) occurs when
\begin{equation}
    \langle\Gamma\rangle = 4/3 + k\frac{M}{R}
\end{equation}
where $\langle\rangle$ is an average defined by 
\begin{align}
    \langle f \rangle = \frac{\int f \, p \, r^2 \, dr}{\int p \, r^2 \, dr}\,.    
\end{align}  
While the system we introduce here is not well-suited for perturbative calculations (which are necessary to compute the stability of a given configuration), we can show the effect of post-Newtonian corrections on configurations near $n=3$ in our approach.  

We will demonstrate that to leading post-Newtonian order, and to next-to-leading Taylor expansion order, the post-Newtonian calculations are equivalent to a shift of the polytropic index $n$.   We start from the post-Newtonian equations for $\bar u$ and $v$, which can be derived from Eq.\eqref{eq:tov-lindblom-dimful}, with the change of variables in Eqs.~\eqref{eq:lindblom-newtonian-reformulated-u} \& \eqref{eq:lindblom-newtonian-reformulated-v}, namely
\begin{align}
    \deriv{\bar u}{s} &= \frac{2\bar u}{v}\quant{1- (2v + w\bar e \frac{\bar u}{v}) } \\
    \deriv{v}{s} &= \quant{\frac{\bar e\bar u}{v}-1}\quant{1- (2v + w\bar e \frac{\bar u}{v}) }. 
\end{align}
We then insert our approximate solution at leading post-Newtonian order from
Eqs.~\eqref{eq:compressible-ebar-core} \& \eqref{eq:compressible-core-v} and compute the solutions for $\bar u$ and $v$ at second order in $s$.  We evaluate them at $\hat s = s/\ln h_c$ as in Sec.~\ref{sec:compressible-limit} to find 
\begin{align}
    \bar u (\hat s)/\ln h_c &= 6(1-3w_c) \hat s+ \quant{\frac{9n(1-w_c)}{5} - 12\ln h_c }\hat s^2\\
    v (\hat s)/\ln h_c &= 2(1-3w_c)\hat s  + \quant{-\frac{3n(1-w_c)}{5} - 4\ln h_c }\hat s^2.
\end{align}
Taking $n=3 - \epsilon$ and evaluating at the stitching point $s_0$, assuming that $w_c = \ln h_c/(n+1)$, we find
\begin{equation}
    v_{\rm stitch}/\ln h_c \approx 0.55 - 1.57 \ln h_c + 0.22 \epsilon.
\end{equation}
It is thus clear that, to maintain the same compactness, $\epsilon \propto\ln h_c$ is necessary, which implies $\epsilon\propto GM/(Rc^2)$.  Alternatively, if the EoS is not changed, all macroscopic observables will receive corrections of order $\ln h_c$. 
}

\section{When the Universal Relations Fail: Strong Phase Transitions}
\label{app:when-universal-relations-fail}

For the Newtonian variables of Sec.~\ref{sec:newtonian-falling-sound-speeds}, the physical mass can be written in terms of the dimensionless combination $\bar M \equiv v(0)\sqrt{\bar u(0)}$ [Eq.~\eqref{eq:Mbar-def}] as
\begin{equation}
    M
    =
    \frac{c^4}{G}\,
    \frac{\sqrt{\bar u(0)}\,v(0)}{\sqrt{4\pi G\,e_c}}
    \;=\;
    \frac{c^4}{G}\,\frac{\bar M}{\sqrt{4\pi G\,e_c}}.
    \label{eq:M_from_Mbar}
\end{equation}
Differentiating Eq.~\eqref{eq:M_from_Mbar} with respect to the central enthalpy parameter $\ln h_c$ gives
\begin{align}
    \deriv{M}{\ln h_c}
    &=
    \frac{c^4}{G}\,\frac{1}{\sqrt{4\pi G\,e_c}}\,
    \deriv{}{\ln h_c}\!\left[\sqrt{\bar u(0)}\,v(0)\right]
    \nonumber \\
    &-\frac{c^4}{G}\,\frac{\sqrt{\bar u(0)}\,v(0)}{2\sqrt{4\pi G\,e_c}}\,
    \deriv{\ln e_c}{\ln h_c}.
    \label{eq:mass-differential}
\end{align}
Thermodynamics, again, gives $d{\ln e}/d{\ln h}=({1+w})/{c_s^2}$, so at the center $\displaystyle d\ln e_c/d\ln h_c = (1+w_c)/c_{s,c}^2$, which reduces to $1/c_{s,c}^2$ at leading Newtonian order. The first term in Eq.~\eqref{eq:mass-differential} is controlled by the change in the \emph{dimensionless} stellar profile along the sequence.  The second term is controlled purely by how rapidly the central density changes with $\ln h_c$, i.e.\ by the local compressibility through $1/c_{s,c}^2$.

If the sound speed drops to (nearly) zero in the core, then $1/c_{s,c}^2$ becomes very large and the second term dominates.  In that regime, the mass cannot increase with increasing $\ln h_c$ regardless of the detailed structure encoded in the first term, so the sequence can terminate (or plateau) ``abruptly.''  More refined quantitative conditions for the phase transition needed to terminate a given sequence can be read off directly from Eq.~\eqref{eq:mass-differential}, but for present purposes the main point is that the relevant information is already contained there.

\bibliography{References}
\end{document}